 \documentclass[preprint,12pt]{elsarticle}
\usepackage{amssymb,amsmath,latexsym}
\usepackage[dvips]{color}
\usepackage[headings]{fullpage}
\usepackage{epic,epsfig,graphicx}
\usepackage{enumerate}
\usepackage{multirow}
\usepackage{setspace}
\usepackage{keyval}
\usepackage{float}

\usepackage{amssymb}

 \newtheorem{thm}{Theorem}[section]

 \newtheorem{lemma}{Lemma}[section]

 \newtheorem{rem}{Remark}[section]
  \newtheorem{assumption}{Assumption}[section]

 \newtheorem{definition}{Definition}[section]
 \newtheorem{hypothesis}{Hypothesis}[section]
 \numberwithin{equation}{section}
\begin{document}
\begin{frontmatter}
\title{Analyzing the qualitative properties of white noise on a family of infectious disease models in a highly random environment}
\author{Divine Wanduku }
\address{Department of Mathematical Sciences,
Georgia Southern University, 65 Georgia Ave, Room 3042, Statesboro,
Georgia, 30460, U.S.A. E-mail:dwanduku@georgiasouthern.edu;wandukudivine@yahoo.com\footnote{Corresponding author. Tel: +14073009605.
} }
\begin{abstract}
 A class of stochastic vector-borne infectious disease models is derived and studied. The class type is determined by a general nonlinear incidence rate of the disease. The disease spreads in a highly random environment with variability from the disease transmission and natural death rates. Other sources of variability include the random delays of disease incubation inside the vector and the human being, and also the random delay due to the period of effective acquired immunity against the disease. The  basic reproduction number and other threshold conditions for disease eradication are computed. The qualitative behaviors of the disease dynamics are examined under the different  sources of variability in the system. A technique to classify the different levels of the intensities of the noises in the system is presented, and used to investigate the qualitative behaviors of the disease dynamics in the infection-free steady state population under the different intensity levels of the white noises in the system. Moreover, the possibility of population extinction, whenever the intensities of the white noises in the system are high is examined.  Numerical simulation results are presented to elucidate the theoretical results.
\end{abstract}

\end{frontmatter}
\section{Introduction\label{ch1.sec0}}
In the general class of infectious diseases, vector-borne diseases exhibit several unique biological characteristics. For instance, the incubation of the disease requires two hosts - the vector and human hosts,  which may be either  directly involved in a full life cycle of the infectious agent consisting of two separate and independent segments of sub-life cycles which are completed separately inside the two hosts, or directly involved in two separate and independent half-life cycles of the infectious agent in the hosts. Therefore, there exists a total latent time lapse of disease incubation which extends over the two segments of delay incubation times namely:- (1) the incubation period of the infectious agent ( or the half-life cycle) inside the vector, and (2) the incubation period of the infectious agent (or the other half-life cycle) inside the human being. For example, the dengue fever virus transmitted primarily by the \textit{Aedes aegypti and Aedes albopictus} mosquitos undergoes two delay incubation periods:- (1) about 8-12 days incubation period inside the female mosquito vector, which starts immediately after the ingestion of a dengue fever virus infected blood meal, which is successfully taken from a dengue fever infectious human being via a mosquito bite, and (2) another delay incubation period of about 2-7 days in the human being when the hosting female infectious vector  bites a susceptible human being, whereby the virus is successfully transmitted  from the infectious mosquito to the susceptible person. See \cite{WHO,CDC}.

  Indeed, for dengue fever transmission, a susceptible vector acquires infected blood meal from a dengue fever infectious person via a mosquito bite. The virus incubates in the mosquito for about 8-12 days, and at the end of the first incubation period the exposed mosquito becomes infectious. The virus is transferred to a susceptible human being after another successful mosquito bite, and it subsequently undergoes a second incubation phase in the exposed human being. The second incubation phase is mostly  a viremia phase that involves the complete circulation of the virus in the human blood stream, and at the end of the phase the exposed person develops  full blown fever. See \cite{WHO,CDC}.

   While the infectious dengue fever vector is known to stay infectious for the rest of the life span, it is important to note in the modelling of the vector-borne disease dynamics that no relationship between vector survival and viral invasion of the mosquito has been determined\cite{WHO,CDC}.

 For the vector-borne disease, malaria, the parasite undergoes a first half-life cycle called the \textit{sporogonic cycle} in the female \textit{Anopheles} mosquito lasting approximately $10-18$ days after the first successful  mosquito bite from a malaria infectious person. The parasite  further completes the remaining half-life cycle called the \textit{exo-erythrocytic cycle} lasting about 7-30 days inside the exposed human being\cite{WHO,CDC}, whenever the parasite is transferred to a susceptible person after another successful infectious mosquito bite by the hosting female mosquito.

 Several vector-borne diseases  induce or confer natural immunity against the disease after infection and recovery. The effectiveness and duration of the natural protective immunity varies depending on the type of disease and also on other biological factors. For example, the exposure and successful recovery from one dengue fever viral strain confers lifelong immunity against the particular viral serotype\cite{WHO}.

Also, the exposure and successful recovery from a malaria parasite, for example, \textit{falciparum vivae} induces natural immunity against the disease which can protect against subsequent severe outbreaks of the disease. Moreover, the effectiveness and duration of the naturally acquired immunity against malaria is determined by several factors such as the  species and the frequency of  exposure to the parasites. Furthermore, it has been determined that  the naturally acquired immunity against malaria  has bearings on   other biological factors such as the genetic characteristics of the human being\cite{CDC,lars,denise}.

 Many infectious diseases have been investigated utilizing compartmental mathematical epidemic dynamic models. For instance, malaria and dengue fever are studied in \cite{eric,sya}, and measles is studied in \cite{pang}.  In general, these models are largely classified as SIS, SIR, SIRS, SEIRS,  and  SEIR etc.\cite{qun,qunliu, nguyen,joaq, sena,wanduku-fundamental,Wanduku-2017, zhica} epidemic dynamic models depending on the compartments of the disease classes directly involved in the general disease dynamics.

  Several studies devote interest to SEIRS and SEIR
 models\cite{joaq,sena,cesar,sen,zhica} which allow the inclusion of the compartment of individuals who are exposed to the disease, $E$, that is, infected but noninfectious individuals. This natural inclusion of the exposed class of individuals allows for more insight about the disease dynamics during the incubation stage of the disease. For example, the existence of periodic solutions are investigated in the SEIRS epidemic study\cite{joaq,zhica}. And the effects of seasonal changes on the disease dynamics are investigated in the SEIRS epidemic study in \cite{zheng}.

 Many epidemic dynamic models are modified and  improved in reality by including the time delays that occur in the disease dynamics. Generally, two distinct classes of delays are studied namely:-disease latency and immunity delay. The disease latency has been represented as the infected but noninfectious period of disease incubation, and also as the period of infectiousness which nonetheless is studied as a delay in the dynamics of the disease. The immunity delay represents the period of effective naturally acquired immunity against the disease after exposure and successful recovery from infection.  Whereas, some authors  study diseases and disease scenarios under the realistic assumption of one form of these two classes of delays in the disease dynamics\cite{Wanduku-2017,wanduku-delay,kyrychko,qun}, other authors study one or more forms of the classes of delays  represented as two separate  delay times\cite{zhica,cooke-driessche,shuj,Sampath}. The occurrence of delays in the disease dynamics  may influence the dynamics of the disease in many important  ways. For instance, in \cite{zhica}, the presence of delays in the epidemic dynamic system leads to the existence of periodic solutions. In \cite{cooke, baretta-takeuchi1}, the occurrence of a delay in the vector-borne disease dynamics  destabilizes the equilibrium population state of the system.

 Stochastic epidemic dynamic models more realistically represent epidemic dynamic processes because they include the randomness which naturally occurs during a disease outbreak, owing to the presence of constant random environmental fluctuations in the disease dynamics. The presence of stochastic white noise process in the epidemic dynamic system may directly impact the density of the system or indirectly influence other driving parameters of the infectious system such as the disease transmission, natural death, birth and disease related death rates etc. In \cite{Wanduku-2017,wanduku-fundamental,wanduku-delay},  the stochastic white noise process represents the random fluctuations in the disease transmission process. In \cite{qun}, the white noise process represents the variability in the natural death rate of the population. In \cite{Baretta-kolmanovskii},  the white noise process represents the random fluctuations in the system which deviate the state of the system from the equilibrium state, that is, the  white noise process is proportional to the difference between the state and equilibrium of the system.

  A stochastic white noise process driven infectious system generally exhibits  more complex behavior in the disease dynamics, than would be observed in the corresponding deterministic system. For instance, the presence of stochastic white noise process in the disease dynamics  may destabilize a disease free steady state population and drive the system into an endemic state. In other  cases, the presence of white noise with high intensity in the disease dynamics may continuously  decrease the population size over time,  and eventually  lead to the extinction of the population.   For example, in \cite{qun,Wanduku-2017,wanduku-fundamental,wanduku-delay,zhuhu,yanli}, the occurrence of stochastic noise in the system destabilizes the disease free steady population state. In \cite{qun}, it is observed that the disease free steady state fails to exist when the intensity of the noise in the system from the natural death rate of the susceptible population is high.

 The interaction between susceptible, $S$, and infectious individuals, $I$, during the disease transmission process  can sometimes generate complex nonlinear responses from the susceptible population as the infectious population increases. Such complex nonlinear responses can no longer be represented by the  the frequently used bilinear incidence rate (or force of infection) given as $\beta S(t)I(t-T)$ for vector-borne diseases, or $\beta S(t)I(t)$, for infectious diseases that involve direct human-to-human disease transmission, where $\beta$ is the effective contact rate, and $T$ is the incubation  period for the vector-borne disease. Some examples of nonlinear complex responses from the susceptible population include- psychological or crowding effects stemming from behavioral change of susceptible individuals when the infectious population increases significantly over time.  These nonlinear response behaviors exist for certain types of infectious diseases and disease scenarios,  where the contact between the susceptible and infectious classes are regulated, and consequently  prevent unboundedness in the disease transmission rate.

 For instance, in  \cite{yakui,xiao,huo,kyrychko,qun,muroya,liu,capasso-serio,capasso,hethcote,koro} several different functional forms for the force of infection are used to represent the nonlinear behavior of the disease transmission rate. In \cite{yakui,xiao,capasso-serio,huo} the authors consider a Holling Type II functional form, $\beta S(t)G(I(t))=\frac{\beta S(t)I(t)}{1+\alpha I(t)}$, that saturates for large values of $I$. In \cite{muroya,xiao, capasso}, a bounded  Holling Type II function,  $\beta S(t)G(I(t))=\frac{\beta S(t)I^{p}(t)}{1+\alpha I^{p}(t)}, p\geq 0$,  is used to represent the force of infection of the disease. In \cite{hethcote,koro}, the nonlinear incidence rate is represented by the general functional form, $\beta S(t)G(I(t))=\beta S^{p}(t)I^{q}(t), p,q\geq 0$. And the authors in \cite{yakui,huo, capasso-serio, muroya,capasso, qun} studied vector-borne diseases with several different functional forms for the nonlinear incidence rates of the disease.

Cooke\cite{cooke} presented a deterministic epidemic dynamic model for  vector-borne diseases, where the bilinear incidence rate defined as $\beta S(t)I(t-T)$ represents the number of new infections occurring per unit time during the disease transmission process. It is assumed in the formulation of this incidence rate that the number of infectious vectors at time $t$ interacting and effectively transmitting infection to susceptible individuals, $S$, after $\beta$  number of effective contacts per unit time per infective is proportional to the infectious human population, $I$,  at earlier time $t-T$.
%

 This paper employs similar reasoning in \cite{cooke,shuj}, to derive a class of SEIRS stochastic epidemic dynamic models with three delays for vector-borne diseases. The three delays are classified under the two general forms-disease latency and immunity delay. Two of the delays represent the incubation period of the infectious agent inside the vector  and human hosts, and the third delay represents the period of effective naturally acquired immunity against the vector-borne disease conferred after recovery from infection. Furthermore, the delays are random variables. In addition, the general vector-borne disease dynamics is  driven by stochastic white noise processes originating from the random environmental fluctuations in  the natural death and disease transmission rates in the population. The epidemic dynamic model is represented as a system of Ito-Doob type stochastic differential equations.

 It is important to note that this study is part of the broader project investigating vector-borne diseases in the human population. As part of this project, a deterministic study of malaria will appear in  Wanduku\cite{wanduku-biomath}. Some specialized stochastic extensions of this project addressing the impacts of noise on the persistence of malaria in the endemic equilibrium population will appear in Wanduku\cite{wanduku-comparative}. Moreover, the stochastic permanence of malaria and existence of stationary distribution will appear in Wanduku\cite{wanduku-permanence}. The primary focus of the current study is to develop and study the fundamental properties of the class of stochastic models for vector-borne diseases in a very noisy environment comprising of variability from the disease transmission and natural death rates. In this study, the intensities of the noises in the infectious system are classified, and their qualitative impacts on the eradication of the vector-borne disease in the steady state population are characterized.

   This work is presented as follows:- In section~\ref{ch1.sec0}, the epidemic dynamic model is derived. In section~\ref{ch1.sec1}, the model validation results are presented. In section~\ref{ch1.sec2}, the existence and asymptotic stochastic stability of the disease free equilibrium population is investigated. In Section~\ref{ch1.sec2-2}, the asymptotic behavior of the stochastic system under the influence of the various intensity levels of the white noise processes in the system is characterized. In section~\ref{ch1.sec4},  numerical simulation results are given.
\section{Derivation of Model}\label{ch1.sec0}
 A generalized class of stochastic SEIRS delayed epidemic dynamic models for  vector-borne diseases is presented. The delays represent the incubation period of the infectious agents in the vector $T_{1}$, and in the human host $T_{2}$. The third delay represents the naturally acquired immunity period of the disease $T_{3}$, where the delays are random variables with density functions $f_{T_{1}}, t_{0}\leq T_{1}\leq h_{1}, h_{1}>0$, and $f_{T_{2}}, t_{0}\leq T_{2}\leq h_{2}, h_{2}>0$ and $f_{T_{3}}, t_{0}\leq T_{3}<\infty$. Furthermore, the joint density of $T_{1}$ and $T_{2}$ is given by $f_{T_{1},T_{2}}, t_{0}\leq T_{1}\leq h_{1} , t_{0}\leq T_{2}\leq h_{2}$. Moreover, it is assumed that the random variables $T_{1}$ and $T_{2}$ are independent (i.e. $f_{T_{1},T_{2}}=f_{T_{1}}.f_{T_{2}}, t_{0}\leq T_{1}\leq h_{1} , t_{0}\leq T_{2}\leq h_{2}$). Indeed, the independence between $T_{1}$ and $T_{2}$ is justified from the understanding that the duration of incubation of  the infectious agent for the vector-borne disease depends only on the suitable  biological environmental requirements for incubation inside the vector and the human body which are unrelated. Furthermore, the independence between $T_{1}$ and $T_{3}$ follows from the lack of any real biological evidence to justify the interconnection between the incubation of  the infectious agent inside the vector and the acquired natural immunity conferred to the human being. But $T_{2}$ and $T_{3}$ may be dependent as biological evidence suggests that the naturally acquired immunity is induced by exposure to the infectious agent.

By employing similar reasoning in \cite{cooke,qun,capasso,huo}, the expected incidence rate of the disease or force of infection of the disease at time $t$ due to the disease transmission process between the infectious vectors and susceptible humans, $S(t)$, is given by the expression $\beta \int^{h_{1}}_{t_{0}}f_{T_{1}}(s) e^{-\mu s}S(t)G(I(t-s))ds$, where $\mu$ is the natural death rate of individuals in the population, and it is assumed for simplicity that the natural death rate for the vectors and human beings are the same. Assuming exponential lifetime for the random incubation period $T_{1}$, the probability rate, $0<e^{-\mu s}\leq 1, s\in [t_{0}, h_{1}], h_{1}>0$,  represents the survival probability rate of  exposed vectors over the incubation period, $T_{1}$, of the infectious agent inside the vectors with the length of the period given as $T_{1}=s, \forall s \in [t_{0}, h_{1}]$, where the vectors acquired infection at the earlier time $t-s$  from an infectious human via for instance, biting and collecting an infected blood meal, and  become infectious at time $t$.  Furthermore, it is assumed that the survival of the vectors over the incubation period of length  $s\in [t_{0}, h_{1}]$ is independent of the age of the vectors. In addition, $I(t-s)$, is the infectious human population at earlier time $t-s$, $G$ is a nonlinear incidence function for the disease dynamics,  and $\beta$ is the average number of effective contacts per infectious individual per unit time. Indeed, the force of infection,  $\beta \int^{h_{1}}_{t_{0}}f_{T_{1}}(s) e^{-\mu s}S(t)G(I(t-s))ds$  signifies the expected rate of new infections at time $t$ between the infectious vectors and the susceptible human population $S(t)$ at time $t$, where the infectious agent is transmitted per infectious vector per unit time at the rate $\beta$. Furthermore, it is assumed that the number of infectious vectors at time $t$ is proportional to the infectious human population at earlier time $t-s$. Moreover, it is further assumed that the interaction between the infectious vectors and  susceptible humans exhibits nonlinear behavior, for instance, psychological and overcrowding effects,  which is characterized by the nonlinear incidence  function $G$. Therefore, the force of infection given by
 \begin{equation}\label{ch1.sec0.eqn0}
   \beta \int^{h_{1}}_{t_{0}}f_{T_{1}}(s) e^{-\mu s}S(t)G(I(t-s))ds,
 \end{equation}
  represents the expected rate at which infected individuals leave the susceptible state and become exposed at time $t$.

\begin{figure}[H]
\includegraphics[height=8cm]{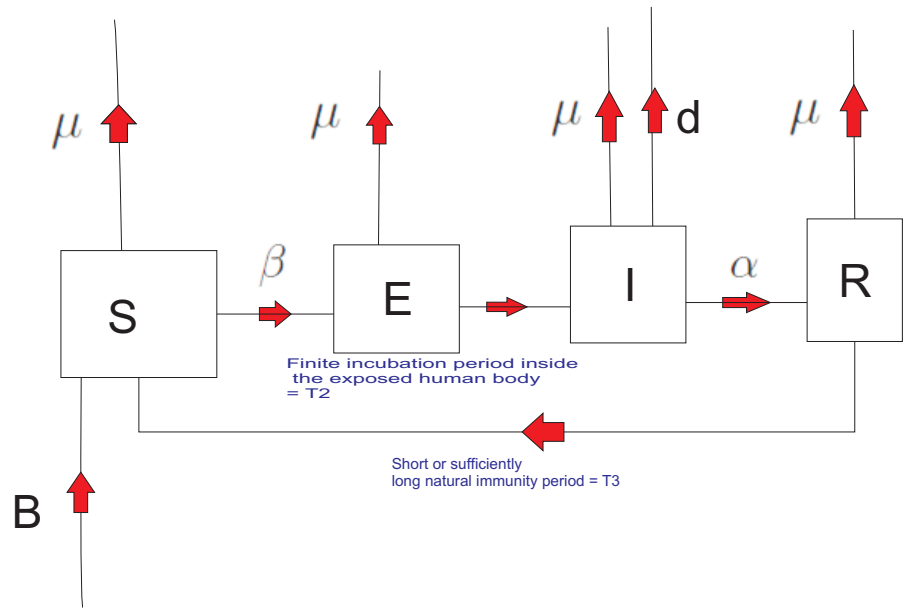}
\caption{The compartmental framework illustrates the transition rates between the states $S,E,I,R$ of the system. It also shows the incubation delay $T_{2}$ and the naturally acquired immunity $T_{3}$ periods. \label{ch1.sec4.figure 1}}
\end{figure}

 The susceptible individuals who have acquired infection from infectious vectors but are non infectious form the exposed class $E$. The population of exposed individuals at time $t$ is denoted $E(t)$. After the incubation period, $T_{2}=u\in [t_{0}, h_{2}]$, of the infectious agent in the exposed human host, the individual becomes infectious, $I(t)$, at time $t$. Applying similar reasoning in  \cite{cooke-driessche},
 the exposed population, $E(t)$, at time $t$ can be written as follows
  \begin{equation}\label{ch1.sec0.eqn1a}
    E(t)=E(t_{0})e^{-\mu (t-t_{0})}p_{1}(t-t_{0})+\int^{t}_{t_{0}}\beta S(\xi)e^{-\mu T_{1}}G(I(\xi-T_{1}))e^{-\mu(t-\xi)}p_{1}(t-\xi)d \xi,
   \end{equation}
   where
   \begin{equation}\label{ch1.seco.eqn1b}
     p_{1}(t)=\left\{\begin{array}{l}0,t\geq T_{2},\\
 1, t< T_{2} \end{array}\right.
   \end{equation}
   represents the probability that an individual remains exposed over the time interval $[0, t]$.
   It is easy to see from (\ref{ch1.sec0.eqn1a}) that under the assumption that the disease has been in the population for at least a time $t>\max_{t_{0}\leq T_{1}\leq h_{1}, t_{0}\leq T_{2}\leq h_{2}} {( T_{1}+ T_{2})}$, in fact, $t>h_{1}+h_{2}$, so that all initial perturbations have died out, the expected number of exposed individuals at time $t$ is given  by
\begin{equation}\label{ch1.sec0.eqn1}
E(t)=\int_{t_{0}}^{h_{2}}f_{T_{2}}(u)\int_{t-u}^{t}\beta \int^{h_{1}}_{t_{0}} f_{T_{1}}(s) e^{-\mu s}S(v)G(I(v-s))e^{-\mu(t-u)}dsdvdu.
\end{equation}
    Similarly, for the removal population, $R(t)$, at time $t$, individuals recover from the infectious state $I(t)$  at the per capita rate $\alpha$  and acquire natural immunity.  The natural immunity wanes after the varying immunity period $T_{3}=r\in [ t_{0},\infty]$, and removed individuals become susceptible again to the disease. Therefore, at time $t$, individuals leave the infectious state at the rate $\alpha I(t)$  and become part of the removal population $R(t)$. Thus, at time $t$ the removed population is given by the following equation
  \begin{equation}\label{ch1.sec0.eqn2a}
    R(t)=R(t_{0})e^{-\mu (t-t_{0})}p_{2}(t-t_{0})+\int^{t}_{t_{0}}\alpha I(\xi)e^{-\mu(t-\xi)}p_{2}(t-\xi)d \xi,
  \end{equation}
    where
    \begin{equation}\label{ch1.sec0.eqn2b}
      p_{2}(t)=\left\{\begin{array}{l}0,t\geq T_{3},\\
 1, t< T_{3} \end{array}\right.
    \end{equation}
 represents the probability that an individual remains naturally immune to the disease over the time interval $[0, t]$.
 But it follows from  (\ref{ch1.sec0.eqn2a}) that under the assumption that the disease has been in the population for at least a time $t> \max_{t_{0}\leq T_{1}\leq h_{1}, t_{0}\leq T_{2}\leq h_{2}, T_{3}\geq t_{0}}{(T_{1}+ T_{2}, T_{3})}\geq \max_{ T_{3}\geq t_{0}}{(T_{3})}$, in fact, the disease has been in the population for sufficiently large  amount of time so that all initial perturbations have died out,  then the expected number of removal individuals at time $t$ can be written as
  \begin{equation}\label{ch1.sec0.eqn2}
R(t)=\int_{t_{0}}^{\infty}f_{T_{3}}(r)\int_{t-r}^{t}\alpha I(v)e^{-\mu (t-v)}dvdr.
\end{equation}
There is also constant birth rate $B$ of susceptible individuals in the population. Furthermore, individuals die additionally due to disease related causes at the rate $d$. A compartmental framework illustrating the transition rates between the different states in the system and also showing the delays in the disease dynamics is given in Figure~\ref{ch1.sec4.figure 1}.
It follows from (\ref{ch1.sec0.eqn0}), (\ref{ch1.sec0.eqn1}), (\ref{ch1.sec0.eqn2}) and the transition rates illustrated in the compartmental framework in Figure~\ref{ch1.sec4.figure 1} above that the family of SEIRS epidemic dynamic models for a vector-borne diseases in the absence of any random environmental fluctuations can be written as follows:
\begin{eqnarray}
dS(t)&=&\left[ B-\beta S(t)\int^{h_{1}}_{t_{0}}f_{T_{1}}(s) e^{-\mu s}G(I(t-s))ds - \mu S(t)+ \alpha \int_{t_{0}}^{\infty}f_{T_{3}}(r)I(t-r)e^{-\mu r}dr \right]dt,\nonumber\\
&&\label{ch1.sec0.eq3}\\
dE(t)&=& \left[ \beta S(t)\int^{h_{1}}_{t_{0}}f_{T_{1}}(s) e^{-\mu s}G(I(t-s))ds - \mu E(t)\right.\nonumber\\
&&\left.-\beta \int_{t_{0}}^{h_{2}}f_{T_{2}}(u)S(t-u)\int^{h_{1}}_{t_{0}}f_{T_{1}}(s) e^{-\mu s-\mu u}G(I(t-s-u))dsdu \right]dt,\label{ch1.sec0.eq4}\\
&&\nonumber\\
dI(t)&=& \left[\beta \int_{t_{0}}^{h_{2}}f_{T_{2}}(u)S(t-u)\int^{h_{1}}_{t_{0}}f_{T_{1}}(s) e^{-\mu s-\mu u}G(I(t-s-u))dsdu- (\mu +d+ \alpha) I(t) \right]dt,\nonumber\\
&&\label{ch1.sec0.eq5}\\
dR(t)&=&\left[ \alpha I(t) - \mu R(t)- \alpha \int_{t_{0}}^{\infty}f_{T_{3}}(r)I(t-r)e^{-\mu s}dr \right]dt.\label{ch1.sec0.eq6}
\end{eqnarray}
 It should be noted that the deterministic model (\ref{ch1.sec0.eq3})-(\ref{ch1.sec0.eq6}) with the initial conditions in (\ref{ch1.sec0.eq12})-(\ref{ch1.sec0.eq13}) has been studied in the special case of malaria in Wanduku\cite{wanduku-biomath}.

 It is assumed in the current study that the  effects of random environmental fluctuations lead to variability in the disease transmission and natural death rates.
 For $t\geq t_{0}$, let $(\Omega, \mathfrak{F}, P)$ be a complete probability space, and $\mathfrak{F}_{t}$ be a filtration (that is, sub $\sigma$- algebra $\mathfrak{F}_{t}$ that satisfies the following: given $t_{1}\leq t_{2} \Rightarrow \mathfrak{F}_{t_{1}}\subset \mathfrak{F}_{t_{2}}; E\in \mathfrak{F}_{t}$ and $P(E)=0 \Rightarrow E\in \mathfrak{F}_{0} $ ).
 The variability in the disease transmission and natural death rates are represented by independent white noise processes, and the rates are expressed as follows:
 \begin{equation}\label{ch1.sec0.eq7}
 \mu  \rightarrow \mu  + \sigma_{i}\xi_{i}(t),\quad \xi_{i}(t) dt= dw_{i}(t),i=S,E,I,R, \quad  \beta \rightarrow \beta + \sigma_{\beta}\xi_{\beta}(t),\quad \xi_{\beta}(t)dt=dw_{\beta}(t),
 \end{equation}
 where $\xi_{i}(t)$ and $w_{i}(t)$ represent the  standard white noise and normalized wiener processes  for the  $i^{th}$ state at time $t$, with the following properties: $w(0)=0, E(w(t))=0, var(w(t))=t$.  Furthermore,  $\sigma_{i},i=S,E,I,R $, represents the intensity value of the white noise process due to the natural death rate in the $i^{th}$ state, and $\sigma_{\beta}$ is the intensity value of the white noise process due to the disease transmission rate.

  The ideas behind the formulation of the expressions in (\ref{ch1.sec0.eq7}) are given in the following. The constant parameters $\mu$ and $\beta$ represent the natural death and disease transmission rates per unit time, respectively. In reality, random environmental fluctuations impact these rates turning them into random variables $\tilde{\mu}$ and $\tilde{\beta}$. Thus, the natural death and disease transmission rates over an infinitesimally small interval of time $[t, t+dt]$  with length $dt$ is given by the expressions  $\tilde{\mu}(t)=\tilde{\mu}dt$ and $\tilde{\beta}(t)=\tilde{\beta}dt$, respectively. It is assumed that there are independent and identical random impacts acting upon these rates at times $t_{j+1}$ over $n$ subintervals $[t_{j}, t_{j+1}]$ of length $\triangle t=\frac{dt}{n}$, where $t_{j}=t_{0}+j\triangle t, j=0,1,\cdots,n$, and  $t_{0}=t$. Furthermore,  it is assumed that $\tilde{\mu}(t_{0})=\tilde{\mu}(t)=\mu dt$ is  constant or deterministic, and $\tilde{\beta}(t_{0})=\tilde{\beta}(t)=\beta dt$ is also a constant. It follows that by letting the independent identically distributed random variables $Z_{i},i=1,\cdots,n $ represent the random effects acting on the natural death rate, then it follows further that the rate at time $t_{n}=t+dt$, that is,
   \begin{equation}\label{ch1.sec0.eq7.eq1}
    \tilde{\mu}(t+dt)=\tilde{\mu}(t)+\sum_{j=1}^{n}Z_{j},
  \end{equation}
  where $E(Z_{j})=0$,and $Var(Z_{j})=\sigma^{2}_{i}\triangle t, i\in \{S, E, I, R\}$.
    Note that $\tilde{\beta}(t+dt)$ can similarly be expressed as (\ref{ch1.sec0.eq7.eq1}). And for sufficient large value of $n$, the summation in (\ref{ch1.sec0.eq7.eq1}) converges in distribution by the central limit theorem to a random variable which is identically distributed as the wiener process $\sigma_{i}(w_{i}(t+dt)-w_{i}(t))=\sigma_{i}dw_{i}(t)$, with mean $0$ and variance $\sigma^{2}_{i}dt, i\in \{S, E, I, R\}$. It follows easily from (\ref{ch1.sec0.eq7.eq1}) that
  \begin{equation}\label{ch1.sec0.eq7.eq2}
   \tilde{\mu}dt =\mu dt+ \sigma_{i}dw_{i}(t), i\in \{S, E, I, R\}.
  \end{equation}
  Similarly, it can be easily seen that
  \begin{equation}\label{ch1.sec0.eq7.eq2}
   \tilde{\beta}dt =\beta dt+ \sigma_{\beta}dw_{\beta}(t).
  \end{equation}
    Note that the intensities $\sigma^{2}_{i},i=S,E,I,R, \beta $ of the independent white noise processes in the expressions   $\tilde{\mu}(t)=\mu dt  + \sigma_{i}\xi_{i}(t)$ and $\tilde{\beta} (t)=\beta dt + \sigma_{\beta}\xi_{\beta}(t)$ that represent the  natural death rate, $\tilde{\mu}(t)$, and disease transmission rate, $\tilde{\beta} (t)$,  at time $t$,  measure the average deviation of the random variable disease transmission rate, $\tilde{\beta}$,  and natural death rate, $\tilde{\mu}$, about their constant mean values $\beta$ and $\mu$, respectively, over the infinitesimally small time interval $[t, t+dt]$. These measures reflect the force of the random fluctuations that occur during the disease outbreak at anytime, and which lead to oscillations in the natural death and disease transmission rates overtime, and consequently lead to oscillations of the susceptible, exposed, infectious and removal states of the system over time during the disease outbreak. Thus, in this study the words "strength" and "intensity" of the white noise are used synonymously. Also, the constructions "strong noise" and "weak noise" are used to refer to white noise with high and low intensities, respectively.

    Under the assumptions in the formulation of the natural death rate per unit time $\tilde{\mu}$ as a brownian motion process above, it can also be seen  easily that under further assumption that the number of natural deaths $N(t)$ over an interval $[t_{0}, t_{0}+t]$ of length $t$ follows a poisson process $\{N(t),t\geq t_{0}\}$ with intensity of the process $E(\tilde{\mu})=\mu$, and mean $E(N(t))=E(\tilde{\mu}t)=\mu t$, then the lifetime is exponentially distributed with mean $\frac{1}{\mu}$ and survival function
    \begin{equation}\label{ch1.sec0.eq7.eq3}
      \mathfrak{S}(t)=e^{-\mu t},t>0.
    \end{equation}
 Substituting (\ref{ch1.sec0.eq7})-(\ref{ch1.sec0.eq7.eq3}) into the deterministic system (\ref{ch1.sec0.eq3})-(\ref{ch1.sec0.eq6}) leads to the following generalized system of Ito-Doob stochastic differential equations describing the dynamics of  vector-borne diseases in the human population.
 \begin{eqnarray}
dS(t)&=&\left[ B-\beta S(t)\int^{h_{1}}_{t_{0}}f_{T_{1}}(s) e^{-\mu s}G(I(t-s))ds - \mu S(t)+ \alpha \int_{t_{0}}^{\infty}f_{T_{3}}(r)I(t-r)e^{-\mu r}dr \right]dt\nonumber\\
&&-\sigma_{S}S(t)dw_{S}(t)-\sigma_{\beta} S(t)\int^{h_{1}}_{t_{0}}f_{T_{1}}(s) e^{-\mu s}G(I(t-s))dsdw_{\beta}(t) \label{ch1.sec0.eq8}\\
dE(t)&=& \left[ \beta S(t)\int^{h_{1}}_{t_{0}}f_{T_{1}}(s) e^{-\mu s}G(I(t-s))ds - \mu E(t)\right.\nonumber\\
&&\left.-\beta \int_{t_{0}}^{h_{2}}f_{T_{2}}(u)S(t-u)\int^{h_{1}}_{t_{0}}f_{T_{1}}(s) e^{-\mu s-\mu u}G(I(t-s-u))dsdu \right]dt\nonumber\\
&&-\sigma_{E}E(t)dw_{E}(t)+\sigma_{\beta} S(t)\int^{h_{1}}_{t_{0}}f_{T_{1}}(s) e^{-\mu s}G(I(t-s))dsdw_{\beta}(t)\nonumber\\
&&-\sigma_{\beta} \int_{t_{0}}^{h_{2}}f_{T_{2}}(u)S(t-u)\int^{h_{1}}_{t_{0}}f_{T_{1}}(s) e^{-\mu s-\mu u}G(I(t-s-u))dsdudw_{\beta}(t)\label{ch1.sec0.eq9}\\
dI(t)&=& \left[\beta \int_{t_{0}}^{h_{2}}f_{T_{2}}(u)S(t-u)\int^{h_{1}}_{t_{0}}f_{T_{1}}(s) e^{-\mu s-\mu u}G(I(t-s-u))dsdu- (\mu +d+ \alpha) I(t) \right]dt\nonumber\\
&&-\sigma_{I}I(t)dw_{I}(t)+\sigma_{\beta} \int_{t_{0}}^{h_{2}}f_{T_{2}}(u)S(t-u)\int^{h_{1}}_{t_{0}}f_{T_{1}}(s) e^{-\mu s-\mu u}G(I(t-s-u))dsdudw_{\beta}(t)\nonumber\\
&&\label{ch1.sec0.eq10}\\
dR(t)&=&\left[ \alpha I(t) - \mu R(t)- \alpha \int_{t_{0}}^{\infty}f_{T_{3}}(r)I(t-r)e^{-\mu s}dr \right]dt-\sigma_{R}R(t)dw_{R}(t),\label{ch1.sec0.eq11}
\end{eqnarray}
where the initial conditions are given in the following:- where ever necessary, we  let $h= h_{1}+ h_{2}$ and define
\begin{eqnarray}
&&\left(S(t),E(t), I(t), R(t)\right)
=\left(\varphi_{1}(t),\varphi_{2}(t), \varphi_{3}(t),\varphi_{4}(t)\right), t\in (-\infty,t_{0}],\nonumber\\
&&\varphi_{k}\in \mathcal{C}((-\infty,t_{0}],\mathbb{R}_{+}),\forall k=1,2,3,4, \nonumber\\
&&\varphi_{k}(t_{0})>0,\forall k=1,2,3,4,\nonumber\\
 \label{ch1.sec0.eq12}
\end{eqnarray}
where $\mathcal{C}((-\infty,t_{0}],\mathbb{R}_{+})$ is the space of continuous functions with  the supremum norm
\begin{equation}\label{ch1.sec0.eq13}
||\varphi||_{\infty}=\sup_{ t\leq t_{0}}{|\varphi(t)|}.
\end{equation}
Furthermore, the random continuous functions $\varphi_{k},k=1,2,3,4$ are
$\mathfrak{F}_{0}-measurable$, or  independent of $w(t)$
for all $t\geq t_{0}$.

Several epidemiological studies \cite{gumel,zhica,joaq,kyrychko,qun} have been conducted involving families of SIR, SEIRS, SIS etc. epidemic dynamic  models, where the family type is determined by a set of general assumptions which characterize the nonlinear behavior of the incidence function $G(I)$ of the disease. Some general properties of the incidence function $G$ assumed in this study include the following:
\begin{assumption}\label{ch1.sec0.assum1}
\begin{enumerate}
  \item [$A1$]$G(0)=0$.
  \item [$A2$]$G(I)$ is strictly monotonic on $[0,\infty)$.
  \item [$A3$] $G''(I)<0$ $\Leftrightarrow$ $G(I)$ is differentiable concave on $[0,\infty)$.
  \item [$A4$] $\lim_{I\rightarrow}G(I)=C, 0\leq C<\infty$ $\Leftrightarrow$  $G(I)$ has a horizontal asymptote $0\leq C<\infty$.
  \item [$A5$] $G(I)\leq I, \forall I>0$ $\Leftrightarrow$ $G(I)$ is at most as large as the identity function $f:I\mapsto I$ over the positive all $I\in (0,\infty)$.
\end{enumerate}
\end{assumption}
An incidence function $G$ that satisfies  Assumption~\ref{ch1.sec0.assum1} $A1$-$A5$ can be used to describe the disease transmission process of a vector-borne disease scenario, where the disease dynamics is represented by the system (\ref{ch1.sec0.eq8})-(\ref{ch1.sec0.eq11}), and the disease transmission rate between the vectors and the human beings initially increases or decreases for relatively small values of the infectious population size, and is bounded or steady for sufficiently large size of the infectious  individuals in the population.  It is noted that Assumption~\ref{ch1.sec0.assum1} is a generalization of some subcases of the assumptions $A1$-$A5$ investigated in \cite{gumel,zhica,kyrychko, qun}. Some examples of frequently used incidence functions in the literature that  satisfy Assumption~\ref{ch1.sec0.assum1}$A1$-$A5$ include:  $G(I(t))=\frac{I(t)}{1+\alpha I(t)}, \alpha>0$, $G(I(t))=\frac{I(t)}{1+\alpha I^{2}(t)}, \alpha>0$, $G(I(t))=I^{p}(t),0<p<1$ and $G(I)=1-e^{-aI}, a>0$.

It can be observed that (\ref{ch1.sec0.eq9}) and (\ref{ch1.sec0.eq11}) decouple from the other equations for $S$ and $I$ in the system (\ref{ch1.sec0.eq8})-(\ref{ch1.sec0.eq11}). It is customary to show the results for this kind of decoupled system using the simplified system containing only the non-decoupled system equations for $S$ and $I$, and then infer the results to the states $E$ and $R$, since these states depend exclusively on $S$ and $I$.  Nevertheless, for convenience, the existence and stability results of the system (\ref{ch1.sec0.eq8})-(\ref{ch1.sec0.eq11}) will be shown for the vector $X(t)=(S(t), E(t), I(t))$. The following notations will be used throughout this study:
\begin{equation}\label{ch1.sec0.eq13b}
\left\{
  \begin{array}{lll}
    Y(t)&=&(S(t), E(t), I(t), R(t)) \\
   X(t)&=&(S(t), E(t), I(t)) \\
   N(t)&=&S(t)+ E(t)+ I(t)+ R(t).
  \end{array}
  \right.
\end{equation}
\section{Model Validation Results\label{ch1.sec1}}
In this section, the existence and uniqueness results for the solutions of the stochastic system  (\ref{ch1.sec0.eq8})-(\ref{ch1.sec0.eq11}) are presented.  The standard method  utilized in the earlier studies\cite{Wanduku-2017,wanduku-delay,divine5} is applied to establish the results.
  It should be noted that the existence and qualitative behavior of the positive solutions of the system (\ref{ch1.sec0.eq8})-(\ref{ch1.sec0.eq11}) depend on the sources (natural death or disease transmission rates) of variability in the system. As it is shown below, certain sources of variability lead to very complex uncontrolled behavior of the solutions of the system.
  The following Lemma describes the behavior of the positive local solutions for the system (\ref{ch1.sec0.eq8})-(\ref{ch1.sec0.eq11}). This result will be useful in   establishing the existence and uniqueness results for the global solutions of the stochastic system (\ref{ch1.sec0.eq8})-(\ref{ch1.sec0.eq11}).
\begin{lemma}\label{ch1.sec1.lemma1}
Suppose for some $\tau_{e}>t_{0}\geq 0$ the system (\ref{ch1.sec0.eq8})-(\ref{ch1.sec0.eq11}) with initial condition in (\ref{ch1.sec0.eq12}) has a unique positive solution $Y(t)\in \mathbb{R}^{4}_{+}$, for all $t\in (-\infty, \tau_{e}]$, then  if $N(t_{0})\leq \frac{B}{\mu}$, it follows that
\item[(a.)] if the intensities of the independent white noise processes in the system satisfy  $\sigma_{i}=0$, $i\in \{S, E, I\}$ and $\sigma_{\beta}\geq 0$, then $N(t)\leq \frac{B}{\mu}$, and in addition, the set denoted by
\begin{equation}\label{ch1.sec1.lemma1.eq1}
  D(\tau_{e})=\left\{Y(t)\in \mathbb{R}^{4}_{+}: N(t)=S(t)+ E(t)+ I(t)+ R(t)\leq \frac{B}{\mu}, \forall t\in (-\infty, \tau_{e}] \right\}=\bar{B}^{(-\infty, \tau_{e}]}_{\mathbb{R}^{4}_{+},}\left(0,\frac{B}{\mu}\right),
\end{equation}
is locally self-invariant with respect to the system (\ref{ch1.sec0.eq8})-(\ref{ch1.sec0.eq11}), where $\bar{B}^{(-\infty, \tau_{e}]}_{\mathbb{R}^{4}_{+},}\left(0,\frac{B}{\mu}\right)$ is the closed ball in $\mathbb{R}^{4}_{+}$ centered at the origin with radius $\frac{B}{\mu}$ containing the local positive solutions defined over $(-\infty, \tau_{e}]$.
\item[(b.)] If the intensities of the independent white noise processes in the system satisfy  $\sigma_{i}>0$, $i\in \{S, E, I\}$ and $\sigma_{\beta}\geq 0$, then $N(t)\geq 0$, for all $t\in (-\infty, \tau_{e}]$.
\end{lemma}
Proof:\\
 It follows directly from (\ref{ch1.sec0.eq8})-(\ref{ch1.sec0.eq11}) that when $\sigma_{i}=0$, $i\in \{S, E, I\}$ and $\sigma_{\beta}\geq 0$, then
\begin{equation}\label{ch1.sec1.lemma1.eq2}
dN(t)=[B-\mu N(t)-dI(t)]dt
\end{equation}
The result in (a.) follows easily by observing that for $Y(t)\in \mathbb{R}^{4}_{+}$, the equation (\ref{ch1.sec1.lemma1.eq2}) leads to  $N(t)\leq \frac{B}{\mu}-\frac{B}{\mu}e^{-\mu(t-t_{0})}+N(t_{0})e^{-\mu(t-t_{0})}$. And under the assumption that $N(t_{0})\leq \frac{B}{\mu}$, the result follows immediately. The result in (b.) follows directly from Theorem~\ref{ch1.sec1.thm1}.
 The following theorem presents the existence and uniqueness results for the global solutions  of the stochastic system (\ref{ch1.sec0.eq8})-(\ref{ch1.sec0.eq11}). 
\begin{thm}\label{ch1.sec1.thm1}
  Given the initial conditions (\ref{ch1.sec0.eq12}) and (\ref{ch1.sec0.eq13}), there exists a unique solution process $X(t,w)=(S(t,w),E(t,w), I(t,w))^{T}$ satisfying (\ref{ch1.sec0.eq8})-(\ref{ch1.sec0.eq11}), for all $t\geq t_{0}$. Moreover,
  \item[(a.)] the solution process is positive for all $t\geq t_{0}$ a.s. and lies in $D(\infty)$, whenever  the intensities of the independent white noise processes in the system satisfy  $\sigma_{i}=0$, $i\in \{S, E, I\}$ and $\sigma_{\beta}\geq 0$.
        That is, $S(t,w)>0,E(t,w)>0,  I(t,w)>0, \forall t\geq t_{0}$ a.s. and $X(t,w)\in D(\infty)=\bar{B}^{(-\infty, \infty)}_{\mathbb{R}^{4}_{+},}\left(0,\frac{B}{\mu}\right)$, where $D(\infty)$ is defined in Lemma~\ref{ch1.sec1.lemma1}, (\ref{ch1.sec1.lemma1.eq1}).
        \item[(b.)] Also, the solution process is positive for all $t\geq t_{0}$ a.s. and lies in $\mathbb{R}^{4}_{+}$, whenever  the intensities of the independent white noise processes in the system satisfy  $\sigma_{i}>0$, $i\in \{S, E, I\}$ and $\sigma_{\beta}\geq 0$.
        That is, $S(t,w)>0,E(t,w)>0,  I(t,w)>0, \forall t\geq t_{0}$ a.s. and $X(t,w)\in \mathbb{R}^{4}_{+}$.
\end{thm}
Proof:\\
It is easy to see that the coefficients of (\ref{ch1.sec0.eq8})-(\ref{ch1.sec0.eq11}) satisfy the local Lipschitz condition for the given initial data (\ref{ch1.sec0.eq12}). Therefore there exist a unique maximal local solution $X(t,w)=(S(t,w), E(t,w), I(t,w))$ on $t\in (-\infty,\tau_{e}(w)]$, where $\tau_{e}(w)$ is the first hitting time or the explosion time\cite{mao}. The following shows that $X(t,w)\in D(\tau_{e})$ almost surely, whenever $\sigma_{i}=0$, $i\in \{S, E, I\}$ and $\sigma_{\beta}\geq 0$,  where $D(\tau_{e}(w))$ is defined in Lemma~\ref{ch1.sec1.lemma1} (\ref{ch1.sec1.lemma1.eq1}), and also that $X(t,w)\in \mathbb{R}^{4}_{+}$, whenever  $\sigma_{i}>0$, $i\in \{S, E, I\}$ and $\sigma_{\beta}\geq 0$.
Define the following stopping time
\begin{equation}\label{ch1.sec1.thm1.eq1}
\left\{
\begin{array}{lll}
\tau_{+}&=&sup\{t\in (t_{0},\tau_{e}(w)): S|_{[t_{0},t]}>0,\quad E|_{[t_{0},t]}>0,\quad and\quad I|_{[t_{0},t]}>0 \},\\
\tau_{+}(t)&=&\min(t,\tau_{+}),\quad for\quad t\geq t_{0}.\\
\end{array}
\right.
\end{equation}
and lets show that $\tau_{+}(t)=\tau_{e}(w)$ a.s. Suppose on the contrary that $P(\tau_{+}(t)<\tau_{e}(w))>0$. Let $w\in \{\tau_{+}(t)<\tau_{e}(w)\}$, and $t\in [t_{0},\tau_{+}(t))$. Define
\begin{equation}\label{ch1.sec1.thm1.eq2}
\left\{
\begin{array}{ll}
V(X(t))=V_{1}(X(t))+V_{2}(X(t))+V_{3}(X(t)),\\
V_{1}(X(t))=\ln(S(t)),\quad V_{2}(X(t))=\ln(E(t)),\quad V_{3}(X(t))=\ln(I(t)),\forall t\leq\tau_{+}(t).
\end{array}
\right.
\end{equation}
It follows from (\ref{ch1.sec1.thm1.eq2}) that
\begin{equation}\label{ch1.sec1.thm1.eq3}
  dV(X(t))=dV_{1}(X(t))+dV_{2}(X(t))+dV_{3}(X(t)),
\end{equation}
where
\begin{eqnarray}
  dV_{1}(X(t)) &=& \frac{1}{S(t)}dS(t)-\frac{1}{2}\frac{1}{S^{2}(t)}(dS(t))^{2}\nonumber \\
   &=&\left[ \frac{B}{S(t)}-\beta \int^{h_{1}}_{t_{0}}f_{T_{1}}(s) e^{-\mu s}G(I(t-s))ds - \mu + \frac{\alpha}{S(t)} \int_{t_{0}}^{\infty}f_{T_{3}}(r)I(t-r)e^{-\mu r}dr \right.\nonumber\\
   &&\left.-\frac{1}{2}\sigma^{2}_{S}-\frac{1}{2}\sigma^{2}_{\beta}\left(\int^{h_{1}}_{t_{0}}f_{T_{1}}(s) e^{-\mu s}G(I(t-s))ds\right)^{2}\right]dt\nonumber\\
&&-\sigma_{S}dw_{S}(t)-\sigma_{\beta} \int^{h_{1}}_{t_{0}}f_{T_{1}}(s) e^{-\mu s}G(I(t-s))dsdw_{\beta}(t), \label{ch1.sec1.thm1.eq4}
\end{eqnarray}
\begin{eqnarray}
  dV_{2}(X(t)) &=& \frac{1}{E(t)}dE(t)-\frac{1}{2}\frac{1}{E^{2}(t)}(dE(t))^{2} \nonumber\\
  &=& \left[ \beta \frac{S(t)}{E(t)}\int^{h_{1}}_{t_{0}}f_{T_{1}}(s) e^{-\mu s}G(I(t-s))ds - \mu \right.\nonumber\\
&&\left.-\beta\frac{1}{E(t)} \int_{t_{0}}^{h_{2}}f_{T_{2}}(u)S(t-u)\int^{h_{1}}_{t_{0}}f_{T_{1}}(s) e^{-\mu s-\mu u}G(I(t-s-u))dsdu \right.\nonumber\\
&&\left.-\frac{1}{2}\sigma^{2}_{E}-\frac{1}{2}\sigma^{2}_{\beta}\frac{S^{2}(t)}{E^{2}(t)}\left(\int^{h_{1}}_{t_{0}}f_{T_{1}}(s) e^{-\mu s}G(I(t-s))ds\right)^{2}\right.\nonumber\\
&&\left.-\frac{1}{2}\sigma^{2}_{\beta}\frac{1}{E^{2}(t)}\left(\int_{t_{0}}^{h_{2}}f_{T_{2}}(u)S(t-u)\int^{h_{1}}_{t_{0}}f_{T_{1}}(s) e^{-\mu s-\mu u}G(I(t-s-u))dsdu \right)^{2}\right]dt\nonumber\\
&&-\sigma_{E}dw_{E}(t)+\sigma_{\beta} \frac{S(t)}{E(t)}\int^{h_{1}}_{t_{0}}f_{T_{1}}(s) e^{-\mu s}G(I(t-s))dsdw_{\beta}(t)\nonumber\\
&&-\sigma_{\beta}\frac{1}{E(t)} \int_{t_{0}}^{h_{2}}f_{T_{2}}(u)S(t-u)\int^{h_{1}}_{t_{0}}f_{T_{1}}(s) e^{-\mu s-\mu u}G(I(t-s-u))dsdudw_{\beta}(t),\nonumber\\
\label{ch1.sec1.thm1.eq5}
\end{eqnarray}
and
\begin{eqnarray}
  dV_{3}(X(t)) &=& \frac{1}{I(t)}dI(t)-\frac{1}{2}\frac{1}{I^{2}(t)}(dI(t))^{2}\nonumber \\
  &=& \left[\beta \frac{1}{I(t)}\int_{t_{0}}^{h_{2}}f_{T_{2}}(u)S(t-u)\int^{h_{1}}_{t_{0}}f_{T_{1}}(s) e^{-\mu s-\mu u}G(I(t-s-u))dsdu- (\mu +d+ \alpha)\right.  \nonumber\\
  &&\left.-\frac{1}{2}\sigma^{2}_{I}-\frac{1}{2}\sigma^{2}_{\beta}\left(\int_{t_{0}}^{h_{2}}f_{T_{2}}(u)S(t-u)\int^{h_{1}}_{t_{0}}f_{T_{1}}(s) e^{-\mu s-\mu u}G(I(t-s-u))dsdu \right)^{2}\right]dt\nonumber\\
&&-\sigma_{I}dw_{I}(t)+\sigma_{\beta}\frac{1}{I(t)} \int_{t_{0}}^{h_{2}}f_{T_{2}}(u)S(t-u)\int^{h_{1}}_{t_{0}}f_{T_{1}}(s) e^{-\mu s-\mu u}G(I(t-s-u))dsdudw_{\beta}(t)\nonumber\\
&&\label{ch1.sec1.thm1.eq6}
\end{eqnarray}
It follows from (\ref{ch1.sec1.thm1.eq3})-(\ref{ch1.sec1.thm1.eq6}) that for $t<\tau_{+}(t)$,
\begin{eqnarray}
  V(X(t))-V(X(t_{0})) &\geq& \int^{t}_{t_{0}}\left[-\beta \int^{h_{1}}_{t_{0}}f_{T_{1}}(s) e^{-\mu s}G(I(\xi-s))ds-\frac{1}{2}\sigma^{2}_{S}\right.\nonumber\\
   &&\left.-\frac{1}{2}\sigma^{2}_{\beta}\left(\int^{h_{1}}_{t_{0}}f_{T_{1}}(s) e^{-\mu s}G(I(\xi-s))ds\right)^{2}\right]d\xi\nonumber\\
   &&+ \int_{t}^{t_{0}}\left[-\beta\frac{1}{E(\xi)} \int_{t_{0}}^{h_{2}}f_{T_{2}}(u)S(\xi-u)\int^{h_{1}}_{t_{0}}f_{T_{1}}(s) e^{-\mu s-\mu u}G(I(\xi-s-u))dsdu
\right.\nonumber\\
&&\left.-\frac{1}{2}\sigma^{2}_{E}-\frac{1}{2}\sigma^{2}_{\beta}\frac{S^{2}(\xi)}{E^{2}(\xi)}\left(\int^{h_{1}}_{t_{0}}f_{T_{1}}(s) e^{-\mu s}G(I(\xi-s))ds\right)^{2}\right.\nonumber\\
&&\left.-\frac{1}{2}\sigma^{2}_{\beta}\frac{1}{E^{2}(\xi)}\left(\int_{t_{0}}^{h_{2}}f_{T_{2}}(u)S(\xi-u)\int^{h_{1}}_{t_{0}}f_{T_{1}}(s) e^{-\mu s-\mu u}G(I(\xi-s-u))dsdu \right)^{2}\right]d\xi\nonumber\\
   &&+ \int_{t}^{t_{0}}\left[- (3\mu +d+ \alpha)-\frac{1}{2}\sigma^{2}_{I}\right.  \nonumber\\
  &&\left.-\frac{1}{2}\sigma^{2}_{\beta}\left(\int_{t_{0}}^{h_{2}}f_{T_{2}}(u)S(\xi-u)\int^{h_{1}}_{t_{0}}f_{T_{1}}(s) e^{-\mu s-\mu u}G(I(\xi-s-u))dsdu \right)^{2}\right]d\xi\nonumber\\
&&+\int_{t}^{t_{0}}\left[-\sigma_{S}dw_{S}(\xi)-\sigma_{\beta} \int^{h_{1}}_{t_{0}}f_{T_{1}}(s) e^{-\mu s}G(I(\xi-s))dsdw_{\beta}(\xi)\right]\nonumber \\
  &&+\int_{t}^{t_{0}}\left[-\sigma_{E}dw_{E}(\xi)+\sigma_{\beta} \frac{S(\xi)}{E(\xi)}\int^{h_{1}}_{t_{0}}f_{T_{1}}(s) e^{-\mu s}G(I(\xi-s))dsdw_{\beta}(\xi)\right]\nonumber\\
&&-\int_{t}^{t_{0}}\left[\sigma_{\beta}\frac{1}{E(\xi)} \int_{t_{0}}^{h_{2}}f_{T_{2}}(u)S(\xi-u)\int^{h_{1}}_{t_{0}}f_{T_{1}}(s) e^{-\mu s-\mu u}G(I(\xi-s-u))dsdudw_{\beta}(\xi)\right]\nonumber\\
&&+\int_{t_{0}}^{t}\left[-\sigma_{I}dw_{I}(\xi)\right.\nonumber\\
&&\left.+\sigma_{\beta}\frac{1}{I(\xi)} \int_{t_{0}}^{h_{2}}f_{T_{2}}(u)S(\xi-u)\int^{h_{1}}_{t_{0}}f_{T_{1}}(s) e^{-\mu s-\mu u}G(I(\xi-s-u))dsdudw_{\beta}(\xi)\right].\nonumber\\
&&\label{ch1.sec1.thm1.eq7}
\end{eqnarray}
Taking the limit on (\ref{ch1.sec1.thm1.eq7}) as $t\rightarrow \tau_{+}(t)$, it follows from (\ref{ch1.sec1.thm1.eq1})-(\ref{ch1.sec1.thm1.eq2}) that the left-hand side $V(X(t))-V(X(t_{0}))\leq -\infty$. This contradicts the finiteness of the right-handside of the inequality (\ref{ch1.sec1.thm1.eq7}). Hence $\tau_{+}(t)=\tau_{e}(w)$ a.s., that is, $X(t,w)\in D(\tau_{e})$,   whenever  $\sigma_{i}=0$, $i\in \{S, E, I\}$ and $\sigma_{\beta}\geq 0$, and $X(t,w)\in \mathbb{R}^{4}_{+}$, whenever  $\sigma_{i}>0$, $i\in \{S, E, I\}$ and $\sigma_{\beta}\geq 0$.

The following shows that $\tau_{e}(w)=\infty$. Let $k>0$ be a positive integer such that $||\vec{\varphi}||_{1}\leq k$, where $\vec{\varphi}=\left(\varphi_{1}(t),\varphi_{2}(t), \varphi_{3}(t)\right), t\in (-\infty,t_{0}]$ defined in (\ref{ch1.sec0.eq12}), and $||.||_{1}$ is the $p-sum$ norm defined on $\mathbb{R}^{3}$, when $p=1$. Define the stopping time
\begin{equation}\label{ch1.sec1.thm1.eq8}
\left\{
\begin{array}{ll}
\tau_{k}=sup\{t\in [t_{0},\tau_{e}): ||X(s)||_{1}=S(s)+E(s)+I(s)\leq k, s\in[t_{0},t] \}\\
\tau_{k}(t)=\min(t,\tau_{k}).
\end{array}
\right.
\end{equation}
It is easy to see that as $k\rightarrow \infty$, $\tau_{k}$ increases. Set $\lim_{k\rightarrow \infty}\tau_{k}(t)=\tau_{\infty}$. Then it follows that $\tau_{\infty}\leq \tau_{e}$ a.s.
 We show in the following that: (1.) $\tau_{e}=\tau_{\infty}\quad a.s.\Leftrightarrow P(\tau_{e}\neq \tau_{\infty})=0$, (2.)  $\tau_{\infty}=\infty\quad a.s.\Leftrightarrow P(\tau_{\infty}=\infty)=1$.

Suppose on the contrary that $P(\tau_{\infty}<\tau_{e})>0$. Let $w\in \{\tau_{\infty}<\tau_{e}\}$ and $t\leq \tau_{\infty}$.
 Define
\begin{equation}\label{ch1.sec1.thm1.eq9}
\left\{
\begin{array}{ll}
\hat{V}_{1}(X(t))=e^{\mu t}(S(t)+E(t)+I(t)),\\
\forall t\leq\tau_{k}(t).
\end{array}
\right.
\end{equation}
The Ito-Doob differential $d\hat{V}_{1}$ of (\ref{ch1.sec1.thm1.eq9}) with respect to the system (\ref{ch1.sec0.eq8})-(\ref{ch1.sec0.eq11}) is given as follows:
\begin{eqnarray}
 d\hat{V}_{1} &=& \mu e^{\mu t}(S(t)+E(t)+I(t)) dt + e^{\mu t}(dS(t)+dE(t)+dI(t))  \\
   &=& e^{\mu t}\left[B+\alpha \int_{t_{0}}^{\infty}f_{T_{3}}(r)I(t-r)e^{-\mu r}dr-(\alpha + d)I(t)\right]dt\nonumber\\
   &&-\sigma_{S}e^{\mu t}S(t)dw_{S}(t)-\sigma_{E}e^{\mu t}E(t)dw_{E}(t)-\sigma_{I}e^{\mu t}I(t)dw_{I}(t)\label{ch1.sec1.thm1.eq10}
\end{eqnarray}
Integrating (\ref{ch1.sec1.thm1.eq9}) over the interval $[t_{0}, \tau]$, and applying some algebraic manipulations and  simplifications  it follows that
\begin{eqnarray}
  V_{1}(X(\tau)) &=& V_{1}(X(t_{0}))+\frac{B}{\mu}\left(e^{\mu \tau}-e^{\mu t_{0}}\right)\nonumber\\
  &&+\int_{t_{0}}^{\infty}f_{T_{3}}(r)e^{-\mu r}\left(\int_{t_{0}-r}^{t_{0}}\alpha I(\xi)d\xi-\int_{\tau-r}^{\tau}\alpha I(\xi)d\xi\right)dr-\int_{t_{0}}^{\tau}d I(\xi)d\xi \nonumber\\
  &&+\int^{\tau}_{t_{0}}\left[-\sigma_{S}e^{\mu \xi}S(\xi)dw_{S}(\xi)-\sigma_{E}e^{\mu \xi}E(\xi)dw_{E}(\xi)-\sigma_{I}e^{\mu \xi}I(\xi)dw_{I}(\xi)\right]\label{ch1.sec1.thm1.eq11}
\end{eqnarray}
Removing negative terms from (\ref{ch1.sec1.thm1.eq11}), it implies from (\ref{ch1.sec0.eq12}) that
\begin{eqnarray}
  V_{1}(X(\tau)) &\leq& V_{1}(X(t_{0}))+\frac{B}{\mu}e^{\mu \tau}\nonumber\\
  &&+\int_{t_{0}}^{\infty}f_{T_{3}}(r)e^{-\mu r}\left(\int_{t_{0}-r}^{t_{0}}\alpha \varphi_{3}(\xi)d\xi\right)dr \nonumber\\
  &&+\int^{\tau}_{t_{0}}\left[-\sigma_{S}e^{\mu \xi}S(\xi)dw_{S}(\xi)-\sigma_{E}e^{\mu \xi}E(\xi)dw_{E}(\xi)-\sigma_{I}e^{\mu \xi}I(\xi)dw_{I}(\xi)\right]\label{ch1.sec1.thm1.eq12}
\end{eqnarray}
But from (\ref{ch1.sec1.thm1.eq9}) it is easy to see that for $\forall t\leq\tau_{k}(t)$,
\begin{equation}\label{ch1.sec1.thm1.eq12a}
  ||X(t)||_{1}=S(t)+E(t)+I(t)\leq V(X(t)).
\end{equation}
 Thus setting $\tau=\tau_{k}(t)$, then it follows from
(\ref{ch1.sec1.thm1.eq8}), (\ref{ch1.sec1.thm1.eq12}) and  (\ref{ch1.sec1.thm1.eq12a}) that
\begin{equation}\label{ch1.sec1.thm1.eq13}
  k=||X(\tau_{k}(t))||_{1}\leq V_{1}(X(\tau_{k}(t)))
\end{equation}
Taking the limit on (\ref{ch1.sec1.thm1.eq13}) as $k\rightarrow \infty$ leads to a contradiction because the left-hand-side of the inequality (\ref{ch1.sec1.thm1.eq13}) is infinite, but following the right-hand-side  from (\ref{ch1.sec1.thm1.eq12}) leads to a finite value. Hence $\tau_{e}=\tau_{\infty}$ a.s. The following shows that $\tau_{e}=\tau_{\infty}=\infty$ a.s.

  Let $\ w\in \{\tau_{e}<\infty\}$. It follows from (\ref{ch1.sec1.thm1.eq11})-(\ref{ch1.sec1.thm1.eq12}) that
  \begin{eqnarray}
  I_{\{\tau_{e}<\infty\}}V_{1}(X(\tau)) &\leq& I_{\{\tau_{e}<\infty\}}V_{1}(X(t_{0}))+I_{\{\tau_{e}<\infty\}}\frac{B}{\mu}e^{\mu \tau}\nonumber\\
  &&+I_{\{\tau_{e}<\infty\}}\int_{t_{0}}^{\infty}f_{T_{3}}(r)e^{-\mu r}\left(\int_{t_{0}-r}^{t_{0}}\alpha \varphi_{3}(\xi)d\xi\right)dr\nonumber\\
  &&+I_{\{\tau_{e}<\infty\}}\int^{\tau}_{t_{0}}\left[-\sigma_{S}e^{\mu \xi}S(\xi)dw_{S}(\xi)-\sigma_{E}e^{\mu \xi}E(\xi)dw_{E}(\xi)-\sigma_{I}e^{\mu \xi}I(\xi)dw_{I}(\xi)\right].
  \nonumber\\
  \label{ch1.sec1.thm1.eq14}
\end{eqnarray}
Suppose $\tau=\tau_{k}(t)\wedge T$, where $ T>0$ is arbitrary, then taking the expected value of (\ref{ch1.sec1.thm1.eq14}) follows that
\begin{equation}\label{ch1.sec1.thm1.eq14a}
  E(I_{\{\tau_{e}<\infty\}}V_{1}(X(\tau_{k}(t)\wedge T))) \leq V_{1}(X(t_{0}))+\frac{B}{\mu}e^{\mu T}
\end{equation}
But from (\ref{ch1.sec1.thm1.eq12a}) it is easy to see that
\begin{equation}\label{ch1.sec1.thm1.eq15}
 I_{\{\tau_{e}<\infty,\tau_{k}(t)\leq T\}}||X(\tau_{k}(t))||_{1}\leq I_{\{\tau_{e}<\infty\}}V_{1}(X(\tau_{k}(t)\wedge T))
\end{equation}
It follows from (\ref{ch1.sec1.thm1.eq14})-(\ref{ch1.sec1.thm1.eq15}) and
   (\ref{ch1.sec1.thm1.eq8}) that
 \begin{eqnarray}
 P(\{\tau_{e}<\infty,\tau_{k}(t)\leq T\})k&=&E\left[I_{\{\tau_{e}<\infty,\tau_{k}(t)\leq T\}}||X(\tau_{k}(t))||_{1}\right]\nonumber\\
 &\leq& E\left[I_{\{\tau_{e}<\infty\}}V_{1}(X(\tau_{k}(t)\wedge T))\right]\nonumber\\
 &\leq& V_{1}(X(t_{0}))+\frac{B}{\mu}e^{\mu T}.
\label{ch1.sec1.thm1.eq16}
 \end{eqnarray}
  It follows immediately from (\ref{ch1.sec1.thm1.eq16}) that
 $P(\{\tau_{e}<\infty,\tau_{\infty}\leq T\})\rightarrow 0$ as $k\rightarrow \infty$. Furthermore, since $T<\infty$ is arbitrary, we conclude that $P(\{\tau_{e}<\infty,\tau_{\infty}< \infty\})= 0$.
Finally,  by the total probability principle,
 \begin{eqnarray}
 P(\{\tau_{e}<\infty\})&=&P(\{\tau_{e}<\infty,\tau_{\infty}=\infty\})+P(\{\tau_{e}<\infty,\tau_{\infty}<\infty\})\nonumber\\
 &\leq&P(\{\tau_{e}\neq\tau_{\infty}\})+P(\{\tau_{e}<\infty,\tau_{\infty}<\infty\})\nonumber\\
 &=&0.\label{ch1.sec1.thm1.eq17}
 \end{eqnarray}
 Thus from (\ref{ch1.sec1.thm1.eq17}), $\tau_{e}=\tau_{\infty}=\infty$ a.s.. In addition, $X(t)\in D(\infty)$, whenever  $\sigma_{i}=0$, $i\in \{S, E, I\}$ and $\sigma_{\beta}\geq 0$, and $X(t,w)\in \mathbb{R}^{4}_{+}$, whenever  $\sigma_{i}>0$, $i\in \{S, E, I\}$ and $\sigma_{\beta}\geq 0$.
\begin{rem}\label{ch1.sec0.remark1}
Theorem~\ref{ch1.sec1.thm1} and Lemma~\ref{ch1.sec1.lemma1} signify that the stochastic system (\ref{ch1.sec0.eq8})-(\ref{ch1.sec0.eq11}) has a unique positive solution  $Y(t)\in \mathbb{R}^{4}_{+}$ globally for all $t\in (-\infty, \infty)$. Furthermore, it follows that a positive solution of the stochastic system that starts in the closed ball centered at the origin with a radius of $\frac{B}{\mu}$, $D(\infty)=\bar{B}^{(-\infty, \infty)}_{\mathbb{R}^{4}_{+},}\left(0,\frac{B}{\mu}\right)$, will continue to oscillate and remain bounded in the closed ball for all time $t\geq t_{0}$, whenever the intensities of the independent white noise processes in the system satisfy  $\sigma_{i}=0$, $i\in \{S, E, I\}$ and $\sigma_{\beta}\geq 0$. Hence, the set $D(\infty)=\bar{B}^{(-\infty, \infty)}_{\mathbb{R}^{4}_{+},}\left(0,\frac{B}{\mu}\right)$ is a positive self-invariant set for the stochastic system (\ref{ch1.sec0.eq8})-(\ref{ch1.sec0.eq11}). In the case where the intensities of the independent white noise processes in the system satisfy  $\sigma_{i}>0$, $i\in \{S, E, I\}$ and $\sigma_{\beta}\geq 0$, the solution are positive and unique, and continue to oscillate in the unbounded space of positive real numbers $\mathbb{R}^{4}_{+}$. In other words, the positive solutions of the system are bounded, whenever $\sigma_{i}=0$, $i\in \{S, E, I\}$ and $\sigma_{\beta}\geq 0$ and unbounded,  whenever $\sigma_{i}>0$, $i\in \{S, E, I\}$ and $\sigma_{\beta}\geq 0$.

The implication of this result to the disease dynamics represented by (\ref{ch1.sec0.eq8})-(\ref{ch1.sec0.eq11}) is that the occurrence of noise exclusively from the disease transmission rate allows a controlled situation for the disease dynamics, since the positive solutions exist within a positive self invariant space. The additional source of variability from the natural death rate can lead to more complex and uncontrolled situations for the disease dynamics, since it is obvious that the intensities of the white noise processes from the natural death rates of the different states in the system are driving the positive solutions of the system unbounded. Some examples of uncontrolled disease situations that can occur when the positive solutions are unbounded include:-  (1) extinction of the population, (2) failure to find an infection-free steady population state, wherein the disease be controlled by bringing the population into that state, and (3)  a sudden significant random flip of a given state such as the infectious state from a low to high value, or vice versa over a short time interval etc.  These facts become more apparent in the subsequent sections where conditions for disease eradication are derived.
\end{rem}
\section{Existence and Asymptotic Behavior of Disease Free Equilibrium \label{ch1.sec2}}
In this section, the existence and the general asymptotic properties of the disease free equilibrium for the stochastic system  (\ref{ch1.sec0.eq8})-(\ref{ch1.sec0.eq11})  are investigated. Indeed, the disease-free equilibrium for  the  stochastic system (\ref{ch1.sec0.eq8})-(\ref{ch1.sec0.eq11}) is obtained by solving the system of  algebraic equations produced after setting the drift and the diffusion parts of the stochastic system to zero.
 In addition,  the condition $ E= I = R = 0$ is utilized in the event
when there is no disease in the population.  The equilibria of the delayed stochastic system  (\ref{ch1.sec0.eq8})-(\ref{ch1.sec0.eq11}) are denoted generally by $E=(S^{*}, E^{*}, I^{*})$.

 It is shown in the following result that the existence of an infection-free steady state for the population with the disease dynamics given by (\ref{ch1.sec0.eq8})-(\ref{ch1.sec0.eq11}) depends on the sources (disease transmission or natural death rates) of the noises in the system which are reflected  by the intensities of the white noise processes in the system $\sigma_{i}, i=S, E, I, \beta$. In other words, the following result presents the existence of the disease-free steady state solution for the stochastic system (\ref{ch1.sec0.eq8})-(\ref{ch1.sec0.eq11}) for different restrictions of the intensities of the independent white noise processes in the system  $\sigma_{i}, i=S, E, I, \beta$.
  In the next section, the qualitative behavior of the infection-free steady state solution of the system (\ref{ch1.sec0.eq8})-(\ref{ch1.sec0.eq11}) will be examined under different growth rates or growth orders for the intensities of the white noise processes in the system.
\begin{thm}\label{ch1.sec2.thm0}
\item[1.] Let $\sigma_{i}=0, i= E, I, \beta$ and $\sigma_{S}=0$. There exists a disease-free steady state solution $E_{0}=(S^{*}_{0}, 0, 0)$ for the stochastic  system (\ref{ch1.sec0.eq8})-(\ref{ch1.sec0.eq11}), where $S^{*}_{0}=\frac{B}{\mu}$. And this infection-free steady state $E_{0}$ is exactly the same infection-free steady state for the corresponding deterministic system (\ref{ch1.sec0.eq3})-(\ref{ch1.sec0.eq6})(see \cite{wanduku-biomath}).
    \item[2.] Let $\sigma_{i}> 0, i= E, I, \beta$ and $\sigma_{S}=0$. There exists a disease-free steady state solution $E_{0}=(S^{*}_{0}, 0, 0)$ for the stochastic  system (\ref{ch1.sec0.eq8})-(\ref{ch1.sec0.eq11}), where $S^{*}_{0}=\frac{B}{\mu}$. And this infection-free steady state $E_{0}$ is again exactly the same infection-free steady state for the corresponding deterministic system (\ref{ch1.sec0.eq3})-(\ref{ch1.sec0.eq6}).
    \item[3.] Let $\sigma_{i}\geq 0, i= E, I, \beta$ and $\sigma_{S}>0$. The system (\ref{ch1.sec0.eq8})-(\ref{ch1.sec0.eq11}) does not have a disease-free steady state solution.
\end{thm}
 Proof:\\
The results follow immediately by applying the method of finding the equilibria of the system described above.
\begin{rem}
In results of Theorem~\ref{ch1.sec2.thm0} can be interpreted in a physical context of a vector-borne disease in a population following the disease dynamic model  (\ref{ch1.sec0.eq8})-(\ref{ch1.sec0.eq11}) as follows:-  Theorem~\ref{ch1.sec2.thm0}[1.] asserts that the population has an infection-free population steady state,  whenever the strengths of the noises in the system from all the different sources considered in this study namely:-  (1) the disease transmission rate, and (2) the natural death rates for all the different classes in the population (susceptible, exposed, infectious or removal) are weak, that is, whenever the intensities $\sigma_{i}, i= S, E, I, \beta$ are all  infinitesimally small.

The results in Theorem~\ref{ch1.sec2.thm0}[2.] and Theorem~\ref{ch1.sec2.thm0}[3.] signify that regardless of the strength of the noises in the population from the disease transmission rate between the infectious vectors and the susceptible class, or from the natural death rate of the exposed, infectious and removal classes in the population, there always exist an infection-free population steady state, provided that the strength of the noise in the population from the natural death rate of susceptible individuals is infinitesimally small, that is, whenever $\sigma_{S}=0$. In the event where the strength of the noise in the population from the natural death rate of susceptible individuals is significant, that is, when $\sigma_{S}>0$, then the infection-free state ceases to exist.

     Furthermore, it can be deduced from the results in Theorem~\ref{ch1.sec2.thm0}[1.] and Theorem~\ref{ch1.sec2.thm0}[2.] that the sources and strengths of the noises in the system would not have any consequences on the existence of the infection-free population steady state provided that the noises in the population do not originate from  the natural death rate of the susceptible class, since the same steady state $E_{0}$ is obtained in both cases where (1) there is complete absence of noise in the population given by the deterministic system (\ref{ch1.sec0.eq3})-(\ref{ch1.sec0.eq6}), and (2) there is some presence of noise from every other source namely:- the disease transmission rate, and the natural death rates of the exposed, infectious and removal classes in the population, except from the natural death rate of the susceptible class.

       Another important observation from Theorem~\ref{ch1.sec2.thm0} is that the intensity or strength of the noise from the natural death rate of the susceptible class controls the existence of the infection-free population steady state for the population primarily. This observation suggests that the strength of the noise in the population from the natural death rate of susceptible individuals plays a major role to determine the qualitative stability character of the infection-free state, and consequently on controlling the disease in the population.

       In this section, the stochastic stability of the disease-free steady state $E_{0}$ will be examined in the general case for which the intensities of the white noise processes satisfy $\sigma_{i}\geq 0$, $i\in \{S, E, I, \beta\}$. In Section~\ref{ch1.sec2-2}, special cases for the intensities of the white noise processes in the system will be examined.
\end{rem}
In the following, the asymptotic stability  of the disease free equilibrium, $E_{0}$, of the system (\ref{ch1.sec0.eq8})-(\ref{ch1.sec0.eq11}) is investigated.  The  stochastic version of the Lyapunov functional technique\cite{Wanduku-2017,wanduku-delay} is utilized to establish the stability results. In order to study the qualitative properties of (\ref{ch1.sec0.eq8})-(\ref{ch1.sec0.eq11}) with respect to the equilibrium state $E_{0}=(S^{*}_{0},0,0), S^{*}_{0}=\frac{B}{\mu} $, first the following transformation of the variables of the system (\ref{ch1.sec0.eq8})-(\ref{ch1.sec0.eq11}) which shifts the equilibrium state of the system  to the origin is used:
\begin{equation}\label{ch1.sec2.eq1a}
\left\{
\begin{array}{lll}
U(t)&=&S(t)-S^{*}_{0}\\
V(t)&=&E(t)\\
W(t)&=&I(t).
\end{array}
\right.
\end{equation}
By employing the  transformation in (\ref{ch1.sec2.eq1a}) to the system (\ref{ch1.sec0.eq8})-(\ref{ch1.sec0.eq10}), the following system is obtained:
 \begin{eqnarray}
dU(t)&=&\left[ -\beta U(t)\int^{h_{1}}_{t_{0}}f_{T_{1}}(s) e^{-\mu s}G(W(t-s))ds - \mu U(t)+ \alpha \int_{t_{0}}^{\infty}f_{T_{3}}(r)W(t-r)e^{-\mu r}dr \right]dt\nonumber\\
&&-\sigma_{S}(S^{*}_{0}+U(t))dw_{S}(t)-\sigma_{\beta} (S^{*}_{0}+U(t))\int^{h_{1}}_{t_{0}}f_{T_{1}}(s) e^{-\mu s}G(W(t-s))dsdw_{\beta}(t) \label{ch1.sec2.eq1}\\
dV(t)&=& \left[ \beta (S^{*}_{0}+U(t))\int^{h_{1}}_{t_{0}}f_{T_{1}}(s) e^{-\mu s}G(W(t-s))ds - \mu V(t)\right.\nonumber\\
&&\left.-\beta \int_{t_{0}}^{h_{2}}f_{T_{2}}(u)(S^{*}_{0}+U(t-u))\int^{h_{1}}_{t_{0}}f_{T_{1}}(s) e^{-\mu s-\mu u}G(W(t-s-u))dsdu \right]dt\nonumber\\
&&-\sigma_{E}V(t)dw_{E}(t)+\sigma_{\beta} (S^{*}_{0}+U(t))\int^{h_{1}}_{t_{0}}f_{T_{1}}(s) e^{-\mu s}G(W(t-s))dsdw_{\beta}(t)\nonumber\\
&&-\sigma_{\beta} \int_{t_{0}}^{h_{2}}f_{T_{2}}(u)(S^{*}_{0}+U(t-u))\int^{h_{1}}_{t_{0}}f_{T_{1}}(s) e^{-\mu s-\mu u}G(W(t-s-u))dsdudw_{\beta}(t)\label{ch1.sec2.eq2}\\
dW(t)&=& \left[\beta \int_{t_{0}}^{h_{2}}f_{T_{2}}(u)(S^{*}_{0}+U(t-u))\int^{h_{1}}_{t_{0}}f_{T_{1}}(s) e^{-\mu s-\mu u}G(W(t-s-u))dsdu- (\mu +d+ \alpha) W(t) \right]dt\nonumber\\
&&-\sigma_{I}W(t)dw_{I}(t)+\sigma_{\beta} \int_{t_{0}}^{h_{2}}f_{T_{2}}(u)(S^{*}_{0}+U(t-u))\int^{h_{1}}_{t_{0}}f_{T_{1}}(s) e^{-\mu s-\mu u}G(W(t-s-u))dsdudw_{\beta}(t)\nonumber\\
&&\label{ch1.sec2.eq3}
\end{eqnarray}
Furthermore, the lemmas  that follow in this section will be utilized to establish the stability results for the disease-free steady state $E_{0}$ of the system (\ref{ch1.sec0.eq8})-(\ref{ch1.sec0.eq11}). These results also characterize the long-term behavior of the sample-paths of the stochastic system (\ref{ch1.sec0.eq8})-(\ref{ch1.sec0.eq11}) in the neighborhood of the infection-free steady state $E_{0}$.
 Let us recall the following result obtained in the earlier study [\cite{wanduku-fundamental}, Lemma~4.1].
\begin{lemma}\label{ch1.sec2.lemma2a-1}
 Suppose   $V_{1}\in\mathcal{C}^{2, 1}(\mathbb{R}^{3}\times \mathbb{R}_{+}, \mathbb{R}_{+})$ is defined by
\begin{eqnarray}
V_{1}(x,t)&=&(S(t)-S^{*}+E(t))^{2}+c(E(t))^{2}+(I(t))^{2}\\
x(t)&=&(S(t)-S^{*},E(t),I(t))^{T},\label{ch2.sec2.thm2a.eq2}
\end{eqnarray}
where  $ c$ is a positive constant. There exists two increasing positive real valued functions $\phi_{1}$, and $\phi_{2}$, such that $V_{1}$ satisfies the inequality
 \begin{eqnarray}
\phi_{1}(||x||)&\leq& V_{1}(x,(t))
\leq \phi_{2}(||x||).\label{ch2.sec2.thm2a.eq3}
\end{eqnarray}
\end{lemma}
Proof:
The result follows directly from Lemma~4.1 in \cite{wanduku-fundamental}.

It should be noted that the result in Lemma~\ref{ch1.sec2.lemma2a-1} asserts that the function $V_{1}$ is positive definite, decrescent and radially unbounded in  $\mathbb{R}^{3}\times \mathbb{R}_{+}$. These properties of the function $V_{1}$ are required, whenever the objective in applying the Lyapunov techniques is to establish the stochastic asymptotic stability in the large of the steady state of the system as stated in \cite{mao}.  The next result presents an upper bound estimate for the drift part of the Ito-derivative of $V_{1}$.
\begin{lemma}\label{ch1.sec2.lemma2a-2}
Let the hypothesis of Theorem~\ref{ch1.sec1.thm1}  be satisfied.
The differential operator\cite{wanduku-fundamental,wanduku-determ} applied to the Lyapunov function $V_{1}$ in (\ref{ch2.sec2.thm2a.eq2}) with
respect to the system  of stochastic differential equation (\ref{ch1.sec0.eq8})-(\ref{ch1.sec0.eq11}) is given by
\begin{eqnarray}
 &&dV_{1}=LV_{1}dt-2\sigma_{S}(U(t)+V(t))(S^{*}_{0}+U(t))dw_{S}(t)\nonumber\\
 &&-2\sigma_{E}(U(t)V(t)+(c+1)V^{2}(t))dw_{E}(t)-2\sigma_{I}W^{2}(t))dw_{I}(t)\nonumber\\
 &&-2c\sigma_{\beta}(S^{*}_{0}+U(t))V(t)\int_{t_{0}}^{h_{1}}f_{T_{1}}(s)e^{-\mu s}G(W(t-s))dsdw_{\beta}\nonumber\\
 &&-2\sigma_{E}[U(t)+(c+1)V(t)+W(t)]\times\nonumber\\
 &&\times\int_{t_{0}}^{h_{2}}\int_{t_{0}}^{h_{1}}f_{T_{2}}(u)f_{T_{1}}(s)e^{-\mu (s+u)}(S^{*}_{0}+U(t-u))G(W(t-s-u))dsdu dw_{\beta}(t),\label{ch2.sec2.thm2a.eq4}
\end{eqnarray}
where for some positive valued function $\tilde{K}(\mu)$ that depends on $\mu$, the drift part $LV_{1}$ of $dV_{1}$ in (\ref{ch2.sec2.thm2a.eq4}),  satisfies the inequality
\begin{eqnarray}
  LV_{1}(x,t) &\leq&(2\beta S^{*}_{0}+\beta +\alpha + 2\frac{\mu}{\tilde{K}(\mu)^{2}} -2\mu ) U^{2}(t)\nonumber\\
  &&+\left[2\mu \tilde{K}(\mu)^{2} + \alpha + \beta (2S^{*}_{0}+1 ) + c\beta (3S^{*}_{0}+1) -2(1+c)\mu \right]V^{2}(t)\nonumber\\
  &&+2[\beta S^{*}_{0}-(\mu+d+ \alpha)]W^{2}(t) \nonumber \\
   &&+2\alpha \int_{t_{0}}^{\infty}f_{T_{3}}(r)e^{-2\mu r} W^{2}(t-r)dr  \nonumber\\
   &&+[2\beta S^{*}_{0}\left(1+c\right)+ {\sigma}^{2}_{\beta}(S^{*}_{0})^{2}(4c+2(1-c)^{2})]\int_{t_{0}}^{h_{1}}f_{T_{1}}(s)e^{-2\mu s}G^{2}(W(t-s))ds\nonumber\\
   &&+\left[\beta S^{*}_{0}(4+c)+\beta (S^{*}_{0})^{2}(2+c)+{\sigma}^{2}_{\beta}(S^{*}_{0})^{2}(4c+10)\right]\times\nonumber\\
   &&\times\int_{t_{0}}^{h_{2}}\int_{t_{0}}^{h_{1}}f_{T_{2}}(u)f_{T_{1}}(s)e^{-2\mu (s+u)}G^{2}(W(t-s-u))dsdu\nonumber\\
   &&+{\sigma}^2_{S}\left(S^{*}_{0}+U(t)\right)^{2}+{\sigma}^2_{E}(c+1)V^{2}(t)+{\sigma}^2_{I}W^{2}(t),\label{ch2.sec2.thm2a.eq5}
\end{eqnarray}
\end{lemma}
Proof:\\
 The computation of the drift part $LV$ ( see the references \cite{wanduku-determ,wanduku-delay}) of the differential operator $dV$   applied to the Lyapunov function $V_{1}$ in (\ref{ch2.sec2.thm2a.eq2}) with
respect to the system  of stochastic differential equation (\ref{ch1.sec0.eq8})-(\ref{ch1.sec0.eq11}) gives the following:
\begin{eqnarray}
  LV_{1}(x,t) &=&
  -4\mu U(t)V(t)-2\mu U^{2}(t)-2(1+c)\mu V^{2}(t)-2(\mu+d+ \alpha)W^{2}(t) \nonumber \\
   &&+2\alpha( U(t)+V(t))\int_{t_{0}}^{\infty}f_{T_{3}}(r)e^{-\mu r} W(t-r)dr  \nonumber\\
   &&+2\beta \left[S^{*}_{0} U(t)+ (1+c)S^{*}_{0} V(t) + cV(t) U(t)\right]\int_{t_{0}}^{h_{1}}f_{T_{1}}(s)e^{-\mu s}G(W(t-s))ds\nonumber\\
   &&-2\beta \left[ U(t)+(1+c)V(t)-W(t)\right]\times\nonumber\\
   &&\times\int_{t_{0}}^{h_{2}}\int_{t_{0}}^{h_{1}}f_{T_{2}}(u)f_{T_{1}}(s)e^{-\mu (s+u)}(S^{*}_{0}+U(t-u))G(W(t-s-u))dsdu
   \nonumber\\
   && +{\sigma}^{2}_{\beta}c\left(S^{*}_{0}+U(t)\right)^{2}\left(\int_{t_{0}}^{h_{1}}f_{T_{1}}(s)e^{-\mu s}G(W(t-s))ds\right)^{2}\nonumber\\
   &&+{\sigma}^2_{\beta}(c+2)\left(\int_{t_{0}}^{h_{2}}\int_{t_{0}}^{h_{1}}f_{T_{2}}(u)f_{T_{1}}(s)e^{-\mu (s+u)}(S^{*}_{0}+U(t-u))G(W(t-s-u))dsdu \right)^{2}\nonumber\\
   &&+{\sigma}^2_{\beta}(1-c) \left(S^{*}_{0}+U(t)\right)\left(\int_{t_{0}}^{h_{1}}f_{T_{1}}(s)e^{-\mu s}G(W(t-s))ds\right)\nonumber\\
     &&\times\left(\int_{t_{0}}^{h_{2}}\int_{t_{0}}^{h_{1}}f_{T_{2}}(u)f_{T_{1}}(s)e^{-\mu (s+u)}(S^{*}_{0}+U(t-u))G(W(t-s-u))dsdu \right)\nonumber\\
   &&+{\sigma}^2_{S}\left(S^{*}_{0}+U(t)\right)^{2}+{\sigma}^2_{E}(c+1)V^{2}(t)+{\sigma}^2_{E}W^{2}(t).\label{ch2.sec2.thm2.proof.eq1a}
\end{eqnarray}
Applying Theorem~\ref{ch1.sec1.thm1}, $Cauchy-Swartz$, $H\ddot{o}lder$ inequalities,  and the following algebraic inequality
\begin{equation}\label{ch2.sec2.thm2.proof.eq2a}
2ab\leq \frac{a^{2}}{g(c)}+b^{2}g(c),
\end{equation}
where $a,b,c\in \mathbb{R}$,  and the function $g$ is such that $g(c)> 0$,  to estimate the terms with integral signs  in  (\ref{ch2.sec2.thm2.proof.eq1a}), one can see the following:
\begin{equation}\label{ch2.sec2.thm2.proof.eq3a}
  2\alpha( U(t)+V(t))\int_{t_{0}}^{\infty}f_{T_{3}}(r)e^{-\mu r} W(t-r)dr\leq \alpha U^{2}(t)+\alpha V^{2}(t)+2\alpha \int_{t_{0}}^{\infty}f_{T_{3}}(r)e^{-2\mu r} W^{2}(t-r)dr.   \\
  \end{equation}
\begin{eqnarray}
  &&2\beta \left[S^{*}_{0} U(t)+ (1+c)S^{*}_{0} V(t) + cV(t) U(t)\right]\int_{t_{0}}^{h_{1}}f_{T_{1}}(s)e^{-\mu s}G(W(t-s))ds\nonumber\\
  &&\leq \beta S^{*}_{0}U^{2}(t)+\beta S^{*}_{0}\left(1+2c \right)V^{2}(t) + 2\beta S^{*}_{0} \left(1+ c \right)\int_{t_{0}}^{h_{1}}f_{T_{1}}(s)e^{-2\mu s}G^{2}(W(t-s))ds\nonumber\\
  &&\label{ch2.sec2.thm2.proof.eq4a}
\end{eqnarray}
\begin{eqnarray}
 &&-2\beta \left[ U(t)+(1+c)V(t)-W(t)\right]\times\nonumber\\
   &&\times\int_{t_{0}}^{h_{2}}\int_{t_{0}}^{h_{1}}f_{T_{2}}(u)f_{T_{1}}(s)e^{-\mu (s+u)}(S^{*}_{0}+U(t-u))G(W(t-s-u))dsdu\nonumber\\
   && \leq  \beta (S^{*}_{0}+1) U^{2}(t)+(1+c) \beta (S^{*}_{0}+1) V^{2}(t)+2\beta S^{*}_{0} W^{2}(t)\nonumber\\
  &&+\left[\beta S^{*}_{0} (4+c) + \beta (S^{*}_{0})^{2}(2+c)\right]\int_{t_{0}}^{h_{2}}\int_{t_{0}}^{h_{1}}f_{T_{2}}(u)f_{T_{1}}(s)e^{-2\mu (s+u)}G^{2}(W(t-s-u))dsdu.\nonumber\\
  &&\label{ch2.sec2.thm2.proof.eq5ab}
\end{eqnarray}
\begin{eqnarray}
{\sigma}^{2}_{\beta}c\left(S^{*}_{0}+U(t)\right)^{2}\left(\int_{t_{0}}^{h_{1}}f_{T_{1}}(s)e^{-\mu s}G(W(t-s))ds\right)^{2}\leq4c{\sigma}^{2}_{\beta}(S^{*}_{0})^{2}\int_{t_{0}}^{h_{1}}f_{T_{1}}(s)e^{-2\mu s}G^{2}(W(t-s))ds\nonumber\\
&&\label{ch2.sec2.thm2.proof.eq5a}
\end{eqnarray}
\begin{eqnarray}
  &&{\sigma}^2_{\beta}(c+2)\left(\int_{t_{0}}^{h_{2}}\int_{t_{0}}^{h_{1}}f_{T_{2}}(u)f_{T_{1}}(s)e^{-\mu (s+u)}(S^{*}_{0}+U(t-u))G(W(t-s-u))dsdu \right)^{2}\nonumber\\
&&\leq 4(c+2){\sigma}^{2}_{\beta}(S^{*}_{0})^{2}\int_{t_{0}}^{h_{2}}\int_{t_{0}}^{h_{1}}f_{T_{2}}(u)f_{T_{1}}(s)e^{-2\mu (s+u)}G^{2}(W(t-s-u))dsdu.\nonumber\\
&&\label{ch2.sec2.thm2.proof.eq5b}
\end{eqnarray}
\begin{eqnarray}
  &&{\sigma}^2_{\beta}(1-c) \left(S^{*}_{0}+U(t)\right)\left(\int_{t_{0}}^{h_{1}}f_{T_{1}}(s)e^{-\mu s}G(W(t-s))ds\right)\nonumber\\
     &&\times\left(\int_{t_{0}}^{h_{2}}\int_{t_{0}}^{h_{1}}f_{T_{2}}(u)f_{T_{1}}(s)e^{-\mu (s+u)}(S^{*}_{0}+U(t-u))G(W(t-s-u))dsdu \right)\nonumber\\
   &&\leq 2{\sigma}^2_{\beta}(1-c)^{2}(S^{*}_{0})^{2}\int_{t_{0}}^{h_{1}}f_{T_{1}}(s)e^{-2\mu s}G^{2}(W(t-s))ds \nonumber\\
   &&+ 2{\sigma}^{2}_{\beta}(S^{*}_{0})^{2}\int_{t_{0}}^{h_{2}}\int_{t_{0}}^{h_{1}}f_{T_{2}}(u)f_{T_{1}}(s)e^{-2\mu (s+u)}G^{2}(W(t-s-u))dsdu.\nonumber\\
&&\label{ch2.sec2.thm2.proof.eq5c}
\end{eqnarray}
The result (\ref{ch2.sec2.thm2a.eq5}) follows by applying  (\ref{ch2.sec2.thm2.proof.eq3a})-(\ref{ch2.sec2.thm2.proof.eq5c}) and the inequality (\ref{ch2.sec2.thm2.proof.eq2a}) into (\ref{ch2.sec2.thm2.proof.eq1a}). That is, $ LV_{1}(x,t)$ becomes
\begin{eqnarray}
  LV_{1}(x,t) &\leq&(2\beta S^{*}_{0}+\beta +\alpha + 2\frac{\mu}{\tilde{K}(\mu)^{2}} -2\mu ) U^{2}(t)\nonumber\\
  &&+\left[2\mu \tilde{K}(\mu)^{2} + \alpha + \beta (2S^{*}_{0}+1 ) + c\beta (3S^{*}_{0}+1) -2(1+c)\mu \right]V^{2}(t)\nonumber\\
  &&+2[\beta S^{*}_{0}-(\mu+d+ \alpha)]W^{2}(t) \nonumber \\
   &&+2\alpha \int_{t_{0}}^{\infty}f_{T_{3}}(r)e^{-2\mu r} W^{2}(t-r)dr  \nonumber\\
   &&+[2\beta S^{*}_{0}\left(1+c\right)+ {\sigma}^{2}_{\beta}(S^{*}_{0})^{2}(4c+2(1-c)^{2})]\int_{t_{0}}^{h_{1}}f_{T_{1}}(s)e^{-2\mu s}G^{2}(W(t-s))ds\nonumber\\
   &&+\left[\beta S^{*}_{0}(4+c)+\beta (S^{*}_{0})^{2}(2+c)+{\sigma}^{2}_{\beta}(S^{*}_{0})^{2}(4c+10)\right]\times\nonumber\\
   &&\times\int_{t_{0}}^{h_{2}}\int_{t_{0}}^{h_{1}}f_{T_{2}}(u)f_{T_{1}}(s)e^{-2\mu (s+u)}G^{2}(W(t-s-u))dsdu\nonumber\\
   &&+{\sigma}^2_{S}\left(S^{*}_{0}+U(t)\right)^{2}+{\sigma}^2_{E}(c+1)V^{2}(t)+{\sigma}^2_{E}W^{2}(t),\label{ch2.sec2.thm2.proof.eq6}
\end{eqnarray}
where $\tilde{K}(\mu)=g(\mu)$ and $ g$ is defined in  (\ref{ch2.sec2.thm2.proof.eq2a}).

The following result characterizes the stochastic asymptotic stability of the infection-free equilibrium $E_{0}$ of the system (\ref{ch1.sec0.eq8})-(\ref{ch1.sec0.eq11}), whenever the intensity of the white noise process from the natural death rate of the susceptible population is zero, that is, when $\sigma_{S}=0 $.
\begin{lemma}\label{ch1.sec2.lemma2}
 Let the hypotheses of  Theorem~\ref{ch1.sec1.thm1}, Theorem~\ref{ch1.sec2.thm0}[2.] and Lemma~\ref{ch1.sec2.lemma2a-2} be satisfied. There exists a Lyapunov functional
 \begin{equation}\label{ch2.sec2.thm2.eq1a}
V=V_{1}+V_{2},
\end{equation}
where
  $V_{1}\in\mathcal{C}^{2, 1}(\mathbb{R}^{3}\times \mathbb{R}_{+}, \mathbb{R}_{+})$ is defined by (\ref{ch2.sec2.thm2a.eq2})
and $V_{2}$ is defined as follows:
%
 \begin{eqnarray}
   &&V_{2}(x,t)=2\alpha \int_{t_{0}}^{\infty}f_{T_{3}}(r)e^{-2\mu r} \int_{t-r}^{t}I^{2}(v)dvdr  \nonumber\\
   &&+[2\beta S^{*}_{0}\left(1+c\right)+ {\sigma}^{2}_{\beta}(S^{*}_{0})^{2}(4c+2(1-c)^{2})]\int_{t_{0}}^{h_{1}}f_{T_{1}}(s)e^{-2\mu s}\int^{t}_{t-s}G^{2}(I(v))dvds\nonumber\\
   &&+\left[\beta S^{*}_{0}(4+c)+\beta (S^{*}_{0})^{2}(2+c)+{\sigma}^{2}_{\beta}(S^{*}_{0})^{2}(4c+10)\right]\times\nonumber\\
   &&\times\left[\int_{t_{0}}^{h_{2}}\int_{t_{0}}^{h_{1}}f_{T_{2}}(u)f_{T_{1}}(s)e^{-2\mu (s+u)}\int^{t}_{t-u}G^{2}(I(v-s))dvdsdu\right.\nonumber\\
   &&\left.+\int_{t_{0}}^{h_{2}}\int_{t_{0}}^{h_{1}}f_{T_{2}}(u)f_{T_{1}}(s)e^{-2\mu (s+u)}\int^{t}_{t-s}G^{2}(I(v))dvdsdu\right]\nonumber\\
   \label{{ch2.sec2.thm2.eq4a}}
\end{eqnarray}
Furthermore, there exists threshold values $R_{1}$, $U_{0}$ and $V_{0}$  defined as follows:
\begin{equation}\label{ch2.sec2.thm1.eq5a}
R_{1}=\frac{\beta S^{*}_{0} \hat{K}_{1}}{(\mu+d+\alpha)}+\frac{\alpha}{(\mu+d+\alpha)}+\frac{\hat{k}_{1}\sigma^{2}_{\beta}+\frac{1}{2}\sigma^{2}_{I}}{(\mu+d+\alpha)},
\end{equation}
\begin{equation}\label{ch2.sec2.thm1.eq5b}
U_{0}=\frac{2\beta S^{*}_{0}+\beta +\alpha + 2\frac{\mu}{\tilde{K}(\mu)^{2}}}{2\mu},
\end{equation}
 and
 \begin{equation}\label{ch2.sec2.thm1.eq5c}
V_{0}=\frac{(2\mu \tilde{K}(\mu)^{2} + \alpha + \beta (2S^{*}_{0}+1 ) )}{2\mu}+\frac{\sigma^{2}_{E}}{2\mu},
\end{equation}
with some constants $\hat{K}_{1}>0,\hat{k}_{1}>0 $  that depends on $S^{*}_{0}$  (in fact, $\hat{k}_{1}=6S^{*}_{0}$ and $\hat{K}_{1}=4+ S^{*}_{0} $), and 
   under the assumptions that $R_{1}\leq 1$, $U_{0}\leq 1$, and $V_{0}\leq 1$,  the drift part $LV$ of the differential operator $dV$ applied to $V$ with respect to the stochastic dynamic system (\ref{ch1.sec0.eq8})-(\ref{ch1.sec0.eq11}) satisfies the following inequality:
 \begin{equation}\label{ch2.sec2.thm1.eq6}
  L{V}(x,t)\leq  -\left(\phi U^{2}(t)+\psi V^{2}(t)+ \varphi W^{2}(t)\right).
\end{equation}
\end{lemma}
Proof:\\
The drift part $LV$ of the differential operator $dV$ applied to the Lyapunov functional defined in (\ref{ch2.sec2.thm2.eq1a}), (\ref{ch2.sec2.thm2a.eq2}) and (\ref{{ch2.sec2.thm2.eq4a}}) with respect to system (\ref{ch1.sec0.eq8})-(\ref{ch1.sec0.eq11}) leads to the following:
\begin{eqnarray}
  L{V}(x,t)&=&L{V}_{1}(x,t)\nonumber\\
   &&+2\alpha \int_{t_{0}}^{\infty}f_{T_{3}}(r)e^{-2\mu r} W^{2}(t)dr  \nonumber\\
   &&+[2\beta S^{*}_{0}\left(1+c\right)+ {\sigma}^{2}_{\beta}(S^{*}_{0})^{2}(4c+2(1-c)^{2})]\int_{t_{0}}^{h_{1}}f_{T_{1}}(s)e^{-2\mu s}G^{2}(W(t))ds\nonumber\\
   &&+\left[\beta S^{*}_{0}(4+c)+\beta (S^{*}_{0})^{2}(2+c)+{\sigma}^{2}_{\beta}(S^{*}_{0})^{2}(4c+10)\right]\times\nonumber\\
   &&\times\int_{t_{0}}^{h_{2}}\int_{t_{0}}^{h_{1}}f_{T_{2}}(u)f_{T_{1}}(s)e^{-2\mu (s+u)}G^{2}(W(t))dsdu\nonumber\\
   &&-2\alpha \int_{t_{0}}^{\infty}f_{T_{3}}(r)e^{-2\mu r} W^{2}(t-r)dr  \nonumber\\
   &&-[2\beta S^{*}_{0}\left(1+c\right)+ {\sigma}^{2}_{\beta}(S^{*}_{0})^{2}(4c+2(1-c)^{2})]\int_{t_{0}}^{h_{1}}f_{T_{1}}(s)e^{-2\mu s}G^{2}(W(t-s))ds\nonumber\\
   &&-\left[\beta S^{*}_{0}(4+c)+\beta (S^{*}_{0})^{2}(2+c)+{\sigma}^{2}_{\beta}(S^{*}_{0})^{2}(4c+10)\right]\times\nonumber\\
   &&\times\int_{t_{0}}^{h_{2}}\int_{t_{0}}^{h_{1}}f_{T_{2}}(u)f_{T_{1}}(s)e^{-2\mu (s+u)}G^{2}(W(t-s-u))dsdu.\nonumber\\
  \label{ch2.sec2.thm1.proof.eq7}
\end{eqnarray}
Under the assumptions for $\sigma_{i},i=S,E,I, \beta $ in Theorem~\ref{ch1.sec2.thm0}[2.], and for some suitable choice of the positive constant $ c$,  it follows from  (\ref{ch2.sec2.thm2a.eq5}), (\ref{ch2.sec2.thm1.proof.eq7}), the statements of Assumption~\ref{ch1.sec0.assum1}, $A5$ (i.e. $G^{2}(x)\leq x^{2}, x\geq 0$)  and some further algebraic manipulations and simplifications that
\begin{equation}\label{ch2.sec2.thm1.proof.eq8}
  L{V}(x,t)\leq  -\left(\phi U^{2}(t)+\psi V^{2}(t)+ \varphi W^{2}(t)\right),
\end{equation}
where,
 \begin{eqnarray}
 \phi &=&2\mu (1-U_{0}),\label{ch2.sec2.thm1.proof.eq8a}\\
 \psi &=&2\mu (1-V_{0})-2\mu c\left(1-\frac{\beta (3S^{*}_{0}+1)+\sigma^{2}_{E}}{2\mu}\right),\label{ch2.sec2.thm1.proof.eq8b}\\
\varphi &=&2(\mu +d+ \alpha) (1-R_{1})-c(3\beta S^{*}_{0}+\beta (S^{*}_{0})^{2}+4\sigma^{2}_{\beta}(S^{*}_{0})^{2})-2c^{2}\sigma^{2}_{\beta}(S^{*}_{0})^{2},\label{ch2.sec2.thm1.proof.eq8c}
 \end{eqnarray}
 and  $
R_{1}, V_{0}$ and $U_{0}$ are defined in (\ref{ch2.sec2.thm1.eq5a})-(\ref{ch2.sec2.thm1.eq5c}).
  In addition, under the assumptions of $R_{1}$, $U_{0}$, and $V_{0}$ in the hypothesis of the theorem, and for a suitable choice of the positive constant $c$ it follows that $\phi$, $\psi$, and  $\varphi$ are positive constants and (\ref{ch2.sec2.thm1.eq6}) follows immediately.

The Lemma~\ref{ch1.sec2.lemma2} asserts that there exists a positive-definite decrescent radially unbounded functional $V$ defined in (\ref{ch2.sec2.thm2.eq1a}) which satisfy the conditions of the stochastic Lyapunov stability characterizations (see \cite{mao}). This result implies that the zero steady state of the transformed system (\ref{ch1.sec2.eq1})-(\ref{ch1.sec2.eq3}), and consequently the nonzero infection-free steady state $E_{0}$ of the original stochastic system (\ref{ch1.sec0.eq8})-(\ref{ch1.sec0.eq11}) is stochastically asymptotically stable in the large. The following theorem formally states the asymptotic stability result for the  disease free equilibrium $E_{0}$, whenever it exists.
\begin{thm}\label{ch1.sec2.theorem1}
Suppose Theorem~\ref{ch1.sec2.thm0}[2.] and the hypotheses of Lemma~\ref{ch1.sec2.lemma2a-1} and Lemma~\ref{ch1.sec2.lemma2} are satisfied, then the
  disease-free equilibrium $E_{0}$ of the stochastic dynamic system (\ref{ch1.sec0.eq8})-(\ref{ch1.sec0.eq11}) is stochastically  asymptotically stable in the large in the set $D(\infty)$. Moreover, the steady state solution $E_{0}$ is  exponentially mean square stable.
\end{thm}
Proof\\
The result follows directly from comparative stochastic stability results.
 Moreover, the disease free equilibrium state is exponentially mean square stable from [Corollary~3.4, or Theorem~4.4 of \cite{mao}].

The results in Lemma~\ref{ch1.sec2.lemma2} and Theorem~\ref{ch1.sec2.theorem1} hold the key to understand the underlying factors of the infectious dynamic system  (\ref{ch1.sec0.eq8})-(\ref{ch1.sec0.eq11}) that are controlling the eradication of disease from the system. Since the primary goal of this paper is to provide mathematical techniques to model and to interpret the impacts of noise in the infectious disease system, the different observations will be titled below and elaborated on in the following:
\subsection{{The disease eradication conditions}}
Theorem~\ref{ch1.sec2.theorem1} signifies that in the absence of the  noise in the system from the random fluctuations in the natural death rate of the susceptible population which is reflected by the condition that $\sigma_{S}=0$, then regardless of whether there is strong or weak noise in the system from the random fluctuations in (1.) the natural death rates of the other disease related subclasses namely- the exposed, the infectious, and the removal populations, which is also reflected by the values of the intensities $\sigma_{i}\geq 0, i= E, I, R, $ or from random fluctuation in (2.) the disease transmission rate of the vector-borne disease which is again reflected by the value of the intensity of the noise given by  $\sigma^{2}_{ \beta}\geq 0$,  there is always an infection-free steady state for the population exhibiting the disease dynamics given by (\ref{ch1.sec0.eq8})-(\ref{ch1.sec0.eq11}), where the infection-free steady state is given by  $E_{0}$. Moreover, the infection-free steady state is stochastically asymptotically stable in the large, whenever the following threshold conditions are satisfied, that is, $R_{1}\leq 1$, $U_{0}\leq 1$, and $V_{0}\leq 1$, where the threshold values $R_{1}$, $U_{0}$, and $V_{0}$ are defined in (\ref{ch2.sec2.thm1.eq5a})-(\ref{ch2.sec2.thm1.eq5c}).

 In other words, if the threshold conditions $R_{1}\leq 1$, $U_{0}\leq 1$, and $V_{0}\leq 1$ hold, then every sample path of the system (\ref{ch1.sec0.eq8})-(\ref{ch1.sec0.eq11}) that starts in the neighborhood of the disease-free steady state  $E_{0}$ has a high chance to stay in the neighborhood of $E_{0}$, and ultimately becomes $E_{0}$. That is, the sample paths of the disease related states $E, I$ and $R$ ultimately become zero, whenever the threshold conditions hold, and as a consequence, the disease is eliminated from the population.

It should be noted that the threshold values $R_{1}$, $U_{0}$, and $V_{0}$ defined in (\ref{ch2.sec2.thm1.eq5a})-(\ref{ch2.sec2.thm1.eq5c}) are explicit in terms of the parameters of the system (\ref{ch1.sec0.eq8})-(\ref{ch1.sec0.eq11}), and  also computationally attractive, whenever the specific values for the parameters of the system are given. This observation  suggests in theory that  in a physical disease outbreak, if the physical characteristics of the disease scenario can be mathematically expressed in terms of the parameters of the system (\ref{ch1.sec0.eq8})-(\ref{ch1.sec0.eq11}), wherein the threshold values, $R_{1}$, $U_{0}$, and $V_{0}$ can be computed,  then the disease eradication conditions $R_{1}\leq 1$, $U_{0}\leq 1$, and $V_{0}\leq 1$ can be easily checked.
  Moreover, the disease will be eradicated whenever the strength of the random fluctuations in the natural death rate of the susceptible individuals is very weak or unnoticeable, and this is also true regardless of whether the strengths of the random fluctuations in the natural death rate of the other disease related classes, or the strength of the fluctuation in the disease transmission rate are strong or weak.

The threshold value $R_{1}$ is called the basic reproduction number ( also called the noise-modified basic reproduction number) for the disease dynamics described by the stochastic system (\ref{ch1.sec0.eq8})-(\ref{ch1.sec0.eq11}), and it exists only under the condition that the intensities of the independent white noise processes in the system satisfy  the assumptions that $\sigma_{i}\geq 0, i= E, I, R, \beta$, and $\sigma_{S}=0$. This important threshold value is defined in theory as the expected number of secondary infectious cases that result from one infectious individual in a completely susceptible population. Moreover, this parameter signals the stability of the infection-free steady state, and consequently signal the eradication of disease from the system, whenever $R_{1}\leq 1$. It also signals the existence of an endemic equilibrium, and consequently signals that the disease persists in the population, since an infectious steady state for the population exists, whenever the condition $R_{1}>1$ holds.   This parameter can also be modified by the presence of noise in the stochastic system, and as a result certain scenarios wherein the disease would be eradicated in the absence of noise in the system, would tend to favor the persistence of the disease, that is, the existence of a stable infectious state, whenever noise is introduced in the system. This is the case in this study as it is shown below.

The noise modified basic reproduction number given by
\begin{equation}\label{ch1.sec2.theorem1.rem1.rem1.eq1}
R_{1}=\frac{\beta S^{*}_{0} \hat{K}_{1}}{(\mu+d+\alpha)}+\frac{\alpha}{(\mu+d+\alpha)}+\frac{\hat{k}_{1}\sigma^{2}_{\beta}+\frac{1}{2}\sigma^{2}_{I}}{(\mu+d+\alpha)},
\end{equation}
can be interpreted  as follows:- firstly, since $\beta$ is the average number of effective infections contacts that every infectious individual can make in the population, whenever homogeneous mixing is assumed, therefore the term $\frac{\beta S^{*}_{0} \hat{K}_{1}}{(\mu+d+\alpha)}$ represents  a constant multiple $\hat{K}_{1}$ of the disease transmission rate (also defined as the average number of new infections per unit time) given by the term $\beta S^{*}_{0}$ in the infection-free steady state population $N^{*}=S^{*}_{0}$ defined in (\ref{ch1.sec0.eq13b}) that contains just one infectious individual over the average life-span of an infectious individual in the population given by $\frac{1}{(\mu+d+\alpha)}$. Note that in a population where people can either die naturally or die from disease related causes, or recover from the disease, the average life-spans is affected by these sources of mortality, and is given by $\frac{1}{(\mu+d+\alpha)}$.

Furthermore, the term  $\frac{\alpha}{(\mu+d+\alpha)}$ in (\ref{ch1.sec2.theorem1.rem1.rem1.eq1}) is the recovery rate in the  infection-free steady state population $N^{*}=S^{*}_{0}$ containing just one infectious individual over the average life-span of an infectious individual in the population. Let us note that the recovery rate $\alpha$ is a probabilistic rate defined as the probability of recovery from vector-borne disease per unit time per infective. The term $\frac{\alpha}{(\mu+d+\alpha)}$  represents the influence of recovery on the basic reproduction number since the single infectious individual in the infection-free steady state population has a chance to survive from both natural and disease related deaths, and also fully recover from infection.

The last term $\frac{\hat{k}_{1}\sigma^{2}_{\beta}+\frac{1}{2}\sigma^{2}_{I}}{(\mu+d+\alpha)}$ represents the influence of the noise in the system from the disease transmission rate, and the natural death rate of the infectious population on the basic reproduction number. As this last term may no longer exist in the absence of noise in the system, so the name "\textrm{modified-basic reproduction number}" is used to describe the basic reproduction number for a stochastic dynamic system.

It is also easy to see that the basic reproduction number in (\ref{ch1.sec2.theorem1.rem1.rem1.eq1}) is inflated by the noise term in the system, whenever the intensities $\sigma_{\beta}>0$ and $\sigma_{I}>0$. This observation is demonstrated in Figure~\ref{ch1.sec4.figure1} of Section~\ref{ch1.sec4}, and elaborated upon further in the next section.  This observation also suggests that stronger noise in the system from the natural death rate of infectious individuals, or from the disease transmission rate which may lead to higher values of $\sigma_{\beta}>0$ and $\sigma_{I}>0$,  may also inflate the basic reproduction number beyond the threshold bound $R_{1}\leq 1$, thereby causing the disease to establish a stable endemic steady state in the population. The next section considers different growth orders for the intensities of the white noise processes in the system, and examines how they affect the stochastic system.
\subsection{{The effects of the source and intensity of white noise}}\label{ch1.sec2.theorem1.rem1}
In addition to the existence results of Theorem~\ref{ch1.sec2.thm0}, the results of Lemma~\ref{ch1.sec2.lemma2} and Theorem~\ref{ch1.sec2.theorem1}  also suggest that the sources of the  noises in the system, that is, from the disease transmission or natural death rates, and also the intensities of the white noises in the stochastic system (\ref{ch1.sec0.eq8})-(\ref{ch1.sec0.eq11}), which are a result of the random fluctuations the disease dynamics exhibit direct consequences on the qualitative outcome of the vector-borne disease dynamics in the system with respect to the factors that determine disease eradication.

To explain further, while Lemma~\ref{ch1.sec2.lemma2} and Theorem~\ref{ch1.sec2.theorem1} assert that the stochastic system (\ref{ch1.sec0.eq8})-(\ref{ch1.sec0.eq11}) has a stochastically stable disease-free equilibrium, the result in Theorem~\ref{ch1.sec2.thm0}[3.] suggests the contrary. That is, Theorem~\ref{ch1.sec2.thm0}[3.] asserts that introducing the additional source of the random fluctuations in the system  from the natural death rate of susceptible individuals in the population, then the stochastic system (\ref{ch1.sec0.eq8})-(\ref{ch1.sec0.eq11}) no longer has a disease-free steady state wherein the disease can be eradicated, whenever the intensity of the white noise from the natural death rate of the susceptible population  $\sigma_{S}$  is significant, that is, whenever $\sigma_{S}>0$. Consequently, the disease can no longer be eradicated by applying the threshold conditions $R_{1}\leq 1$, $U_{0}\leq 1$, and $V_{0}\leq 1$.

The physical interpretation of this observation, in theory, is that stronger noise in the population from the natural deathrate of susceptible people in the population does no good to ease the disease control process in the population. This is justification for the obvious fact that when more people without the disease die from other causes apart from the disease, then only the disease related classes of people are left to continue to infect the remaining susceptible population.

    Furthermore, just like the basic reproduction number $R_{1}$, the other threshold values   $U_{0}$ and $V_{0}$  defined in (\ref{ch2.sec2.thm1.eq5a})-(\ref{ch2.sec2.thm1.eq5c}) also depend on the intensities of the white noise processes $\sigma_{i}\geq 0, i= E, I, R, \beta $  in the stochastic dynamic system (\ref{ch1.sec0.eq8})-(\ref{ch1.sec0.eq11}). It can also be observed from (\ref{ch2.sec2.thm1.eq5a})-(\ref{ch2.sec2.thm1.eq5c}) that high values of the intensities $\sigma_{i}\geq 0, i= E, I, R, \beta $ are associated with high values for the threshold values  $R_{1}$, $U_{0}$, and $V_{0}$, and vice versa. This fact is much more visible in Figure~\ref{ch1.sec4.figure1} of Section~\ref{ch1.sec4}.
     Therefore,  the intensities  of the white noise processes in the system tend to inflate all the threshold values for disease eradication $R_{1}$,   $U_{0}$ and $V_{0}$ of the stochastic dynamic system (\ref{ch1.sec0.eq8})-(\ref{ch1.sec0.eq11}), and consequently exert  constraints on the  threshold conditions  $R_{1}\leq 1$,  $U_{0}\leq 1$, and $ V_{0}\leq 1$ for disease eradication.

      In other words, stronger noises in the population from the natural deathrates of the disease related classes - exposed, infectious and removal populations, and also from the disease transmission rate  tend to inflate the threshold values for disease eradication $U_{0}$, and $V_{0}$ beyond the maximum threshold bounds $R_{1}\leq 1$,  $U_{0}\leq 1$, and $ V_{0}\leq 1$, and as a result, destabilize the infection-free steady state $E_{0}$ of the population, and lead to a stable endemic state for the disease in the population.

      Since it is obvious from the above discussion that the qualitative outcome of the disease dynamics represented in the stochastic dynamic model (\ref{ch1.sec0.eq8})-(\ref{ch1.sec0.eq11}) such as the existence and stochastic stability of the disease-free equilibrium $E_{0}$, and consequently disease eradication from the population depend on the intensities of the white noise  processes in the system, it follows in the next section that the growth orders of the intensities of the white noise processes in the system are classified, and their impacts on the stability of the disease-free equilibrium, and also on disease eradication are examined. Moreover, numerical evidence for the effects of intensities of the white noise processes on the stochastic system (\ref{ch1.sec0.eq8})-(\ref{ch1.sec0.eq11}) are presented in Section~\ref{ch1.sec4}.

As earlier mentioned in the introduction of this section, oftentimes the influence of the noise in a random dynamical system is more apparent in a comparative analysis to the stability results of the corresponding deterministic system.  The following result characterizes the stability of the disease-free equilibrium of the stochastic system (\ref{ch1.sec0.eq8})-(\ref{ch1.sec0.eq11}), whenever the intensities of the white noise processes in the system are so small that their existence can be ignored.
\begin{thm}\label{ch1.sec2.theorem1.corollary1}
 Let the hypotheses of  Theorem~\ref{ch1.sec1.thm1}, Theorem~\ref{ch1.sec2.thm0}[1.] and Lemma~\ref{ch1.sec2.lemma2a-2} be satisfied. There exists a Lyapunov functional
 \begin{equation}\label{ch1.sec2.theorem1.corollary1.eq1}
V=V_{1}+V_{2},
\end{equation}
where
  $V_{1}\in\mathcal{C}^{2, 1}(\mathbb{R}^{3}\times \mathbb{R}_{+}, \mathbb{R}_{+})$ is defined by (\ref{ch2.sec2.thm2a.eq2})
and $V_{2}$ is defined as follows:
%
 \begin{eqnarray}
   &&V_{2}(x,t)=2\alpha \int_{t_{0}}^{\infty}f_{T_{3}}(r)e^{-2\mu r} \int_{t-r}^{t}I^{2}(v)dvdr  \nonumber\\
   &&+[2\beta S^{*}_{0}\left(1+c\right)+ {\sigma}^{2}_{\beta}(S^{*}_{0})^{2}(4c+2(1-c)^{2})]\int_{t_{0}}^{h_{1}}f_{T_{1}}(s)e^{-2\mu s}\int^{t}_{t-s}G^{2}(I(v))dvds\nonumber\\
   &&+\left[\beta S^{*}_{0}(4+c)+\beta (S^{*}_{0})^{2}(2+c)+{\sigma}^{2}_{\beta}(S^{*}_{0})^{2}(4c+10)\right]\times\nonumber\\
   &&\times\left[\int_{t_{0}}^{h_{2}}\int_{t_{0}}^{h_{1}}f_{T_{2}}(u)f_{T_{1}}(s)e^{-2\mu (s+u)}\int^{t}_{t-u}G^{2}(I(v-s))dvdsdu\right.\nonumber\\
   &&\left.+\int_{t_{0}}^{h_{2}}\int_{t_{0}}^{h_{1}}f_{T_{2}}(u)f_{T_{1}}(s)e^{-2\mu (s+u)}\int^{t}_{t-s}G^{2}(I(v))dvdsdu\right].\nonumber\\
   \label{ch1.sec2.theorem1.corollary1.eq2}
\end{eqnarray}
Furthermore, define $R_{0}$, $U_{0}$ and $V_{0}$,  as follows:
\begin{equation}\label{ch1.sec2.theorem1.corollary1.eq3}
R_{0}=\frac{\beta S^{*}_{0} \hat{K}_{0}}{(\mu+d+\alpha)}+\frac{\alpha}{(\mu+d+\alpha)},
\end{equation}
\begin{equation}\label{ch1.sec2.theorem1.corollary1.eq4}
\hat{U}_{0}=\frac{2\beta S^{*}_{0}+\beta +\alpha + 2\frac{\mu}{\tilde{K}(\mu)^{2}}}{2\mu},
\end{equation}
 and
 \begin{equation}\label{ch1.sec2.theorem1.corollary1.eq5}
\hat{V}_{0}=\frac{(2\mu \tilde{K}(\mu)^{2} + \alpha + \beta (2S^{*}_{0}+1 ) )}{2\mu},
\end{equation}
where, $\hat{K}_{0}>0$ is a constant that depends only on $S^{*}_{0}$  (in fact, $\hat{K}_{0}=4+ S^{*}_{0} $).
 Assume that $R_{0}\leq 1$, $\hat{U}_{0}\leq 1$, and $\hat{V}_{0}\leq 1$,  then  there exist positive constants $\phi_{1}$, $\psi_{1}$, and $ \varphi_{1} $, such that the differential operator $\dot{V}$ applied to $V$ with respect to the stochastic system (\ref{ch1.sec0.eq8})-(\ref{ch1.sec0.eq11}) satisfies the following inequality:
 \begin{equation}\label{ch1.sec2.theorem1.corollary1.eq6}
  \dot{V}(x,t)\leq  -\left(\phi_{1} U^{2}(t)+\psi_{1} V^{2}(t)+ \varphi_{1} W^{2}(t)\right).
\end{equation}
Moreover,  under the assumptions in the hypothesis of Theorem~\ref{ch1.sec2.thm0}[1.], the disease free equilibrium $E_{0}$ of the resulting system (\ref{ch1.sec0.eq8})-(\ref{ch1.sec0.eq11})  is uniformly globally asymptotically stable in the set $D(\infty)$.
\end{thm}
Proof: \\
 The result follows directly from the Proofs of Lemma~\ref{ch1.sec2.lemma2a-2} and Lemma~\ref{ch1.sec2.lemma2} by applying the conditions that $\sigma_{i}=0,i=S,E,I, \beta $, and also applying the comparison stability results utilized in \cite{wanduku-determ}, where from (\ref{ch2.sec2.thm1.proof.eq8a})-(\ref{ch2.sec2.thm1.proof.eq8c}),  $\phi_{1}=\phi$, $\psi_{1}=\psi$, and $ \varphi_{1}=\varphi $.

The discussion in Subsection~\ref{ch1.sec2.theorem1.rem1} is carried forward in the following subsection that expands on the results of the comparative analysis between the stability of the infection-free steady state of the system (\ref{ch1.sec0.eq8})-(\ref{ch1.sec0.eq11}) in the presence, and also in absence of noise in the system. That is, the results  given in Theorems~[\ref{ch1.sec2.theorem1}-\ref{ch1.sec2.theorem1.corollary1}] are compared.
\subsection{stability in the absence of noise}\label{ch1.sec2.rem1}
Theorem~\ref{ch1.sec2.theorem1.corollary1} asserts that when the noise in the system from all sources, natural death or disease transmission rates, is very weak that can be ignored, that is, when $\sigma_{i}\rightarrow 0, i=S, E, I, R, \beta$, then the behavior of the stochastic system (\ref{ch1.sec0.eq8})-(\ref{ch1.sec0.eq11}) is equivalent to the behavior of the deterministic system (\ref{ch1.sec0.eq3})-(\ref{ch1.sec0.eq6}). That is, there is an infection-free steady state, $E_{0}$,  for the population which similarly to Theorem~\ref{ch1.sec2.theorem1} is globally asymptotically stable, whenever the threshold conditions:  $R_{0}\leq 1$,  $\hat{U}_{0}\leq 1$ and $\hat{V}_{0}\leq 1$ hold, where the threshold values $R_{0}$,  $\hat{U}_{0}$ and $\hat{V}_{0}$ are defined in (\ref{ch1.sec2.theorem1.corollary1.eq3})-(\ref{ch1.sec2.theorem1.corollary1.eq5}).
 In other words, the trajectories of the systems (\ref{ch1.sec0.eq8})-(\ref{ch1.sec0.eq11}) and (\ref{ch1.sec0.eq3})-(\ref{ch1.sec0.eq6}) in this case are the same, and continuously smooth all over time as there is less deflection by the noise in the system. Furthermore, all the trajectories that start in the neighborhood of the infection-free steady state  $E_{0}$ certainly remain in the neighborhood of $E_{0}$ in a deterministic manner, and ultimately become the disease-disease free steady state. That is, the disease is eradicated from the system, whenever the threshold conditions $R_{0}\leq 1$,  $\hat{U}_{0}\leq 1$ and $\hat{V}_{0}\leq 1$ hold.

  Furthermore, it should be noted similarly to Theorem~\ref{ch1.sec2.theorem1} that the threshold values $R_{0}$,  $\hat{U}_{0}$ and $\hat{V}_{0}$ are explicit in terms of the parameters of the system (\ref{ch1.sec0.eq3})-(\ref{ch1.sec0.eq6}), or equivalently in terms of the parameters of  the system (\ref{ch1.sec0.eq8})-(\ref{ch1.sec0.eq11}), whenever $\sigma_{i}\rightarrow 0, i=S, E, I, R, \beta$, and they are also computationally attractive, whenever the specific values for the parameters of the system are given. This observation, in theory, implies that the conditions for disease eradication in a disease scenario which follows the disease dynamics (\ref{ch1.sec0.eq8})-(\ref{ch1.sec0.eq11}), whenever $\sigma_{i}\rightarrow 0, i=S, E, I, R, \beta$, are determined by verifying the threshold conditions $R_{0}\leq 1$,  $\hat{U}_{0}\leq 1$ and $\hat{V}_{0}\leq 1$ for the specific values of the parameters of the system that correspond to the disease scenario.

  It is also easy to see from (\ref{ch1.sec2.theorem1.rem1.rem1.eq1}) and (\ref{ch1.sec2.theorem1.corollary1.eq3}) that the two threshold values $R_{0}$ and $R_{1}$ satisfy $R_{0}=R_{1}$, whenever the conditions of Theorem~\ref{ch1.sec2.theorem1.corollary1} are satisfied. Therefore, these two threshold parameters $R_{0}$ and $R_{1}$ are  the basic reproduction numbers for the system (\ref{ch1.sec0.eq8})-(\ref{ch1.sec0.eq11}) without and with noise in the system, respectively. Moreover, $R_{1}$ is a modification of $R_{0}$ by adding the noise terms $\sigma_{\beta}$ and $\sigma_{I}$. Thus, $R_{0}$ is called the ordinary basic reproduction number for the system, while $R_{1}$ is the noise-modified basic reproduction number for the disease dynamics.

  Also, comparing the threshold values for Theorem~\ref{ch1.sec2.theorem1.corollary1} and  Theorem~\ref{ch1.sec2.theorem1} defined in (\ref{ch1.sec2.theorem1.corollary1.eq3})-(\ref{ch1.sec2.theorem1.corollary1.eq5}) and (\ref{ch2.sec2.thm1.eq5a})-(\ref{ch2.sec2.thm1.eq5c}) respectively, it is easy to see that the basic reproduction numbers satisfy  $R_{0}\leq R_{1}$, and the other threshold values also satisfy  $ U_{0}=\hat{U}_{0}$ and $\hat{V}_{0}\leq V_{0}$,  whenever the intensities of the white noise processes in the system (\ref{ch1.sec0.eq8})-(\ref{ch1.sec0.eq11}) satisfy $\sigma_{i}>0, i=E, I, \beta$. It is also easy to see that the intensities $\sigma_{i}>0, i=E, I, R, \beta$  of the  corresponding white noise processes from the disease transmission rate,  and the natural deathrates of  the exposed, infections and removal classes  inflate the threshold values  $R_{1},  U_{0}$, and $ V_{0}$ defined in (\ref{ch2.sec2.thm1.eq5a})-(\ref{ch2.sec2.thm1.eq5c}), but have no influence on the threshold values $R_{0}$,  $\hat{U}_{0}$ and $\hat{V}_{0}$ defined in (\ref{ch1.sec2.theorem1.corollary1.eq3})-(\ref{ch1.sec2.theorem1.corollary1.eq5}). Moreover, one can see that when the intensities satisfy  $\sigma_{i}>0, i=E, I, R,  \beta$,  the  threshold values - $R_{0}$,  $\hat{U}_{0}$ and $\hat{V}_{0}$  from (\ref{ch1.sec2.theorem1.corollary1.eq3})-(\ref{ch1.sec2.theorem1.corollary1.eq5}) are smaller in magnitude  than the corresponding threshold values $R_{1},  U_{0}$, and $ V_{0}$. This implies that the threshold conditions $R_{0}\leq 1$,  $\hat{U}_{0}\leq 1$ and $\hat{V}_{0}\leq 1$ are much more easily satisfied, when compared to the other threshold conditions $R_{1}\leq 1$, $  U_{0}\leq 1$, and $ V_{0}\leq 1$ for the  set of threshold values  $R_{1},  U_{0}$, and $ V_{0}$  defined in  (\ref{ch2.sec2.thm1.eq5a})-(\ref{ch2.sec2.thm1.eq5c}).
   This observation implies that,  the disease is more easily eradicated when there almost negligible noise in the system, than when the strength of the noise in the system is strong.
Furthermore,  this observation also suggests that the  intensities of the random fluctuations in the disease dynamics  (\ref{ch1.sec0.eq8})-(\ref{ch1.sec0.eq11}) expressed as $\sigma_{i}, i=E, I, R,  \beta$ reflect the weights of the counter barrier effects exerted against the disease eradication process.

In addition, comparing the results of Theorem~\ref{ch1.sec2.theorem1.corollary1} and Theorem~\ref{ch1.sec2.theorem1}, one can guess that the noise from the natural deathrates of the disease related classes-exposed, infectious and removal populations, and also the noise from the disease transmission rates do not influence  a change on the existence and stability of the disease-free steady state $E_{0}$, since the existence and stability of $E_{0}$ does not change under the conditions of both theorems, regardless of whether the intensities satisfy $\sigma_{i}>0, i=E, I, R,  \beta$, or the intensities satisfy $\sigma_{i}\rightarrow 0, i=E, I, R,  \beta$, provided that the threshold values satisfy $R_{1}\leq 1$, $  U_{0}\leq 1$, and $ V_{0}\leq 1$. This implies that the disease will be continuously eradicated from the system, regardless of the strengths of the noises from  the natural deathrates of the disease related classes-exposed, infectious and removal populations, and also regardless of the strength of the noise from the disease transmission rate, provided that the  threshold conditions $R_{1}\leq 1$, $  U_{0}\leq 1$, and $ V_{0}\leq 1$ are satisfied.

  The result in Theorem~\ref{ch1.sec2.theorem1.corollary1} also confirms the earlier observation from Theorem~\ref{ch1.sec2.theorem1} that the source of the environmental white noise processes in the stochastic system owing to (1.) the  disease transmission rate,  or owing to (2.) the  natural death rates of the different disease classes-$S, E, I , R$ exhibit direct impacts on the disease eradication process from the population. This is because comparing the results of Theorem~\ref{ch1.sec2.theorem1.corollary1} and Theorem~\ref{ch1.sec2.thm0}[3.], it is easy to see that  if the stochastic system is influenced by significant random fluctuations, where the source is from the natural death rate of susceptible individuals in the population, that is, $\sigma_{S}>0$, then by Theorem~\ref{ch1.sec2.thm0}[3.], there is no disease-free steady state for the population, and  the disease cannot be eradicated by applying neither the threshold conditions  ($R_{1}\leq 1,  U_{0}\leq 1$, and $ V_{0}\leq 1$) from Theorem~\ref{ch1.sec2.theorem1}, nor  the threshold conditions  ($R_{0}\leq 1,  \hat{U}_{0}\leq 1$, and $ \hat{V}_{0}$) from Theorem~\ref{ch1.sec2.theorem1.corollary1}. More detailed discussions about the disease eradication process under the influence of various intensity levels of the white noise processes in the system are given in Section~\ref{ch1.sec2-2}. Moreover, numerical evidence for the effects of the intensities of the white noise processes in the system on the disease eradication process are presented in Section~\ref{ch1.sec4}.
\subsection{Asymptotic behavior when there is no infection-free steady state }
As earlier remarked in Remark~\ref{ch1.sec0.remark1}, it is expected from the existence of solution results in Theorem~\ref{ch1.sec1.thm1}[b.], that the absence of a positively-self invariant space for the sample paths of the stochastic system (\ref{ch1.sec0.eq8})-(\ref{ch1.sec0.eq11}), whenever the intensities satisfy $\sigma_{i}>0$, $\forall i\in \{S, E, I, R\}$  can lead to complex uncontrolled situations for the disease dynamics in the system. This fact is true as it will be shown later in this section.

 We recall that Theorem~\ref{ch1.sec2.thm0}[3.] asserts that the stochastic system (\ref{ch1.sec0.eq8})-(\ref{ch1.sec0.eq11}) has no disease-free steady state for the population, whenever the intensity of the  white noise process from the  natural death rate of the susceptible population is significant, that is, $\sigma_{S}>0$.  This study will be incomplete unless we know the long-term fate of the disease dynamics in the human population, whenever there is no disease-free steady state wherein the disease can be eradicated. One approach  to investigate nonlinear stochastic dynamic systems (where explicit solutions are nontrivial), whenever the infection-free steady state exists no where, involves estimating the expected distance between the trajectories of the stochastic system, and a potential disease-free steady state for the system.  The estimate obtained generally informs us about the extend to which the noise from the natural deathrate of the susceptible class in the system (\ref{ch1.sec0.eq8})-(\ref{ch1.sec0.eq11}) deviates the trajectories of the stochastic system from the potential disease-free steady state.

 Since the stochastic system (\ref{ch1.sec0.eq8})-(\ref{ch1.sec0.eq11}) has the disease-free steady state $E_{0}=(S^{*}_{0}, 0, 0), S^{*}_{0}=\frac{B}{\mu}$ from Theorem~\ref{ch1.sec2.thm0}[1.- 2.], whenever $\sigma_{S}=0$, and loses the steady state, whenever $\sigma_{S}>0$, therefore $E_{0}$ is always a potential infection-free steady state for the system (\ref{ch1.sec0.eq8})-(\ref{ch1.sec0.eq11}).

  The following result describes the oscillatory behavior of the  trajectories of the stochastic system (\ref{ch1.sec0.eq8})-(\ref{ch1.sec0.eq11}) in the neighborhood of the potential disease-free equilibrium $E_{0}$ obtained in Theorem~\ref{ch1.sec2.thm0}[1.- 2.],  whenever Theorem~\ref{ch1.sec2.thm0}[3.] is satisfied, that is, whenever   the stochastic system (\ref{ch1.sec0.eq8})-(\ref{ch1.sec0.eq11}) does not have a disease-free equilibrium.  This result characterizes the expected average relative distance between the sample paths of   the stochastic system (\ref{ch1.sec0.eq8})-(\ref{ch1.sec0.eq11}) and the potential disease-free steady state $E_{0}$.  Moreover, this result builds the backbone  to gain more insights about the asymptotic oscillatory behavior of the stochastic system (\ref{ch1.sec0.eq8})-(\ref{ch1.sec0.eq11}), whenever the system is subjected  under the influence of various intensity levels of the white noise processes in the system  which is discussed further in Section~\ref{ch1.sec2-2}.
\begin{thm}\label{ch1.sec2.theorem2}
Let the hypothesis of Theorem~\ref{ch1.sec2.thm0}[3.] be satisfied. And define the following threshold values:
\begin{equation}\label{ch2.sec2.thm2.eq1}
\tilde{R}_{1}=\frac{\beta S^{*}_{0} \hat{K}_{1}}{(\mu+d+\alpha)} +\frac{\alpha}{(\mu+d+\alpha)}+ \frac{\hat{k}_{1}\sigma^{2}_{\beta}+\frac{1}{2}\sigma^{2}_{I}}{(\mu+d+\alpha)},
\end{equation}
\begin{equation}\label{ch2.sec2.thm2.eq2}
\tilde{U}_{0}=\frac{2\beta S^{*}_{0}+\beta +\alpha + 2\frac{\mu}{\tilde{K}(\mu)^{2}}}{2\mu}+\frac{\sigma^{2}_{S}}{2\mu},
\end{equation}
 and
 \begin{equation}\label{ch2.sec2.thm2.eq3}
\tilde{V}_{0}=\frac{(2\mu \tilde{K}(\mu)^{2} + \alpha + \beta (2S^{*}_{0}+1 ) )}{2\mu}+\frac{\sigma^{2}_{E} }{2\mu},
\end{equation}
 with some constants $\hat{K}_{1}>0,\hat{k}_{1}>0 $  that depends on $S^{*}_{0}$  (in fact, $\hat{k}_{1}=6S^{*}_{0}$ and $\hat{K}_{1}=4+ S^{*}_{0} $),
 Let $X(t)=(S(t),E(t),I(t))$ be a solution of the decoupled system from (\ref{ch1.sec0.eq8})-(\ref{ch1.sec0.eq11}) with initial conditions (\ref{ch1.sec0.eq12}). Assume that,
       $\tilde{R}_{1}\leq 1$, $\tilde{U}_{0}\leq 1$, and $\tilde{V}_{0}\leq 1$,  then  there exists a positive constant $\mathfrak{m}>0$,  such that  the following inequality holds
 \begin{equation}\label{ch2.sec2.thm2.eq4}
  \limsup_{t\rightarrow \infty}\frac{1}{t}E\int^{t}_{0}\left[ (S(v)-S^{*}_{0})^{2}+ E^{2}(v)+  I^{2}(v)\right]dv\leq \frac{3\sigma^{2}_{S}(S^{*}_{0})^{2}}{\mathfrak{m}}.
\end{equation}
\end{thm}
Proof:
Let Theorem~\ref{ch1.sec2.thm0}[3.] be satisfied.  Applying the differential operator $dV$ to $V$ defined in (\ref{ch2.sec2.thm2.eq1a}), and utilizing  (\ref{ch2.sec2.thm2a.eq4}) and (\ref{ch2.sec2.thm2a.eq5}), it is easy to see that
\begin{eqnarray}
 &&dV=LV dt-2\sigma_{S}(U(t)+V(t))(S^{*}_{0}+U(t))dw_{S}(t)\nonumber\\
 &&-2\sigma_{E}(U(t)V(t)+(c+1)V^{2}(t))dw_{E}(t)-2\sigma_{I}W^{2}(t))dw_{I}(t)\nonumber\\
 &&-2c\sigma_{\beta}(S^{*}_{0}+U(t))V(t)\int_{t_{0}}^{h_{1}}f_{T_{1}}(s)e^{-\mu s}G(W(t-s))dsdw_{\beta}\nonumber\\
 &&-2\sigma_{E}[U(t)+(c+1)V(t)+W(t)]\times\nonumber\\
 &&\times\int_{t_{0}}^{h_{2}}\int_{t_{0}}^{h_{1}}f_{T_{2}}(u)f_{T_{1}}(s)e^{-\mu (s+u)}(S^{*}_{0}+U(t-u))G(W(t-s-u))dsdu dw_{\beta}(t)\label{ch2.sec2.thm2.proof.eq1}
\end{eqnarray}
where for some positive constant valued function $\tilde{K}(\mu)$, the drift part of (\ref{ch2.sec2.thm2.proof.eq1}),  $LV$,  satisfies the inequality
\begin{equation}\label{ch2.sec2.thm2.proof.eq2}
  L{V}(x,t)\leq  -\left(\tilde{\phi} U^{2}(t)+\tilde{\psi} V^{2}(t)+ \tilde{\varphi} W^{2}(t)\right),
\end{equation}
where
 \begin{eqnarray}
 \tilde{\phi} &=&2\mu (1-\tilde{U}_{0})\\
\tilde{ \psi} &=&2\mu (1-\tilde{V}_{0})-(2\mu+\sigma^{2}_{E}) c\left(1-\frac{\beta (3S^{*}_{0}+1)}{(2\mu+\sigma^{2}_{E})}\right)\\
\tilde{\varphi} &=&2(\mu +d+ \alpha) (1-\tilde{R}_{1})-c(3\beta S^{*}_{0}+\beta (S^{*}_{0})^{2}+4\sigma^{2}_{\beta}(S^{*}_{0})^{2})-2c^{2}\sigma^{2}_{\beta}(S^{*}_{0})^{2}.\label{ch2.sec2.thm2.proof.eq3}
 \end{eqnarray}
 Moreover, $
\tilde{R}_{1}=\frac{\beta S^{*}_{0} \hat{K}_{1}+\alpha+\frac{1}{2}\sigma^{2}_{I}}{(\mu+d+\alpha)},
$ where $\hat{K}_{1}=4+ S^{*}_{0}+ 6\frac{1}{\beta}\sigma^{2}_{\beta}$.
Under the assumptions of $\tilde{R}_{1}$, $\tilde{U}_{0}$, and $\tilde{V}_{0}$ in the hypothesis and for suitable choice of the positive constant $c$ it follows that $\tilde{\phi}$, $\tilde{\psi}$, and  $\tilde{\varphi}$ are positive constants. Therefore, by integrating (\ref{ch2.sec2.thm2.proof.eq1}) from 0 to $t$ on both sides and taking the expectation, it follows from (\ref{ch2.sec2.thm2.proof.eq1})-(\ref{ch2.sec2.thm2.proof.eq3}) that
 \begin{eqnarray}\label{ch2.sec2.thm2.proof.eq4}
 E(V(t)-V(0))\leq -\mathfrak{m}E\int^{t}_{0}\left[ (S(v)-S^{*}_{0})^{2}+ E^{2}(v)+  I^{2}(v)\right]dv+3\sigma^{2}_{S}(S^{*}_{0})^{2}t,
\end{eqnarray}
where $V(0)$ is constant and
\begin{equation}
\mathfrak{m}=min(\tilde{\phi},\tilde{ \psi},\tilde{\varphi}).
\end{equation}
 Hence, diving both sides of (\ref{ch2.sec2.thm2.proof.eq4}) by $t$ and $\mathfrak{m}$, and taking the limit supremum as $t\rightarrow \infty$, then (\ref{ch2.sec2.thm2.eq4}) follows immediately.

The results of Theorem~\ref{ch1.sec2.theorem2} are interpreted in the following.
  Theorem~\ref{ch1.sec2.theorem2} signifies that under  conditions that warrant the nonexistence of a disease free steady state for the stochastic system (\ref{ch1.sec0.eq8})-(\ref{ch1.sec0.eq11}),   the asymptotic expected average relative distance between the white noise influenced trajectories of the stochastic system and the potential disease-free steady state, $E_{0}$ obtained in Theorem~\ref{ch1.sec2.thm0}[1.][2.], does not exceed a constant multiple of the intensity, $\sigma_{S}$, of the white noise process from the natural deathrate of the susceptible population, whenever the following threshold conditions $\tilde{R}_{1}\leq 1$, $\tilde{U}_{0}\leq 1$, and $\tilde{V}_{0}\leq 1$ are satisfied.

  That is, asymptotically, when the physical characteristics in a disease scenario allow significant random fluctuations in the natural deathrate of susceptible individuals which lead to a white noise process with intensity value $\sigma_{S}>0$, and consequently lead to  the nonexistence or  no where existence  of an infection-free population steady state, and also allow physical conditions which mathematically can be represented by the parameters of the stochastic system (\ref{ch1.sec0.eq8})-(\ref{ch1.sec0.eq11}), wherein the threshold values $\tilde{R}_{1}$, $\tilde{U}_{0}$, and $\tilde{V}_{0}$  in (\ref{ch2.sec2.thm2.eq1})-(\ref{ch2.sec2.thm2.eq3}) can be computed,  it follows that if the following threshold conditions $\tilde{R}_{1}\leq 1$, $\tilde{U}_{0}\leq 1$, and $\tilde{V}_{0}\leq 1$ are satisfied,  then the  noise influenced population which is affected by the disease outbreak is expected to oscillate over time  near the potential disease-free steady state population $E_{0}$ obtained in Theorem~\ref{ch1.sec2.thm0}[1.][2.].

  Furthermore, the size or amplitude of the oscillations is determined primarily by the size of the intensity,  $\sigma^{2}_{S}$, of the white noise process from the natural death rate of the susceptible individuals in the population. It is easily observed that for infinitesimally small values for the intensity of the white noise from the natural deathrate of the susceptible class, that is, for $\sigma^{2}_{S}\rightarrow 0$, it follows from (\ref{ch2.sec2.thm2.eq4}) that all trajectories of the stochastic system (\ref{ch1.sec0.eq8})-(\ref{ch1.sec0.eq11}) converge on average (in expectation) to the disease-free steady state $E_{0}$ asymptotically. In addition, for continuous increase in the values $\sigma^{2}_{S}$, that is, for $\sigma^{2}_{S}\rightarrow \infty$, it is also easy to see from (\ref{ch2.sec2.thm2.eq4}) that the average distance of the trajectories of the stochastic system from the potential infection-free steady state  $E_{0}$ gets wider apart. This observation signifies that, the stronger the noise in the system from the natural deathrate of the susceptible population gets, the further and further away the system gets from the infection-free steady state wherein the disease can be eradicated.

  Also, the dependence of the size of the random oscillations of the state of the system near the infection-free steady state $E_{0}$, and also the dependence of the magnitude of the threshold values $\tilde{R}_{1}$, $\tilde{U}_{0}$, and $\tilde{V}_{0}$ on the intensities of the white noise processes in the system, $\sigma_{i}, i= S, E, I, R, \beta$, suggests as similarly remarked in  Remark~\ref{ch1.sec2.theorem1.rem1} and  Remark~\ref{ch1.sec2.rem1},  that the source (disease transmission or natural death rates) and intensity levels\footnote{The intensity levels of the white noise processes in the system are described as infinitesimally small, significant in magnitude and small but not infinitesimally small, big in size and finite, and sufficiently large.} of the white noise processes in the system exert influence on the asymptotic oscillatory behavior of the system near the potential disease-free population steady state, $E_{0}$, obtained in Theorem~\ref{ch1.sec2.thm0}[1.][2.]. This is obvious as it is already remarked above that as $\sigma^{2}_{S}\rightarrow \infty$, the sample paths of the stochastic system get further and further away from  $E_{0}$.

      The examination of the influence of the source and intensity levels of the white noise processes on the oscillatory behavior of the system near $E_{0}$ is elaborated in Section~\ref{ch1.sec2-2}. Moreover, numerical evidence for the effects of the intensity levels of the white noise processes in the system (\ref{ch1.sec0.eq8})-(\ref{ch1.sec0.eq11}) on the oscillatory behavior of the system , such as population extinction etc., over time are presented in Section~\ref{ch1.sec4}.

   Furthermore, as similarly remarked in Remark~\ref{ch1.sec2.rem1}[2.], comparing the threshold values from Theorem~\ref{ch1.sec2.theorem1.corollary1},  Theorem~\ref{ch1.sec2.theorem1}, and Theorem~\ref{ch1.sec2.theorem2}, that is, $R_{1}$, $U_{0}$, $V_{0}$ in  (\ref{ch2.sec2.thm1.eq5a})-(\ref{ch2.sec2.thm1.eq5c}),  $R_{0}$, $\hat{U}_{0}$, $\hat{V}_{0}$ in (\ref{ch1.sec2.theorem1.corollary1.eq3})-(\ref{ch1.sec2.theorem1.corollary1.eq5}), and $\tilde{R}_{1}$, $\tilde{U}_{0}$, $\tilde{V}_{0}$ in (\ref{ch2.sec2.thm2.eq1})-(\ref{ch2.sec2.thm2.eq3}), it is easy to see that $R_{0}\leq R_{1}=\tilde{R}_{1}$, $ U_{0}=\hat{U}_{0}\leq \tilde{U}_{0}$ and $V_{0}\leq \hat{V}_{0}=\tilde{V}_{0}$, whenever the intensity values of the white noise processes in the system (\ref{ch1.sec0.eq8})-(\ref{ch1.sec0.eq11}) satisfy the condition  $\sigma_{i}>0, i= S, E, I, R, \beta$. It is also easy to see that the threshold value $\tilde{U}_{0}$ in (\ref{ch2.sec2.thm2.eq2}) has been further constrained by the assumption that $\sigma_{S}>0$, from the corresponding threshold value $U_{0}=\hat{U}_{0}$ in (\ref{ch2.sec2.thm1.eq5b}) and (\ref{ch1.sec2.theorem1.corollary1.eq4}).

   Meanwhile it was remarked in Remark~\ref{ch1.sec2.rem1}[2.] that the threshold values $R_{1}$, $U_{0}$ and $V_{0}$ from Theorem~\ref{ch1.sec2.theorem1} would satisfy the threshold conditions $R_{1}\leq 1$, $U_{0}\leq 1$ and $V_{0}\leq 1$ less easily  compared to the set of threshold values $R_{0}$, $\hat{U}_{0}$ and $\hat{V}_{0}$ from Theorem~\ref{ch1.sec2.theorem1.corollary1}, it is easy to see that the threshold values $\tilde{R}_{1}$, $\tilde{U}_{0}$, and $\tilde{V}_{0}$ from Theorem~\ref{ch1.sec2.theorem2} would satisfy the threshold conditions $\tilde{R}_{1}\leq 1$, $\tilde{U}_{0}\leq 1$, and $\tilde{V}_{0}\leq 1$ also less easily when  compared to he threshold values  $R_{1}$, $U_{0}$ and $V_{0}$ from Theorem~\ref{ch1.sec2.theorem1}, and much less easily compared to the threshold values  $R_{0}$, $\hat{U}_{0}$ and $\hat{V}_{0}$ from Theorem~\ref{ch1.sec2.theorem1.corollary1}.

    This observation suggests that the  sources (natural death or disease transmission rates) and intensity levels of random fluctuations in the disease dynamics exhibit bearings on (1.) the existence of a disease-free population steady  state for the system (\ref{ch1.sec0.eq8})-(\ref{ch1.sec0.eq11}), and also on (2.) the disease eradication conditions for the system. For instance, adding the new source of random fluctuations due to the natural death rate of the susceptible population which leads to the white noise process with intensity $\sigma_{S}$, then for infinitesimally small values for $\sigma_{S}$, there exists a disease free population state, and the disease can be eradicated, whenever the conditions in Theorem~\ref{ch1.sec2.theorem1.corollary1} and  Theorem~\ref{ch1.sec2.theorem1} are satisfied. But for significant in magnitude values of the intensity $\sigma_{S}>0$, the additional source of the white noise from the random fluctuations in the natural deathrate of the susceptible individuals leads to a loss or nonexistence of  the disease-free population steady state. Moreover, in such events where the disease-free steady state ceases to exist, the solutions of the system (\ref{ch1.sec0.eq8})-(\ref{ch1.sec0.eq11}) can oscillate closely to the potential disease-free population steady state $E_{0}$ obtained in Theorem~\ref{ch1.sec2.thm0}[1.][2.],  provided that the conditions $\tilde{R}_{1}\leq 1$, $\tilde{U}_{0}\leq 1$, and $\tilde{V}_{0}\leq 1$ are satisfied, and the value of the intensity $\sigma_{S}$ is also relatively small, that is, $\sigma_{S}\rightarrow 0$.

      Furthermore, as similarly remarked in Remark~\ref{ch1.sec2.rem1}[2.], the results of Theorem~\ref{ch1.sec2.theorem2}  suggest that in a disease outbreak scenario that exhibits random fluctuations with high intensity values  for the noise from the natural deathrate of susceptible population, that is,  $\sigma_{S}>0$, and as a result does not allow the existence of an infection-free steady state population, but exhibit physical characteristics which can be represented mathematically by the parameters of the system (\ref{ch1.sec0.eq8})-(\ref{ch1.sec0.eq11}), wherein the threshold values $\tilde{R}_{1}$, $\tilde{U}_{0}$, and $\tilde{V}_{0}$ in (\ref{ch2.sec2.thm2.eq1})-(\ref{ch2.sec2.thm2.eq3}), $R_{1}$, $U_{0}$, $V_{0}$ in  (\ref{ch2.sec2.thm1.eq5a})-(\ref{ch2.sec2.thm1.eq5c}) and $R_{0}$, $\hat{U}_{0}$, $\hat{V}_{0}$ in (\ref{ch1.sec2.theorem1.corollary1.eq3})-(\ref{ch1.sec2.theorem1.corollary1.eq5})  can all be computed, and satisfy the following relationship between the threshold values and also the threshold conditions given by $R_{0}\leq R_{1}=\tilde{R}_{1}\leq 1$, $ U_{0}=\hat{U}_{0}\leq \tilde{U}_{0}\leq 1$ and $V_{0}\leq \hat{V}_{0}=\tilde{V}_{0}\leq 1$,  then the occurrence of the white noise processes from the random environmental fluctuations in all other sources namely:- from (1.) the natural deathrates of the exposed, infectious and removal  populations,  and also from (2.) the disease transmission rate, exerts additional  counter-positive  constraints against the disease eradication process as determined by the relationships between the threshold values, and the threshold conditions $R_{0}\leq R_{1}=\tilde{R}_{1}\leq 1$, $ U_{0}=\hat{U}_{0}\leq \tilde{U}_{0}\leq 1$ and $V_{0}\leq \hat{V}_{0}=\tilde{V}_{0}\leq 1$.

      That is, whereas the disease can be eradicated much less rapidly when the disease scenario represented by (\ref{ch1.sec0.eq8})-(\ref{ch1.sec0.eq11}) is controlled by the threshold values $R_{1}$, $U_{0}$, $V_{0}$ in  (\ref{ch2.sec2.thm1.eq5a})-(\ref{ch2.sec2.thm1.eq5c}) than when it is controlled by the threshold values $R_{0}$, $\hat{U}_{0}$, $\hat{V}_{0}$ in (\ref{ch1.sec2.theorem1.corollary1.eq3})-(\ref{ch1.sec2.theorem1.corollary1.eq5}), it follows that when the disease scenario is controlled by the threshold values $\tilde{R}_{1}$, $\tilde{U}_{0}$, and $\tilde{V}_{0}$ in (\ref{ch2.sec2.thm2.eq1})-(\ref{ch2.sec2.thm2.eq3}), the disease cannot be eradicated. Nevertheless, the disease population can be maintained close to a potential disease-free population steady state $E_{0}$ obtained in Theorem~\ref{ch1.sec2.thm0}[1.][2.], whenever the  intensity $\sigma_{s}$ is small, and the threshold conditions $\tilde{R}_{1}\leq 1$, $\tilde{U}_{0}\leq 1$, and $\tilde{V}_{0}\leq 1$ are satisfied.

        The oscillatory behavior of the system (\ref{ch1.sec0.eq8})-(\ref{ch1.sec0.eq11}) relative to $E_{0}$ obtained in Theorem~\ref{ch1.sec2.thm0}[1.][2.] under the influence of various intensity levels of the white noise processes in the system is discussed further in  Section~\ref{ch1.sec2-2}.
 \section{Asymptotic behavior of the system subjected under various orders for the intensities of noise}\label{ch1.sec2-2}
 This section exhibits the asymptotic properties of the stochastic system (\ref{ch1.sec0.eq8})-(\ref{ch1.sec0.eq11}), whenever it is subjected under the direct influence of various growth orders of the intensities of the white noise processes in the system. The techniques applied in Wanduku \cite{Wanduku-2017,wanduku-biosci} are used to classify the intensities of the white noises in the system, and to study their impacts on the disease eradication in the steady state population.

  In Section~\ref{ch1.sec2}, several observations were made about the bearings of the source (natural death or disease transmission rates) and intensity levels of the white noise processes in the system on (1)the existence and  stability of the disease free steady state population $E_{0}$ obtained in Theorem~\ref{ch1.sec2.thm0}[1.-2.], and consequently on (2) the disease eradication and also on (3) the expected distance between the  solutions of the system (\ref{ch1.sec0.eq8})-(\ref{ch1.sec0.eq11}) and the potential disease free equilibrium $E_{0}$.
  In this section, several special disease scenarios are characterized to give more insight about the properties - (1), (2) and (3), with respect to the stochastic system (\ref{ch1.sec0.eq8})-(\ref{ch1.sec0.eq11}).

   The special disease scenarios are determined by the  qualitative behaviors of the intensities, $\sigma^{2}_{i}, i=S, E, I, R, \beta$,  of the white noise processes originating from the natural death and disease transmission rates in the system.  Furthermore, the qualitative character of the intensities of the white noise processes are classified in Hypothesis~\ref{ch1.sec2-2.hypothesis1}. The following definitions in \cite{Wanduku-2017,wanduku-biosci} are helpful to understand the  assumptions made about the intensity levels of the white noise processes described in Hypothesis~\ref{ch1.sec2-2.hypothesis1}:
\begin{definition}\label{ch1.sec2-2.defn1}
 Given two real valued functions $f$ and $g$,
 \item[1.] if $\exists k>0$, and $
n_0$, such that $\forall n>n_{0}, |f(n)|\leq k|g(n)|$,  we say that
$f$ is big-o of $g$, and is denoted by $f(n)=0(g(n))$ or $f=0(g)$.
If $f(n)\rightarrow 0$, as $n\rightarrow \infty$, that is, $f$ turns
in the limit to a zero function for sufficiently large $n$, we write
$f=0(\epsilon)$ or $f(n)=0(\frac{1}{n})$, for $\epsilon>0$. Also, if
$f(n)$ is a constant function as $n\rightarrow \infty$, we write
$f(n)=0(1)$.
\item[2.] if $\exists k_{1},k_{2}>0$, and $
n_0$, such that $\forall n>n_{0},k_{1}|g(n)|\leq |f(n)|\leq
k_{2}|g(n)|$, we say that $f$ is big-theta of $g$, and is denoted by
$f(n)=\theta(g(n))$. If $f(n)\rightarrow \infty$ as $n\rightarrow
\infty$, we write $f(n)=\theta(n)$ or
$f=\theta(\frac{1}{\epsilon})$, for  $\epsilon>0$.
\end{definition}
The hypothesis that follows compares the intensity levels of the white noise processes from the  following (1) the disease transmission rate, (2) the natural death rate of the susceptible population, and (3) the natural death rates of the three other subcategories - exposed, infectious and removal populations.
\begin{hypothesis}\label{ch1.sec2-2.hypothesis1}
Using Definition~\ref{ch1.sec2-2.defn1}, we assume that
\item[$H_{1}$:]$\sigma_{\beta} \rightarrow 0$, $\sigma_{S} \rightarrow 0,$  $\sigma_{i} \rightarrow 0, \forall i= E, I, R$,
 $\iff$ $\sigma_{\beta}=0(\epsilon)$,  $\sigma_{S}=0(\epsilon)$, $\sigma_{i}=0(\epsilon), \forall i= E, I, R$;
 \item[$H_{2}$:]$\sigma_{\beta} <\infty$, $\sigma_{S} \rightarrow 0$,  $\sigma_{i} \rightarrow 0, \forall i= E, I, R$,
 $\iff$ $\sigma_{\beta}=0(1)$,  $\sigma_{S}=0(\epsilon)$, $\sigma_{i}=0(\epsilon), \forall i= E, I, R$;
 \item[$H_{3}$:]$\sigma_{\beta} <\infty$, $\sigma_{S} \rightarrow 0$,  $\sigma_{i} <\infty, \forall i= E, I, R$,
 $\iff$ $\sigma_{\beta}=0(1)$,  $\sigma_{S}=0(\epsilon)$, $\sigma_{i}=0(1), \forall i= E, I, R$;
 \item[$H_{4}$:]($\sigma_{\beta} \rightarrow 0$, or $\sigma_{\beta}<\infty$) and $\sigma_{S}<\infty$, and  ($\sigma_{i} \rightarrow 0$ or $\sigma_{i} <\infty , \forall i= E, I, R$),
 $\iff$ ($\sigma_{\beta}=0(\epsilon)$, or $\sigma_{\beta}=0(1)$ ), and  $\sigma_{S}=0(1)$, and ($\sigma_{i}=0(\epsilon)$ or $\sigma_{i}=0(1), \forall i= E, I, R$;
 \item[$H_{5}$:]$\sigma_{\beta} \rightarrow \infty$, $\sigma_{S}\rightarrow 0 $,  $\sigma_{i} \rightarrow 0,  \forall i= E, I, R$,
 $\iff$ $\sigma_{\beta}=\theta(\frac{1}{\epsilon})$,  $\sigma_{S}=0(\epsilon)$, $\sigma_{i}=0(\epsilon), \forall i= E, I, R$;
 \item[$H_{6}$:]$\sigma_{\beta} \rightarrow \infty$, $\sigma_{S}\rightarrow 0 $,  $\sigma_{i} \rightarrow \infty,  \forall i= E, I, R$,
 $\iff$ $\sigma_{\beta}=\theta(\frac{1}{\epsilon})$,  $\sigma_{S}=0(\epsilon)$, $\sigma_{i}=\theta(\frac{1}{\epsilon}), \forall i= E, I, R$;
 \item[$H_{7}$:]$(\sigma_{\beta} \rightarrow \infty$, or $\sigma_{\beta} \rightarrow 0$, or $\sigma_{\beta} <\infty$), and ($\sigma_{S}\rightarrow \infty)$, and  ($\sigma_{i} \rightarrow \infty$, or $\sigma_{i} <\infty$, or $\sigma_{i} \rightarrow 0, \forall i= E, I, R$),
 $\iff$ ($\sigma_{\beta} =\theta(\frac{1}{\epsilon})$, or $\sigma_{\beta} =0(\epsilon)$, or $\sigma_{\beta} =0(1)$), and ($\sigma_{S}=\theta(\frac{1}{\epsilon})$), and  ($\sigma_{i} =\theta(\frac{1}{\epsilon})$, or $\sigma_{i} =0(1)$, or $\sigma_{i} =0(\epsilon), \forall i= E, I, R$).
\end{hypothesis}
Hypothesis~\ref{ch1.sec2-2.hypothesis1}[$H_{1}$] asserts that all the intensities, $\sigma_{i}, i=S, E, I, R, \beta$,  of the white noise processes in the system  continuously decrease in size to infinitesimally small values.

Hypothesis~\ref{ch1.sec2-2.hypothesis1}[$H_{2}$] assumes that the intensity $\sigma_{\beta}$ of the white noise process from the disease transmission rate is finite in size, but the other white noise intensities $\sigma_{i}, i=S, E, I, R$ from the natural deathrates of the susceptible, exposed and infectious populations continuously decrease in size to infinitesimally small values.

Hypothesis~\ref{ch1.sec2-2.hypothesis1}[$H_{3}$] assumes that the intensities $\sigma_{\beta}$,  $\sigma_{i}, i= E, I, R$ of the white noise processes from the disease transmission rate, and the natural deathrates of the exposed, infectious and removal populations are finite in size, meanwhile the  intensity, $\sigma_{S}$, of the white noise process from the natural deathrate of the susceptible population continuously decreases in size to infinitesimally small values.

Hypothesis~\ref{ch1.sec2-2.hypothesis1}[$H_{4}$] asserts that the intensity, $\sigma_{\beta}$, of the white noise process from the disease transmission rate either continuously decreases in size to infinitesimally small values or it is significant and finite in size.  At the same time, the intensities, $\sigma_{i}, i= E, I, R$, of the white noise processes from the natural death rates  in the exposed,  infectious and removal populations are also either significant and finite in size, or they all continuously decrease in size to infinitesimally small values, meanwhile the intensity, $\sigma_{S}$,  of the white noise process from the natural deathrate of the susceptible population is significant and finite in size.

Hypothesis~\ref{ch1.sec2-2.hypothesis1}[$H_{5}$] states that the intensity, $\sigma_{\beta}$, of the white noise process from the disease transmission rate continuously increases in size to sufficiently large values, whereas the intensities, $\sigma_{i}, i= S, E, I, R$, of the white noise processes from the natural deathrates of the susceptible, exposed,  infectious and removal populations  continuously decrease in size to infinitesimally small values.

Hypothesis~\ref{ch1.sec2-2.hypothesis1}[$H_{6}$] assumes that the intensities, $\sigma_{i}, i= E, I, R, \beta$, of the white noise processes from the disease transmission rate and the natural deathrates of the exposed, infectious and removal populations continuously increase in size to sufficiently large values, but the intensity, $\sigma_{S}$, of  the white noise process from the natural deathrate of the susceptible population  continuously decreases in size to infinitesimally small values.

Hypothesis~\ref{ch1.sec2-2.hypothesis1}[$H_{7}$] states that the intensities, $\sigma_{i}, i= E, I, R, \beta$,  of the white noise processes from the disease transmission rate and natural deathrates of the exposed, infectious and removal populations either  continuously increase in size to sufficiently large values, or they are significant and finite in size, or they continuously decrease in size to infinitesimally small values, while the intensity, $\sigma_{S}$, of the white noise process from the natural deathrate of  the susceptible population  continuously increases in size to sufficiently large values.

 The following results characterize the qualitative behavior of the solutions of the stochastic system (\ref{ch1.sec0.eq8})-(\ref{ch1.sec0.eq11}) under the assumptions of Hypothesis~\ref{ch1.sec2-2.hypothesis1}.
 \begin{thm}\label{ch1.sec2-2.thm1}
 If Hypothesis~\ref{ch1.sec2-2.hypothesis1}[$H_{1}$] holds, then there exists a disease free equilibrium population $E_{0}=(S^{*}_{0},0,0), S^{*}_{0}=\frac{B}{\mu} $ for the stochastic system (\ref{ch1.sec0.eq8})-(\ref{ch1.sec0.eq11}). Furthermore,  there exists threshold values $R_{1}$, $U_{0}$ and $V_{0}$  define as follows:
\begin{equation}\label{ch1.sec2-2.thm1.eq1}
R_{1}=\frac{\beta S^{*}_{0} \hat{K}_{1}+\alpha}{(\mu+d+\alpha)}+\frac{\alpha}{(\mu+d+\alpha)},
\end{equation}
\begin{equation}\label{ch1.sec2-2.thm1.eq2}
U_{0}=\frac{2\beta S^{*}_{0}+\beta + \alpha + 2\frac{\mu}{\tilde{K}(\mu)^{2}}}{2\mu},
\end{equation}
 and
 \begin{equation}\label{ch1.sec2-2.thm1.eq3}
V_{0}=\frac{(2\mu \tilde{K}(\mu)^{2} + \alpha + \beta (2S^{*}_{0}+1 ) )}{2\mu},
\end{equation}
with some constants $\hat{K}_{1}>0,\hat{k}_{1}>0 $  that depends on $S^{*}_{0}$  (in fact, $\hat{k}_{1}=6S^{*}_{0}$ and $\hat{K}_{1}=4+ S^{*}_{0} $), such that
 under the assumptions that $R_{1}\leq 1$, $U_{0}\leq 1$, and $V_{0}\leq 1$,
 the disease free equilibrium state is stochastically asymptotically stable in the large in $D(\infty)$. Moreover, it is mean square stable.
 \end{thm}
 Proof:
 Let $\sigma_{\beta}=0(\epsilon)$,  $\sigma_{S}=0(\epsilon)$, $\sigma_{i}=0(\epsilon), \forall i= E, I, R$. It follows from the Theorem~\ref{ch1.sec2.thm0} that (\ref{ch1.sec0.eq8})-(\ref{ch1.sec0.eq11}) has a disease free steady state given by $E_{0}=(S^{*}_{0},0,0), S^{*}_{0}=\frac{B}{\mu} $. Furthermore, the rest of the result follows immediately from Lemma~\ref{ch1.sec2.lemma2} and Theorem~\ref{ch1.sec2.theorem1}.
 \begin{rem}
 Theorem~\ref{ch1.sec2-2.thm1} signifies that when the stochastic system (\ref{ch1.sec0.eq8})-(\ref{ch1.sec0.eq11}) is continuously influenced by white noise processes from the disease transmission and natural death rates that have intensity values that continuously decrease in size to infinitesimally small values, the system has a disease-free steady state population $E_{0}$, and the steady state is stochastically asymptotically stable in the large, whenever the threshold conditions $R_{1}\leq 1$, $U_{0}\leq 1$, and $V_{0}\leq 1$  given in (\ref{ch1.sec2-2.thm1.eq1})-(\ref{ch1.sec2-2.thm1.eq3}) are satisfied.

  This result suggests that in a disease scenario where the disease outbreak results in random fluctuations in the disease transmission rate and in the natural death rates of all the subclasses-susceptible, exposed, infectious and removal populations, there exists a disease-free population steady state for the population given by $E_{0}=(S^{*}_{0},0,0), S^{*}_{0}=\frac{B}{\mu} $, whenever the random environmental fluctuations in the disease transmission rate and also in the natural death processes have infinitesimally small intensity values. Furthermore, when the physical characteristics of the disease scenario can be mathematically represented by the parameters of the system (\ref{ch1.sec0.eq8})-(\ref{ch1.sec0.eq11}), wherein the threshold values $R_{1}$, $U_{0}$ and $V_{1}$ from (\ref{ch1.sec2-2.thm1.eq1})-(\ref{ch1.sec2-2.thm1.eq3}) can be computed, then the disease can be eradicated from the population, whenever  the threshold conditions $R_{1}\leq 1$, $U_{0}\leq 1$, and $V_{0}\leq 1$ are satisfied.
 \end{rem}
 \begin{thm}\label{ch1.sec2-2.thm2}
 If Hypothesis~\ref{ch1.sec2-2.hypothesis1}[$H_{2}$] holds, then there exists a disease free equilibrium population $E_{0}=(S^{*}_{0},0,0), S^{*}_{0}=\frac{B}{\mu} $ for the stochastic system (\ref{ch1.sec0.eq8})-(\ref{ch1.sec0.eq11}). Furthermore,  there exists threshold values $R_{1}$, $U_{0}$ and $V_{0}$  define as follows:
\begin{equation}\label{ch1.sec2-2.thm2.eq1}
R_{1}=\frac{\beta S^{*}_{0} \hat{K}_{1}}{(\mu+d+\alpha)}+\frac{\alpha}{(\mu+d+\alpha)}+\frac{\hat{k}_{1}\sigma^{2}_{\beta}}{(\mu+d+\alpha)},
\end{equation}
\begin{equation}\label{ch1.sec2-2.thm2.eq2}
U_{0}=\frac{2\beta S^{*}_{0}+\beta + \alpha + 2\frac{\mu}{\tilde{K}(\mu)^{2}}}{2\mu},
\end{equation}
 and
 \begin{equation}\label{ch1.sec2-2.thm2.eq3}
V_{0}=\frac{(2\mu \tilde{K}(\mu)^{2} + \alpha + \beta (2S^{*}_{0}+1 ) )}{2\mu},
\end{equation}
with some constant $\hat{K}_{1}>0$  that depends on $S^{*}_{0}$ (in fact, $\hat{K}_{1}=4+ S^{*}_{0}+ \frac{1}{\beta}6\sigma^{2}_{\beta} $) such that,
   under the assumptions that $R_{1}\leq 1$, $U_{0}\leq 1$, and $V_{0}\leq 1$,
 the disease free equilibrium state is stochastically asymptotically stable in the large in $D(\infty)$. Moreover, it is mean square stable.
 \end{thm}
 Proof:
 Let $\sigma_{\beta}=0(1)$,  $\sigma_{S}=0(\epsilon)$, $\sigma_{i}=0(\epsilon), \forall i= E, I, R$. It follows from the Theorem~\ref{ch1.sec2.thm0} that (\ref{ch1.sec0.eq8})-(\ref{ch1.sec0.eq11}) has a disease free steady state given by $E_{0}=(S^{*}_{0},0,0), S^{*}_{0}=\frac{B}{\mu} $. Furthermore, the rest of the result follows immediately from Lemma~\ref{ch1.sec2.lemma2} and Theorem~\ref{ch1.sec2.theorem1}.
 \begin{thm}\label{ch1.sec2-2.thm3}
 If Hypothesis~\ref{ch1.sec2-2.hypothesis1}[$H_{3}$] holds, then there exists a disease free equilibrium population $E_{0}=(S^{*}_{0},0,0), S^{*}_{0}=\frac{B}{\mu} $ for the stochastic system (\ref{ch1.sec0.eq8})-(\ref{ch1.sec0.eq11}). Furthermore,  there exists threshold values $R_{1}$, $U_{0}$ and $V_{0}$  define as follows:
\begin{equation}\label{ch1.sec2-2.thm3.eq1}
R_{1}=\frac{\beta S^{*}_{0} \hat{K}_{1}}{(\mu+d+\alpha)}+\frac{\alpha}{(\mu+d+\alpha)}+\frac{\hat{k}_{1}\sigma^{2}_{\beta}+\frac{1}{2}\sigma^{2}_{I}}{(\mu+d+\alpha)},
\end{equation}
\begin{equation}\label{ch1.sec2-2.thm3.eq2}
U_{0}=\frac{2\beta S^{*}_{0}+\beta + \alpha + 2\frac{\mu}{\tilde{K}(\mu)^{2}}}{2\mu},
\end{equation}
 and
 \begin{equation}\label{ch1.sec2-2.thm3.eq3}
 V_{0}=\frac{(2\mu \tilde{K}(\mu)^{2} + \alpha + \beta (2S^{*}_{0}+1 ) )}{2\mu}+\frac{\sigma^{2}_{E}}{2\mu},
\end{equation}
with some constants $\hat{K}_{1}>0,\hat{k}_{1}>0 $  that depend on $S^{*}_{0}$  (in fact, $\hat{k}_{1}=6S^{*}_{0}$ and $\hat{K}_{1}=4+ S^{*}_{0} $), such that,
   under the assumptions that $R_{1}\leq 1$, $U_{0}\leq 1$, and $V_{0}\leq 1$,
 the disease free equilibrium state is stochastically asymptotically stable in the large in $D(\infty)$. Moreover, it is mean square stable.
 \end{thm}
 Proof:
 Let $\sigma_{\beta}=0(1)$,  $\sigma_{S}=0(\epsilon)$, $\sigma_{i}=0(1), \forall i= E, I, R$. It follows from the Theorem~\ref{ch1.sec2.thm0} that (\ref{ch1.sec0.eq8})-(\ref{ch1.sec0.eq11}) has a disease-free steady state given by $E_{0}=(S^{*}_{0},0,0), S^{*}_{0}=\frac{B}{\mu} $. Furthermore, the rest of the result follows immediately from Lemma~\ref{ch1.sec2.lemma2} and Theorem~\ref{ch1.sec2.theorem1}.
 \begin{rem}
 Theorem~\ref{ch1.sec2-2.thm2} and Theorem~\ref{ch1.sec2-2.thm3} signify that when the stochastic system (\ref{ch1.sec0.eq8})-(\ref{ch1.sec0.eq11}) is continuously influenced by white noise processes from the disease transmission rate and natural deathrates where the intensity $\sigma_{\beta}$ of the white noise from the disease transmission rate is significant and finite in size, and the intensity $\sigma_{S}$ of the white noise from the natural deathrate of the susceptible population continuously decreases in size to infinitesimally small values, whereas the intensities, $\sigma_{i}, i=E, I, R$, of the white noise processes from the natural deathrates of the exposed,  infectious and removal populations are either significant and finite in size, or they continuously decrease in sizes to infinitesimally small values, it follows that the system has a disease-free steady state population $E_{0}$, and the steady state is stochastically asymptotically stable in the large, whenever the threshold conditions $R_{1}\leq 1$, $U_{0}\leq 1$, and $V_{0}\leq 1$  in (\ref{ch1.sec2-2.thm2.eq1})-(\ref{ch1.sec2-2.thm2.eq3}) and (\ref{ch1.sec2-2.thm3.eq1})-(\ref{ch1.sec2-2.thm3.eq3}) are satisfied.

 This result suggests that in a disease scenario where the disease outbreak results in random fluctuations in the disease transmission and natural death rates, there exists a disease-free population steady state for the population given by $E_{0}=(S^{*}_{0},0,0), S^{*}_{0}=\frac{B}{\mu} $, whenever the random environmental fluctuations in the disease transmission rate have significant and finite intensity values, and the environmental fluctuations in the natural deathrate of the susceptible population has infinitesimally small intensity values, regardless of whether there exists  significant and finite in sizes, or there exists infinitesimally small intensity values for the random fluctuations from the natural deathrates of the exposed, infectious and removal populations. Furthermore,  when the physical characteristics of the disease dynamics can be represented by the parameters of the system (\ref{ch1.sec0.eq8})-(\ref{ch1.sec0.eq11}), wherein the threshold values $R_{1}$, $U_{0}$ and $V_{1}$ from (\ref{ch1.sec2-2.thm2.eq1})-(\ref{ch1.sec2-2.thm2.eq3}) and (\ref{ch1.sec2-2.thm3.eq1})-(\ref{ch1.sec2-2.thm3.eq3}) can be computed, then the disease can be eradicated from the population, whenever  the threshold conditions
  $R_{1}\leq 1$, $U_{0}\leq 1$, and $V_{0}\leq 1$ are satisfied.
 \end{rem}
 The following result characterizes the behavior of the solutions of the stochastic system (\ref{ch1.sec0.eq8})-(\ref{ch1.sec0.eq11}) whenever the assumption of Hypothesis~\ref{ch1.sec2-2.hypothesis1}[$H_{4}$] is satisfied. For simplicity, the results are presented only for the subcase when $\sigma_{\beta}=0(\epsilon)$ ,  $\sigma_{S}=0(1)$, and $\sigma_{i}=0(\epsilon), \forall i= E, I, R$. The other subcases can similarly be derived.
 \begin{thm}\label{ch1.sec2-2.thm4}
 If Hypothesis~\ref{ch1.sec2-2.hypothesis1}[$H_{4}$] holds, then there is no  disease free equilibrium population state for the stochastic system (\ref{ch1.sec0.eq8})-(\ref{ch1.sec0.eq11}). But, when $\sigma_{\beta}=0(\epsilon)$ ,  $\sigma_{S}=0(1)$, and $\sigma_{i}=0(\epsilon), \forall i= E, I, R$,  there exists threshold values:
\begin{equation}\label{ch1.sec2-2.thm4.eq1}
\tilde{R}_{1}=\frac{\beta S^{*}_{0} \hat{K}_{1}}{(\mu+d+\alpha)}+\frac{\alpha}{(\mu+d+\alpha)},
\end{equation}
\begin{equation}\label{ch1.sec2-2.thm4.eq2}
\tilde{U}_{0}=\frac{2\beta S^{*}_{0}+\beta +\alpha + 2\frac{\mu}{\tilde{K}(\mu)^{2}}}{2\mu}+\frac{\sigma^{2}_{S}}{2\mu},
\end{equation}
 and
 \begin{equation}\label{ch1.sec2-2.thm4.eq3}
\tilde{V}_{0}=\frac{(2\mu \tilde{K}(\mu)^{2} + \alpha + \beta (2S^{*}_{0}+1 ) )}{2\mu},
\end{equation}
with some constant $\hat{K}_{1}>0$  that depends on $S^{*}_{0}$  (in fact, $\hat{K}_{1}=4+ S^{*}_{0} $), such that letting $X(t)=(S(t),E(t),I(t))$ be a solution of the decoupled system from (\ref{ch1.sec0.eq8})-(\ref{ch1.sec0.eq11}) with initial conditions (\ref{ch1.sec0.eq12})
       then  there exists a positive constant $\mathfrak{m}>0$,  such that  the following inequality holds
 \begin{equation}\label{ch1.sec2-2.thm4.eq4}
  \limsup_{t\rightarrow \infty}\frac{1}{t}E\int^{t}_{0}\left[ (S(v)-S^{*}_{0})^{2}+ E^{2}(v)+  I^{2}(v)\right]dv\leq \frac{3\sigma^{2}_{S}(S^{*}_{0})^{2}}{\mathfrak{m}},
\end{equation}
whenever $\tilde{R}_{1}\leq 1$, $\tilde{U}_{0}\leq 1$, and $\tilde{V}_{0}\leq 1$.
 \end{thm}
 Proof:
 Let ($\sigma_{\beta}=0(\epsilon)$, or $\sigma_{\beta}=0(1)$ ) and  $\sigma_{S}=0(1)$, and ($\sigma_{i}=0(\epsilon)$ or $\sigma_{i}=0(1), \forall i= E, I, R$). It follows from the Theorem~\ref{ch1.sec2.thm0}[3.] that (\ref{ch1.sec0.eq8})-(\ref{ch1.sec0.eq11}) does not have a disease free steady state.  Furthermore, when  $\sigma_{\beta}=0(\epsilon)$ ,  $\sigma_{S}=0(1)$, and $\sigma_{i}=0(\epsilon), \forall i= E, I, R$, the rest of the result follows immediately from Theorem~\ref{ch1.sec2.theorem2}.
 \begin{rem}
 Theorem~\ref{ch1.sec2-2.thm4}  signifies that when the stochastic system (\ref{ch1.sec0.eq8})-(\ref{ch1.sec0.eq11}) is continuously influenced by the white noise processes from the disease transmission and natural death rates, where the intensity value $\sigma_{\beta}$ of the white noise process from the disease transmission process is significant and finite in size, or it continuously decreases in size to an infinitesimally small value, and the intensity value $\sigma_{S}$ of the white noise process from the natural deathrate in the susceptible population is significant and also finite in size, meanwhile the intensities, $\sigma_{i}, i= E, I, R$, of the white noise processes from the natural deathrates of the exposed, infectious and removal populations are either significant and finite in size, or they  continuously decrease in size to infinitesimally small values, it follows that the system (\ref{ch1.sec0.eq8})-(\ref{ch1.sec0.eq11}) does not have any disease-free steady state for the population. Nevertheless, the solutions of the system (\ref{ch1.sec0.eq8})-(\ref{ch1.sec0.eq11}) continue to oscillate near the potential disease-free steady state $E_{0}=(S^{*}_{0},0,0), S^{*}_{0}=\frac{B}{\mu} $ obtained from Theorem~\ref{ch1.sec2.thm0}[1.]-[2.], whenever the threshold conditions $\tilde{R}_{1}\leq 1$, $\tilde{U}_{0}\leq 1$, and $\tilde{V}_{0}\leq 1$  in (\ref{ch1.sec2-2.thm4.eq1})-(\ref{ch1.sec2-2.thm4.eq3}) are satisfied.  Moreover, the size or amplitude of the oscillations of the solutions of (\ref{ch1.sec0.eq8})-(\ref{ch1.sec0.eq11}) relative to the disease free steady state $E_{0}=(S^{*}_{0},0,0), S^{*}_{0}=\frac{B}{\mu} $, is proportional to the size of the intensity $\sigma_{S}$. This implies that small values of $\sigma_{S}$ result in small asymptotic expected average distance between the solutions of (\ref{ch1.sec0.eq8})-(\ref{ch1.sec0.eq11}) and the potential disease free steady state $E_{0}=(S^{*}_{0},0,0), S^{*}_{0}=\frac{B}{\mu} $ and vice versa.

  This result suggests that in a disease scenario where the disease outbreak results in random fluctuations in the disease transmission and natural death rates, when the random environmental fluctuations in the natural deathrate of the susceptible population has a significant and finite intensity value $\sigma_{S}$,    the  disease-free steady state population exists no where. Nevertheless, the states of the population ( which includes all subclasses- $S, E, I, R$) will oscillate over time near the potential disease-free steady state population  $E_{0}=(S^{*}_{0},0,0), S^{*}_{0}=\frac{B}{\mu} $, whenever the threshold conditions $\tilde{R}_{1}\leq 1$, $\tilde{U}_{0}\leq 1$, and $\tilde{V}_{0}\leq 1$  in (\ref{ch1.sec2-2.thm4.eq1})-(\ref{ch1.sec2-2.thm4.eq3}) are satisfied. Moreover, for small  values of the intensity $\sigma_{S}$, the size of the oscillations of the population relative to $E_{0}$ is small  proportionately to the size of  $\sigma_{S}$.

  This implies that in this disease scenario, whereas the disease can not be eradicated by applying the threshold conditions in Theorem~\ref{ch1.sec2.theorem1}, it follows that when the intensity value, $\sigma_{S}$, of the random fluctuations in the natural deathrate  of the susceptible population is small in size, the  population states affected by the disease outbreak can be contained closely to the potential disease-free steady state $E_{0}=(S^{*}_{0},0,0), S^{*}_{0}=\frac{B}{\mu} $, whenever the threshold values $\tilde{R}_{1}$, $\tilde{U}_{0}$, and $\tilde{V}_{0}$   satisfy the  threshold conditions $\tilde{R}_{1}\leq 1$, $\tilde{U}_{0}\leq 1$, and $\tilde{V}_{0}\leq 1$.

   In addition, the states of the population will continue to oscillate over time near the potential disease-free steady state $E_{0}$ , regardless of the size of the  intensities $\sigma_{i}, i= E, I, R, \beta$ of the random fluctuations in the disease transmission rate, or in the natural death rates of the exposed,  infectious and removal individuals in the population, provided that the intensity values  $\sigma_{i}, i= E, I, R, \beta$ are small in magnitude.
  \end{rem}
 \begin{thm}\label{ch1.sec2-2.thm5}
 If Hypothesis~\ref{ch1.sec2-2.hypothesis1}[$H_{5}$] holds, then there exists a disease free equilibrium population $E_{0}=(S^{*}_{0},0,0), S^{*}_{0}=\frac{B}{\mu} $ for the stochastic system (\ref{ch1.sec0.eq8})-(\ref{ch1.sec0.eq11}). But,
 the disease free equilibrium state is stochastically  unstable  in $D(\infty)$.
 \end{thm}
 Proof:
 Let $\sigma_{\beta}=\theta(\frac{1}{\epsilon})$,  $\sigma_{S}=0(\epsilon)$, $\sigma_{i}=0(\epsilon), \forall i= E, I, R$. It follows from the Theorem~\ref{ch1.sec2.thm0}[1.-2.] that (\ref{ch1.sec0.eq8})-(\ref{ch1.sec0.eq11}) has a disease free steady state given by $E_{0}=(S^{*}_{0},0,0), S^{*}_{0}=\frac{B}{\mu} $. Furthermore, for any Lyapunov functional $V$, the differential operator $LV$ associated with the Ito-Doob type system (\ref{ch1.sec0.eq8})-(\ref{ch1.sec0.eq11}) has the following structure:
 \begin{equation}\label{ch1.sec2-2.thm5.eq1}
   LV(x)=V_{t}(x)+V_{x}(x)\mathbf{\vec{f}(x)}+\frac{1}{2}\mathbf{\vec{g}^{T}(x)}V_{xx}(x)\mathbf{\vec{g}(x)},
 \end{equation}
 where $\mathbf{\vec{f}(x)}$ and $\mathbf{\vec{g}(x)}$ are vectors representing the drift and diffusion parts of the system (\ref{ch1.sec0.eq8})-(\ref{ch1.sec0.eq11}). It is easy to see that under the assumption of  $\sigma_{\beta}=\theta(\frac{1}{\epsilon})$, the diffusion part $\mathbf{\vec{g}(x)}=\theta(\frac{1}{\epsilon})$. Consequently, it follows from (\ref{ch1.sec2-2.thm5.eq1}) that $LV(x)=\theta(\frac{1}{\epsilon}), \forall V$. It follows further from Lyapunov stability comparative results that the steady state $E_{0}$ is stochastically unstable.
 \begin{thm}\label{ch1.sec2-2.thm6}
 If Hypothesis~\ref{ch1.sec2-2.hypothesis1}[$H_{6}$] holds, then there exists a disease-free equilibrium population $E_{0}=(S^{*}_{0},0,0), S^{*}_{0}=\frac{B}{\mu} $ for the stochastic system (\ref{ch1.sec0.eq8})-(\ref{ch1.sec0.eq11}). But,
 the disease free equilibrium state is stochastically  unstable  in $D(\infty)$.
 \end{thm}
 Proof:
 Let $\sigma_{\beta}=\theta(\frac{1}{\epsilon})$,  $\sigma_{S}=0(\epsilon)$, $\sigma_{i}=\theta(\frac{1}{\epsilon}), \forall i= E, I, R$. It follows from the Theorem~\ref{ch1.sec2.thm0}[1.-2.] that (\ref{ch1.sec0.eq8})-(\ref{ch1.sec0.eq11}) has a disease free steady state given by $E_{0}=(S^{*}_{0},0,0), S^{*}_{0}=\frac{B}{\mu} $. Furthermore, for any Lyapunov functional $V$, the differential operator $LV$ associated with the Ito-Doob type system (\ref{ch1.sec0.eq8})-(\ref{ch1.sec0.eq11}) has the following structure:
 \begin{equation}\label{ch1.sec2-2.thm6.eq1}
   LV(x)=V_{t}(x)+V_{x}(x)\mathbf{\vec{f}(x)}+\frac{1}{2}\mathbf{\vec{g}^{T}(x)}V_{xx}(x)\mathbf{\vec{g}(x)},
 \end{equation}
 where $\mathbf{\vec{f}(x)}$ and $\mathbf{\vec{g}(x)}$ are vectors representing the drift and diffusion parts of the system (\ref{ch1.sec0.eq8})-(\ref{ch1.sec0.eq11}). It is easy to see that under the assumption of  $\sigma_{\beta}=\theta(\frac{1}{\epsilon})$, the diffusion part $\mathbf{\vec{g}(x)}=\theta(\frac{1}{\epsilon})$. Consequently, it follows from (\ref{ch1.sec2-2.thm6.eq1}) that $LV(x)=\theta(\frac{1}{\epsilon}), \forall V$. Therefore, from Lyapunov stability comparative results,  the steady state $E_{0}$ is stochastically unstable.
 \begin{rem}
 Theorem~\ref{ch1.sec2-2.thm5} and Theorem~\ref{ch1.sec2-2.thm6} signify that when the stochastic system (\ref{ch1.sec0.eq8})-(\ref{ch1.sec0.eq11}) is continuously influenced by white noise processes from the disease transmission and natural death rates, where the intensity value $\sigma_{S}$ of the white noise process from the natural deathrate of the susceptible population continuously decreases in size to infinitesimally small values, then the system  has a disease free steady state population given by $E_{0}=(S^{*}_{0},0,0), S^{*}_{0}=\frac{B}{\mu} $, regardless of the sizes of the intensities $\sigma_{i}, i= E, I, R, \beta$ of the white noise processes from the disease transmission rate and natural deathrates of the exposed, infectious and removal populations. But the disease free steady state population $E_{0}$ that exists is clearly stochastically unstable, whenever  the intensities $\sigma_{i}, i= E, I, R, \beta$ of the white noise processes from the disease transmission rate or the natural deathrates of the exposed, infectious and removal populations continuously increase in size to sufficiently large values.

   This result suggests that in a disease scenario where the disease outbreak results in random fluctuations in the disease transmission rate and natural deathrates, there exists a disease free steady state for the population given by $E_{0}=(S^{*}_{0},0,0), S^{*}_{0}=\frac{B}{\mu} $, whenever the random environmental fluctuations in the natural deathrate  of the susceptible population have infinitesimally small intensity values, regardless of the sizes of the  intensity values of the random environmental fluctuations in the disease transmission rate and the natural deathrates of the exposed, infectious and removal individuals in the population. However,  the sufficiently large intensity values for the random fluctuations in the disease transmission rate or the natural deathrates of the exposed, infectious and removal populations quickly "drive" all sample paths of the different population states  away from the disease free population steady state $E_{0}$, and consequently adversely favoring conditions that allow the disease to establish an endemic stable steady state or a stable significant number of the disease related classes-infectious, exposed and removal individuals in the population. This result further suggests that significantly high intensity values for random fluctuations in the disease transmission rate and the natural deathrates of the exposed, infectious and removal populations exert strong negative conditions against the disease eradication process. Numerical simulation results in Section~\ref{ch1.sec4} show that the high intensity values for the random fluctuations in the disease transmission rate or the natural deathrates of the exposed, infectious and removal populations lead to a general decrease in the average total population size over time, which in some cases may lead to the population extinction for sufficiently large  intensity values.
    \end{rem}
 The following result describes the behavior of the solutions of the system (\ref{ch1.sec0.eq8})-(\ref{ch1.sec0.eq11}) under the assumptions of Hypothesis~\ref{ch1.sec2-2.hypothesis1}[$H_{7}$]. For simplicity only the special case of  $\sigma_{\beta} =0(1)$, and $\sigma_{S}=\theta(\frac{1}{\epsilon})$, and  $\sigma_{i} =\theta(\frac{1}{\epsilon}),  \forall i= E, I, R$. The results for the other cases can be similarly derived.
 \begin{thm}\label{ch1.sec2-2.thm7}
 If Hypothesis~\ref{ch1.sec2-2.hypothesis1}[$H_{7}$] holds, then there is no  disease free equilibrium population for the stochastic system (\ref{ch1.sec0.eq8})-(\ref{ch1.sec0.eq11}). Furthermore, when $\sigma_{\beta} =0(1)$,  $\sigma_{S}=\theta(\frac{1}{\epsilon})$, and  $\sigma_{i} =\theta(\frac{1}{\epsilon}),  \forall i= E, I, R$,  the system does not oscillate in the neighborhood of the potential disease free equilibrium  $E_{0}=(S^{*}_{0},0,0), S^{*}_{0}=\frac{B}{\mu} $ obtained from Theorem~\ref{ch1.sec2.thm0}[1.-2.].
 \end{thm}
 Proof:
 Let  $\sigma_{\beta} =0(1)$, and $\sigma_{S}=\theta(\frac{1}{\epsilon})$, and  $\sigma_{i} =\theta(\frac{1}{\epsilon}),  \forall i= E, I, R$. It follows from the Theorem~\ref{ch1.sec2.thm0}[3.] that (\ref{ch1.sec0.eq8})-(\ref{ch1.sec0.eq11}) does not have a disease free steady state.  Furthermore, the rest of the result follows immediately from the Proof of Theorem~\ref{ch1.sec2.theorem2}.
 \begin{rem}
 Theorem~\ref{ch1.sec2-2.thm7}  signifies that when the stochastic system (\ref{ch1.sec0.eq8})-(\ref{ch1.sec0.eq11}) is continuously influenced by the white noise processes from the disease transmission rate and natural deathrates, where the intensity value $\sigma_{S}$ of the white noise process from the natural deathrate of the susceptible population continuously increases in size to sufficiently large values, then it follows from Theorem~\ref{ch1.sec2.thm0}[3.] that the system (\ref{ch1.sec0.eq8})-(\ref{ch1.sec0.eq11}) does not have a  disease free steady state population. Moreover, the significantly large intensity values $\sigma_{i}, i= E, I, R$ of the white noise processes from the natural deathrates  of the exposed, infectious or removal populations lead to significantly large threshold values $\tilde{R}_{1}$, $\tilde{U}_{0}$, and $\tilde{V}_{0}$ in (\ref{ch2.sec2.thm2.eq1})-(\ref{ch2.sec2.thm2.eq3}), that violate the threshold conditions,  $\tilde{R}_{1}\leq 1$, $\tilde{U}_{0}\leq 1$, and $\tilde{V}_{0}\leq 1$ in Theorem~\ref{ch1.sec2.theorem2}. This implies that the solutions of the system (\ref{ch1.sec0.eq8})-(\ref{ch1.sec0.eq11}) do no longer oscillate near the disease free steady state $E_{0}$ obtained from Theorem~\ref{ch1.sec2.thm0}[1.-2.].

 Furthermore, for $\sigma_{S}=\theta(\frac{1}{\epsilon})$, it follows from (\ref{ch2.sec2.thm2.eq4}) that the asymptotic expected average distance between the solutions of (\ref{ch1.sec0.eq8})-(\ref{ch1.sec0.eq11}) and potential disease free equilibrium state $E_{0}$ obtained from Theorem~\ref{ch1.sec2.thm0}[1.-2.] is of the order $\theta(\frac{1}{\epsilon})$. This also implies that the sufficiently large intensity values of $\sigma_{S}$ lead to oscillations in the disease dynamics with very large oscillation sizes for the sample paths of the different states of the  population over time. In addition, the sample paths of the different states of the population oscillate over time at  significantly large distances away from the potential disease-free equilibrium state $E_{0}=(S^{*}_{0},0,0), S^{*}_{0}=\frac{B}{\mu} $ obtained from Theorem~\ref{ch1.sec2.thm0}[1.-2.].

  Thus, the disease can never be eradicated under the conditions of Theorem~\ref{ch1.sec2-2.thm7}. Moreover, the numerical simulation results in Section~\ref{ch1.sec4} show that the high intensity values, $\sigma_{i}, i=E, I, R, \beta$,  for the random fluctuations in the disease transmission rate or the natural deathrates of the susceptible, exposed, infectious and removal populations lead to a general decrease in the average total population size over time, which in some cases may lead to the population extinction for sufficiently large  intensity values.

  This result suggests that in a disease scenario where the disease outbreak results in random fluctuations in the disease transmission rate and natural deathrates, it follows that when the random environmental fluctuations in the natural deathrate of the susceptible population exhibit sufficiently large intensity values, then     the population does not exhibit  a disease-free population steady state. Furthermore, when the intensity value of the random fluctuations in the natural death rates  of the exposed, or infectious or removal populations is also sufficiently large in size, then the different states of the  population oscillates over time at a  farther  distance away from the potential disease free population steady state  $E_{0}=(S^{*}_{0},0,0), S^{*}_{0}=\frac{B}{\mu} $ wherein the disease would be eradicated,  regardless of the size of the intensity value, $\sigma_{\beta}$, of the random fluctuations in the disease transmission rate. This implies that  in this disease scenario, the population experiencing the disease outbreak cannot contain the disease.
  \end{rem}
 \section{Example}\label{ch1.sec4}
 \subsection{Example 1: The effect of the intensity of the white noise process on disease eradication: }
 This example illustrates the results in Theorem~\ref{ch1.sec2.theorem1} and Lemma~\ref{ch1.sec2.lemma2}, and also provides numerical evidence in support of the results in Section~\ref{ch1.sec2-2} that characterize the effects of the intensity of the white noise processes in the system originating from the random fluctuations in the disease dynamics on the stochastic asymptotic stability of the  system in relation to the disease free equilibrium $E_{0}=(S^{*}_{0},0,0), S^{*}_{0}=\frac{B}{\mu} $. Recall, Theorem~\ref{ch1.sec2.theorem1} and Lemma~\ref{ch1.sec2.lemma2} provide conditions for the threshold values $R_{1}$, $U_{0}$, and $V_{0}$ defined in (\ref{ch2.sec2.thm1.eq5a})-(\ref{ch2.sec2.thm1.eq5c}) which are sufficient for the stochastic stability of $E_{0}$ and consequently for disease eradication.  For simplicity in this example, the following assumptions are considered:- ($a_{1}$) there are no random fluctuations in the disease dynamics due to the natural death of susceptible individuals, that is, the intensity of the white noise due to the random fluctuations in the natural death of susceptible individuals $\sigma_{S}=0$. Indeed, from Theorem~\ref{ch1.sec2.theorem1} and Lemma~\ref{ch1.sec2.lemma2}, there exists a stable disease free equilibrium $E_{0}$, whenever $\sigma_{S}=0$ and the threshold values satisfy $R_{1}\leq 1$, $U_{0}\leq 1$, and $V_{0}\leq 1$. ($a_{2}$) It is also assumed that the intensities of the white noise processes in the system due to the random fluctuations in the natural death  and disease transmission processes for the other disease classes-exposed, infectious and removal are equal,  that is, $\sigma_{E}=\sigma_{I}=\sigma_{R}=\sigma_{\beta}=\sigma$.

\begin{figure}[H]
  \centering
  \includegraphics[height=8cm]{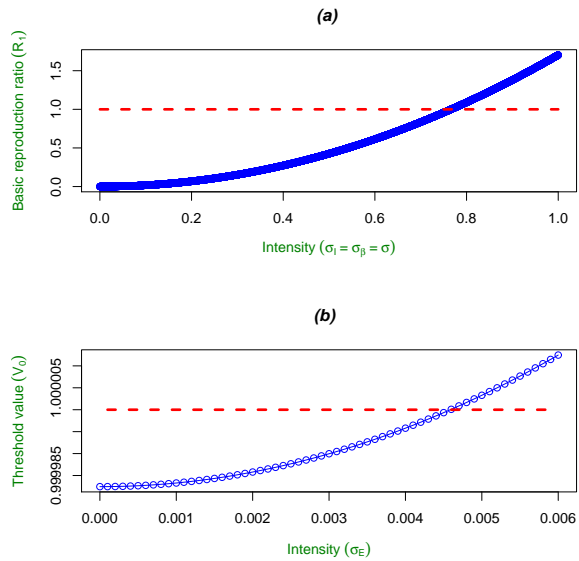}
  \caption{(a) and (b) Show the values of the noise modified basic reproduction number, $R_{1}$, (defined in (\ref{ch2.sec2.thm1.eq5a})) and the threshold parameter $V_{0}$ (defined in (\ref{ch2.sec2.thm1.eq5c})) over continuous changes in the values of the intensities of white noise processes due to random fluctuations in natural death and disease transmission processes of exposed, infectious and removal individuals, that is, $\sigma=\sigma_{E}=\sigma_{\beta}=\sigma_{I}=\sigma_{R}$.  The curves in (a) and (b) show the values of $R_{1}$ and $V_{0}$ respectively. In addition, the broken horizontal lines depict the threshold mark, $1$, for the threshold values  $R_{1}$ and $V_{0}$, where  for the values of $R_{1}$ and $V_{0}$ below the threshold mark $1$, the disease free equilibrium $E_{0}$ is stochastically asymptotically stable, and the  disease can consequently be eradicated.   It is easy to see that low values of $\sigma\in[0, 0.7661]$ lead to $R_{1}\leq 1$, and $R_{1}> 1$ other wise. For $V_{0}$, the low values of $\sigma\in[0, 0.0045]$ lead to $V_{0}\leq 1$, and $V_{0}> 1$ other wise. Therefore, values for $R_{1}$, $U_{0}$, and $V_{0}$ that satisfy $R_{1}\leq 1$, $U_{0}\leq 1$, and $V_{0}\leq 1$ are achieved for very low values of $\sigma$. This observation signifies that for a disease scenario where the physical processes lead to the specific parameter values defined in Table~\ref{ch1.sec4.table1}, the disease can only be eradicated when the random fluctuations in the disease dynamics exhibit very low intensity values of $\sigma\in[0, 0.0045]$. For any intensity values higher than $0.0045$, the disease-free equilibrium $E_{0}$ is unstable, and this signifies that the disease outbreak becomes naturally uncontrollable and establishes a stable endemic population. Note that the observations of this example are consistent with the results in Theorems~[\ref{ch1.sec2-2.thm1}-\ref{ch1.sec2-2.thm3},$\&$ \ref{ch1.sec2-2.thm5}-\ref{ch1.sec2-2.thm6}].}\label{ch1.sec4.figure1}
\end{figure}

  The list of system parameter values in Table~\ref{ch1.sec4.table1} are used to generate different values for $R_{1}$, $U_{0}$, and $V_{0}$ under continuous changes in the values of $\sigma=\sigma_{E}=\sigma_{I}=\sigma_{R}=\sigma_{\beta}$. The Figure~\ref{ch1.sec4.figure1} depicts the results for $R_{1}$  and $V_{0}$. For $U_{0}$, it follows from Table~\ref{ch1.sec4.table1} and (\ref{ch2.sec2.thm1.eq5b}) that $U_{0}\approx 1$, where $\tilde{K}(\mu)=0.999991$.
 \begin{table}[h]
  \centering
  \caption{A list of specific values chosen for the system parameters for Example 1. }\label{ch1.sec4.table1}
  \begin{tabular}{l l l}
  Disease transmission rate&$\beta$& $6.277E-66$\\\hline
  Constant Birth rate&$B$&$ \frac{22.39}{1000}$\\\hline
  Recovery rate& $\alpha$& $5.5067E-07$\\\hline
  Disease death rate& $d$& 0.11838\\\hline
  Natural death rate& $\mu$& $0.6$\\\hline
  \end{tabular}
\end{table}
 \subsection{Example 2: Effect of the intensity of white noise on the trajectories of the system  }
\begin{figure}[H]
\begin{center}
\includegraphics[height=8cm]{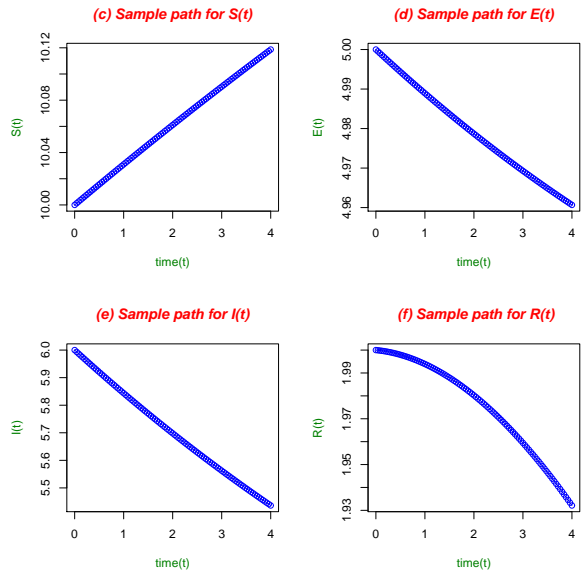}
\caption{(c), (d), (e) and (f) show the trajectories of the disease classes $(S,E,I,R)$ respectively, when there are only infinitesimally small random fluctuations in the disease dynamics, that is,  when the intensities of the white noise processes in the system due to random fluctuations in the natural death  and disease transmission processes in all the classes $(S,E,I,R)$ are described as follows:   $\sigma_{S}=\sigma_{E}=\sigma_{\beta}=\sigma_{I}=\sigma_{R}=0(\epsilon)$.}\label{ch1.sec4.figure 2}
\end{center}
\end{figure}

\begin{figure}[H]
\begin{center}
\includegraphics[height=8cm]{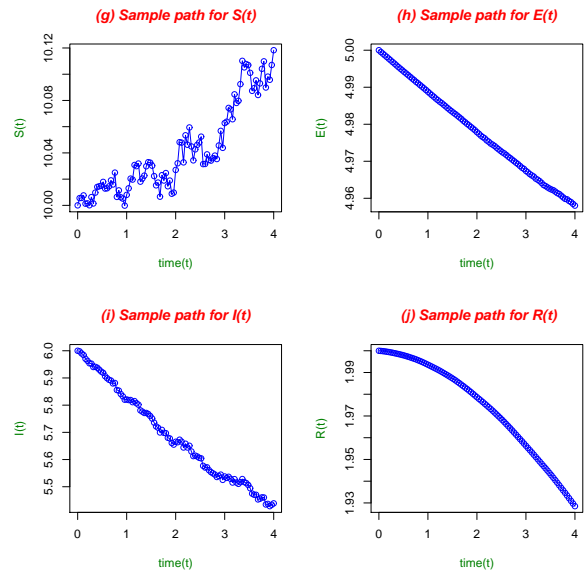}
\caption{(g), (h), (i) and (j) show the trajectories of the disease classes $(S,E,I,R)$ respectively, when there are only infinitesimally small random fluctuations in the disease dynamics from the natural death of the classes (S,E,I,R), that is, when  $\sigma_{S}=\sigma_{E}=\sigma_{I}=\sigma_{R}=0(\epsilon)$, but there are random fluctuations in the disease transmission process with low intensity value of $\sigma_{\beta}=0.5$.}\label{ch1.sec4.figure 3}
\end{center}
\end{figure}

\begin{figure}[H]
\begin{center}
\includegraphics[height=8cm]{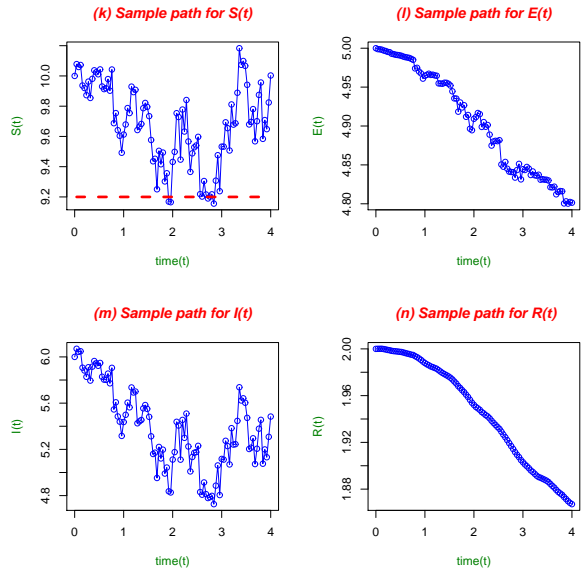}
\caption{(k), (l), (m) and (n) show the trajectories of the disease classes $(S,E,I,R)$ respectively, when there are only infinitesimally small random fluctuations in the disease dynamics from the natural death of the classes (S,E,I,R), that is, when  $\sigma_{S}=\sigma_{E}=\sigma_{I}=\sigma_{R}=0(\epsilon)$, but there are random fluctuations in the disease transmission process with high intensity value of $\sigma_{\beta}=9$. In addition, the broken line on the sample path for $S(t)$ in (k) depicts the $S$-coordinate $S^{*}_{0}=\frac{B}{\mu}=9.199999$ for the disease free steady state $E_{0}=(S^{*}_{0},0,0,0), S^{*}_{0}=\frac{B}{\mu}, E^{*}_{0}=0, I^{*}_{0}=0, R^{*}_{0}=0$. }\label{ch1.sec4.figure 4}
\end{center}
\end{figure}

\begin{figure}[H]
\begin{center}
\includegraphics[height=8cm]{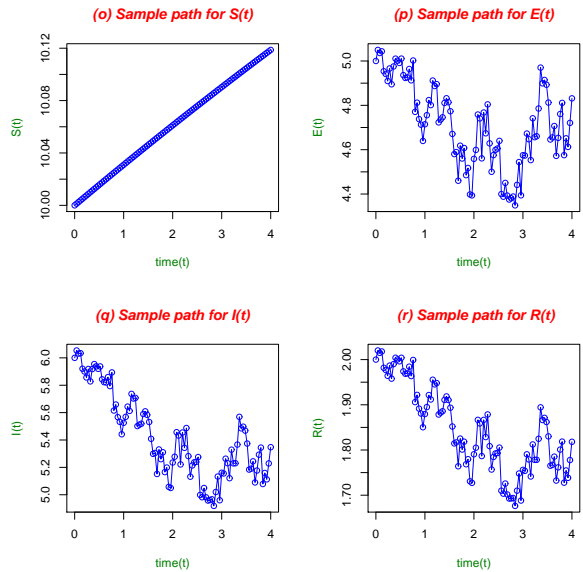}
\caption{(o), (p), (q) and (r) show the trajectories of the disease classes $(S,E,I,R)$ respectively, when there are significant small random fluctuations in the disease dynamics from the natural death process of exposed, infectious and removal classes, with intensity  value $\sigma_{E}=\sigma_{I}=\sigma_{R}=0.5$, but there are only infinitesimally small fluctuations in the disease dynamics due to the  disease transmission and natural death processes of susceptible individuals, that is,  $\sigma_{S}=\sigma_{\beta}=0(\epsilon)$.}\label{ch1.sec4.figure 5}
\end{center}
\end{figure}

\begin{figure}[H]
\begin{center}
\includegraphics[height=8cm]{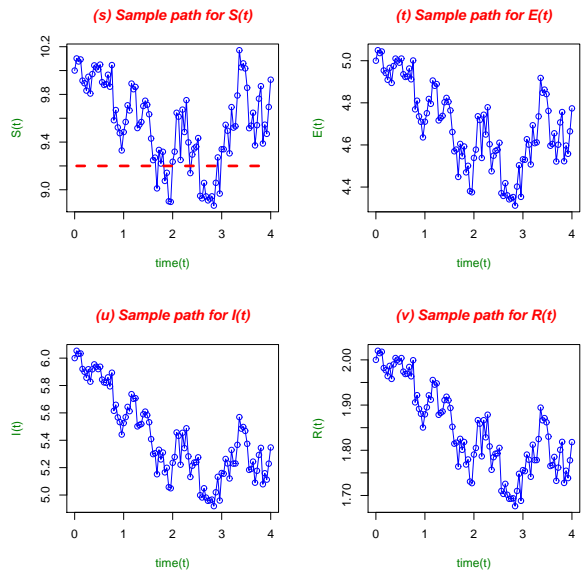}
\caption{(s), (t), (u) and (v) show the trajectories of the disease classes $(S,E,I,R)$ respectively, when there are significant but small random fluctuations in the disease dynamics from the natural death process in all the disease classes- susceptible, exposed, infectious and removal classes with low intensity value of  $\sigma_{S}=\sigma_{E}=\sigma_{I}=\sigma_{R}=0.5$, but there are infinitesimally small fluctuations in the disease dynamics from the disease transmission process,  that is, $\sigma_{\beta}=0(\epsilon)$. In addition, the broken line on the sample path for $S(t)$ in (s) depicts the $S$-coordinate $S^{*}_{0}=\frac{B}{\mu}=9.199999$ for the disease free steady state $E_{0}=(S^{*}_{0},0,0,0), S^{*}_{0}=\frac{B}{\mu}, E^{*}_{0}=0, I^{*}_{0}=0, R^{*}_{0}=0$.}\label{ch1.sec4.figure 6}
\end{center}
\end{figure}

\begin{figure}[H]
\begin{center}
\includegraphics[height=8cm]{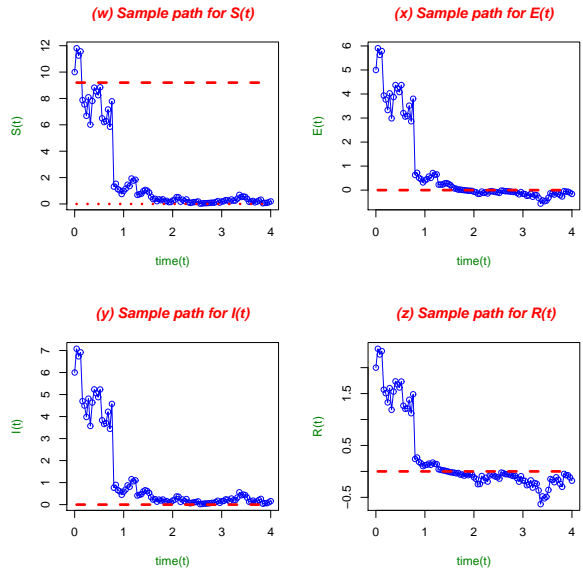}
\caption{(w), (x), (y) and (z) show the trajectories of the disease classes $(S,E,I,R)$ respectively, when there are significant and large random fluctuations in the disease dynamics from the natural death process in all the disease classes- susceptible, exposed, infectious and removal classes with sufficiently high intensity value of  $\sigma_{S}=\sigma_{E}=\sigma_{I}=\sigma_{R}=9$, but there are infinitesimally small fluctuations in the disease dynamics from the disease transmission process,  that is, $\sigma_{\beta}=0(\epsilon)$. In addition, the broken line on the sample paths for $S(t)$, $E(t)$, $I(t)$ and $R(t)$  depict the $S, E, I, R$-coordinates $S^{*}_{0}=\frac{B}{\mu}=9.199999, E^{*}_{0}=0, I^{*}_{0}=0, R^{*}_{0}=0$ for the disease free steady state $E_{0}=(S^{*}_{0},0,0,0), S^{*}_{0}=\frac{B}{\mu}, E^{*}_{0}=0, I^{*}_{0}=0, R^{*}_{0}=0$. (w), (x), (y) and (z) also show that the population goes extinct over time due to the high intensity of the white noise.}\label{ch1.sec4.figure 7}
\end{center}
\end{figure}

\begin{figure}[H]
\begin{center}
\includegraphics[height=8cm]{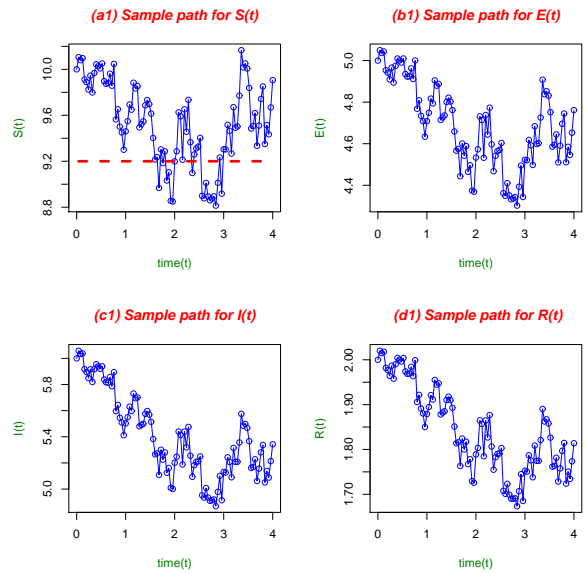}
\caption{(a1), (b1), (c1) and (d1) show the trajectories of the disease classes $(S,E,I,R)$ respectively, when there are significant but small random fluctuations in the disease dynamics from the natural death process in all the disease classes- susceptible, exposed, infectious and removal classes with low intensity value of  $\sigma_{S}=\sigma_{E}=\sigma_{I}=\sigma_{R}=0.5$,  and there are also significant fluctuations in the disease dynamics from the disease transmission with a low intensity value of $\sigma_{\beta}=0.5$. In addition, the broken line on the sample path for $S(t)$ in (a1) depicts the $S$-coordinate $S^{*}_{0}=\frac{B}{\mu}=9.199999$ for the disease free steady state $E_{0}=(S^{*}_{0},0,0,0), S^{*}_{0}=\frac{B}{\mu}, E^{*}_{0}=0, I^{*}_{0}=0, R^{*}_{0}=0$.}\label{ch1.sec4.figure 8}
\end{center}
\end{figure}

\begin{figure}[H]
\begin{center}
\includegraphics[height=8cm]{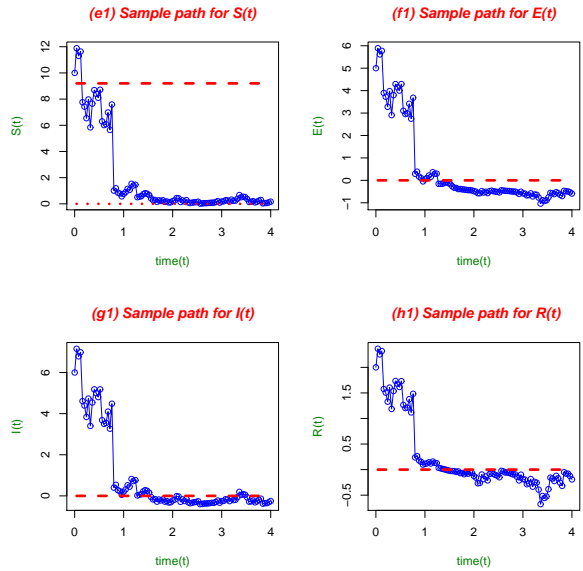}
\caption{(e1), (f1), (g1) and (h1) show the trajectories of the disease classes $(S,E,I,R)$ respectively, when there are significant and large random fluctuations in the disease dynamics from the natural death process in all the disease classes- susceptible, exposed, infectious and removal classes with a sufficiently high intensity value of  $\sigma_{S}=\sigma_{E}=\sigma_{I}=\sigma_{R}=9$,  and there are also significant fluctuations in the disease dynamics from the disease transmission process with a sufficiently high intensity value of $\sigma_{\beta}=9$. In addition, the broken line on the sample paths for $S(t)$, $E(t)$, $I(t)$ and $R(t)$  depict the $S, E, I, R$-coordinates $S^{*}_{0}=\frac{B}{\mu}=9.199999, E^{*}_{0}=0, I^{*}_{0}=0, R^{*}_{0}=0$ for the disease free steady state $E_{0}=(S^{*}_{0},0,0,0), S^{*}_{0}=\frac{B}{\mu}, E^{*}_{0}=0, I^{*}_{0}=0, R^{*}_{0}=0$. Furthermore,  (e1), (f1), (g1) and (h1) show that the population goes extinct over time due to the high intensity of the white noise.}\label{ch1.sec4.figure 9}
\end{center}
\end{figure}

The list of convenient choice of parameter values in Table~\ref{ch1.sec4.table2} are used to generate the trajectories of the stochastic system (\ref{ch1.sec0.eq8})-(\ref{ch1.sec0.eq11}) in order to (1.) illustrate the impact of the source of the white noise process in the system (owing to the random fluctuations in  the natural death or disease transmission processes) and also to (2.) illustrate the effect of the intensity of the  white noise process in the system on the trajectories of the different disease classes in the system, in order to uncover the overall behavior of the system over time.
 \begin{table}[h]
  \centering
  \caption{A list of specific values chosen for the system parameters for Example 2.}\label{ch1.sec4.table2}
  \begin{tabular}{l l l}
  Disease transmission rate&$\beta$& 0.6277\\\hline
  Constant Birth rate&$B$&$ \frac{22.39}{1000}$\\\hline
  Recovery rate& $\alpha$& 0.05067\\\hline
  Disease death rate& $d$& 0.01838\\\hline
  Natural death rate& $\mu$& $0.002433696$\\\hline
  Incubation delay time in vector& $T_{1}$& 2 units \\\hline
  Incubation delay time in host& $T_{2}$& 1 unit \\\hline
  Immunity delay time& $T_{3}$& 4 units\\\hline
  \end{tabular}
\end{table}
The Euer-Maruyama stochastic approximation scheme\footnote{A seed is set on the random number generator to reproduce  the same sequence of random numbers for the Brownian motion in order to generate reliable graphs for the trajectories of the system under different intensity values for the white noise processes, so that comparison can be made to identify differences that reflect the effect of intensity values.} is used to generate trajectories for the different states $S(t), E(t), I(t), R(t)$ over the time interval $[0,T]$, where $T=\max(T_{1}+T_{2}, T_{3})=4$. The special nonlinear incidence  functions $G(I)=\frac{aI}{1+I}, a=0.05$ in \cite{gumel} is utilized to generate the numeric results. Furthermore, the following initial conditions are used
\begin{equation}\label{ch1.sec4.eq1}
\left\{
\begin{array}{l l}
S(t)= 10,\\
E(t)= 5,\\
I(t)= 6,\\
R(t)= 2,
\end{array}
\right.
\forall t\in [-T,0], T=\max(T_{1}+T_{2}, T_{3})=4.
\end{equation}
The sample means for the sample paths of the $S, E, I, R$ states generated over time $t\in [0, T]$  are summarized in Table~\ref{ch1.sec4.table3}, and will be used to compare the effect of the intensity values of the white noise processes in the system on the trajectories of the system over time.
\begin{table}[h]
  \centering
  \caption{ Shows the intensity values of the white noise processes in the system and the corresponding sample means for the trajectories of the $S, E, I, R$ states generated over time $t\in [0, 4]$ in Example 2. The sample means for $S, E, I, R$ are denoted $\bar{S}, \bar{E}, \bar{I}, \bar{R}$ respectively. }\label{ch1.sec4.table3}
  \begin{tabular}{l l l l l l}
  $\sigma_{i}, i= S, E, I, R, \beta$&Figure \# &$\bar{S}$& $\bar{E}$&$\bar{I}$&$\bar{R}$\\\hline
  $\sigma_{i}=O(\epsilon), i= S, E, I, R, \beta$&Figure~\ref{ch1.sec4.figure 2}&10.06048&4.979256&5.704827&1.975407\\\hline
  $\sigma_{i}=O(\epsilon), i= S, E, I, R$, and $\sigma_{\beta}=0.5$&Figure~\ref{ch1.sec4.figure 3}&10.04129&4.978257&5.687113&1.973783\\\hline
  $\sigma_{i}=O(\epsilon), i= S, E, I, R$, and $\sigma_{\beta}=9$&Figure~\ref{ch1.sec4.figure 4}&9.681482&4.906452&5.385973&1.94617\\\hline
  $\sigma_{i}=0.5, i=  E, I, R$, and $\sigma_{S}=\sigma_{\beta}=O(\epsilon)$&Figure~\ref{ch1.sec4.figure 5}&10.06048&4.715779&5.42661&1.845652\\\hline
  $\sigma_{i}=0.5, i= S, E, I, R$, and $\sigma_{\beta}=O(\epsilon)$&Figure~\ref{ch1.sec4.figure 6}&9.553725&4.692877&5.42661&1.845652\\\hline
  $\sigma_{i}=9, i= S, E, I, R$, and $\sigma_{\beta}=O(\epsilon)$&Figure~\ref{ch1.sec4.figure 7}&1.980488&0.8066963&1.200498&0.240599\\\hline
  $\sigma_{i}=0.5, i= S, E, I, R$, and $\sigma_{\beta}=0.5$&Figure~\ref{ch1.sec4.figure 8}&9.529665&4.687529&5.406493&1.843888\\\hline
  $\sigma_{i}=9, i= S, E, I, R$, and $\sigma_{\beta}=9$&Figure~\ref{ch1.sec4.figure 9}&1.88787&0.4633994&0.8659143&0.2315721\\\hline
    \end{tabular}
\end{table}
The following observations can be made from Table~\ref{ch1.sec4.table3}:
\begin{rem}\label{ch1.sec4.rem1}
\item[1.] When  $\sigma_{i}=O(\epsilon), i= S, E, I, R$, there is   moderate decrease in the average values $\bar{S}, \bar{E}, \bar{I}, \bar{R}$ of $S, E, I, R$ from the trajectories in Figures~\ref{ch1.sec4.figure 2}-\ref{ch1.sec4.figure 4} as $\sigma_{\beta}$ increases from $\sigma_{\beta}=O(\epsilon)$ to $\sigma_{\beta}=9$.
    \item[2.] When  $\sigma_{\beta}=O(\epsilon)$, there is  a sharp decrease in the average values $\bar{S}, \bar{E}, \bar{I}, \bar{R}$ of $S, E, I, R$ from the trajectories in Figure~\ref{ch1.sec4.figure 2}, and Figures~\ref{ch1.sec4.figure 6}-\ref{ch1.sec4.figure 7} as $\sigma_{i}, i= S, E, I, R$ increases from $\sigma_{i}=O(\epsilon), i= S, E, I, R$ to $\sigma_{i}=9, i= S, E, I, R$.
        \item[3.] When all the $\sigma_{i}$'s , that is,  $\sigma_{i}, i= S, E, I, R, \beta$ equally increase together from $\sigma_{i}=O(\epsilon), i= S, E, I, R, \beta$ to $\sigma_{i}=9, i= S, E, I, R, \beta$, there is  a sharper decrease in the average values $\bar{S}, \bar{E}, \bar{I}, \bar{R}$ of $S, E, I, R$ from the trajectories in Figure~\ref{ch1.sec4.figure 2},  and Figures~\ref{ch1.sec4.figure 8}-\ref{ch1.sec4.figure 9}.
            \item[4.] When  $\sigma_{S}=\sigma_{\beta}=O(\epsilon) $, there is no change in the average value $\bar{S}$ of $S$  and there is  moderate decrease in the average values $ \bar{E}, \bar{I}, \bar{R}$ of $ E, I, R$  from the trajectories in Figure~\ref{ch1.sec4.figure 2} and Figure~\ref{ch1.sec4.figure 5} as $\sigma_{i}, i=  E, I, R$ increases from $\sigma_{i}=O(\epsilon), i=  E, I, R$ to $\sigma_{i}=0.5, i=  E, I, R$.
    \end{rem}
The Figures~\ref{ch1.sec4.figure 2}-\ref{ch1.sec4.figure 4} can be used to examine the effect of increasing the intensity value of the white noise process, $\sigma_{\beta}$, originating from the random fluctuations in the disease transmission process on the trajectories for $(S,E,I,R)$ in the absence of any significant random fluctuations in the disease dynamics due to the natural death process for all the disease classes $(S,E,I,R)$, that is, $\sigma_{i}=(\epsilon), i=S,E,I,R$ .
It can be observed from Figure~\ref{ch1.sec4.figure 2} that when the intensity value  $\sigma_{\beta}$ is  infinitesimally small, that is, $\sigma_{\beta}=0(\epsilon)$,  no significant oscillations occur over time on the trajectories for $S, E, I, R$ in (a), (b), (c) and (d) respectively. Furthermore, for significant but low intensity values\footnote{That is, $\sigma_{\beta}=O(1)$.} for $\sigma_{\beta}$, that is, $\sigma_{\beta}=0.5$, Figure~\ref{ch1.sec4.figure 3} shows that some significant oscillations  occur on the trajectories for the susceptible (g) and infectious (i) populations. Moreover, the size of the oscillations observed on the trajectories for the susceptible (g) and infectious population (i) seem to be small in value over time compared to Figure~\ref{ch1.sec4.figure 4}. In addition, no significant oscillations are observed on the trajectories for the exposed (h) and removal (j) populations. In Figure~\ref{ch1.sec4.figure 4}, with an increase in the intensity value for $\sigma_{\beta}$ to $\sigma_{\beta}=9$, more disease classes exhibit significant oscillations on their trajectories, for instance, more significant sized oscillations are observed on the trajectory of one additional class- exposed population (l) than is observed in Figure~\ref{ch1.sec4.figure 3} (h). Moreover, it appears that the high intensity value $\sigma_{\beta}=9$  has increased  the size of the oscillations in the susceptible (k) and infectious  (m) populations,  and  further deviating  the trajectories of the system away from the noise-free state in Figure~\ref{ch1.sec4.figure 2}. In addition, the trajectories for the states- $(S,E,I)$ in Figure~\ref{ch1.sec4.figure 4} (k), (l), (m) respectively,  oscillate near the disease free state $E_{0}=(S^{*}_{0},0,0,0)$, where $ S^{*}_{0}=\frac{B}{\mu}=9.199999, E^{*}_{0}=0, I^{*}_{0}=0, R^{*}_{0}=0$.

 One can also observe from Table~\ref{ch1.sec4.table3} and Remark~\ref{ch1.sec4.rem1} that for $\sigma_{i}=O(\epsilon), i= S, E, I, R$, the average values of  $S,E,I,R$ over time on the trajectories in Figures~\ref{ch1.sec4.figure 2}-\ref{ch1.sec4.figure 4} decrease continuously with increase in the intensity value of $\sigma_{\beta}$ from $\sigma_{\beta}=O(\epsilon)$ to $\sigma_{\beta}=9$.
 These observations related to the oscillatory behavior of the system, for example, comparing the trajectory of $S$ in  Figure~\ref{ch1.sec4.figure 2}(c), Figure~\ref{ch1.sec4.figure 3}(g) and Figure~\ref{ch1.sec4.figure 4}(k), and also comparing the trajectory for $I$ in Figure~\ref{ch1.sec4.figure 2}(e), Figure~\ref{ch1.sec4.figure 3}(i) and Figure~\ref{ch1.sec4.figure 4}(m) suggest that continuously increasing the intensity value for $\sigma_{\beta}$ tends to increase the oscillatory behavior of the trajectories of the system that results in an average decrease in the size of the susceptible, exposed, infectious and removal populations over time. Furthermore, the size of the oscillations in the system is proportional to the size of the intensity values of the white noise process as remarked in the previous section.
Figure~\ref{ch1.sec4.figure 2} , Figure~\ref{ch1.sec4.figure 5}, and Figure~\ref{ch1.sec4.figure 6} can be used as an example to examine the effect of the intensity of the white noise process, $\sigma_{i}, i= S, E, I, R$, originating from the random fluctuations in the natural death process of each class-$S, E, I, R$, on the trajectories of the system, in the absence of any significant fluctuation in the disease dynamics owing to the disease transmission process, that is, $\sigma_{\beta}=O(\epsilon)$. For example, to examine the effect of $\sigma_{\beta}$ for the susceptible class, $S$, on the trajectories of the stochastic stochastic system, observe that in Figure~\ref{ch1.sec4.figure 5}, when $\sigma_{S}=\sigma_{\beta}=O(\epsilon)$ and $\sigma_{i}=0.5, i= E, I, R$, no significant oscillations occur on the trajectories of $S$ in Figure~\ref{ch1.sec4.figure 5}(o) and also on  Figure~\ref{ch1.sec4.figure 2}(c). Furthermore, when $\sigma_{S}$ is increased to $\sigma_{S}=0.5$,  Figure~\ref{ch1.sec4.figure 6}(s) depicts significant sized oscillations on the trajectory of $S$. Moreover, the trajectory for $S$ oscillates near the disease free steady state $S^{*}_{0}=9.199999$. It can be further observed using Table~\ref{ch1.sec4.table3} and Remark~\ref{ch1.sec4.rem1} that no major differences have occurred on the trajectories of the other states $E, I, R$ in both  Figure~\ref{ch1.sec4.figure 5}(p),(q),(r) and Figure~\ref{ch1.sec4.figure 6}(t),(u),(v) respectively. In addition, it can be seen from Table~\ref{ch1.sec4.table3} and Remark~\ref{ch1.sec4.rem1} that when $\sigma_{\beta}=O(\epsilon)$,  the increase in the intensity value of $\sigma_{S}$ from $\sigma_{S}=O(\epsilon)$ to $\sigma_{S}=0.5$ on average leads to a decrease in the susceptible population size over time in Figure~\ref{ch1.sec4.figure 6}(s) than it is observed in Figure~\ref{ch1.sec4.figure 5}(o) and  Figure~\ref{ch1.sec4.figure 2}(c).   These observations suggest that in the absence of random fluctuations in the disease dynamics from the  disease transmission process, that is, $\sigma_{\beta}=O(\epsilon)$, the intensity of the white noise process, $\sigma_{S}$, owing to the natural death of the susceptible class $S$, (1.) exhibits a significant effect primarily on its trajectory, and (2.) the effect of increasing the intensity value\footnote{that is $\sigma_{S}=\theta(\frac{1}{\epsilon})$} of $\sigma_{S}$ leads to  an oscillatory behavior on the trajectory of $S$ that  decreases the susceptible population averagely over time. Note that a similar numerical and graphical diagnostic approach can be used to examine the effects of the other classes $E, I, R$, whenever $\sigma_{\beta}=O(\epsilon)$.
Figure~\ref{ch1.sec4.figure 2}, Figure~\ref{ch1.sec4.figure 6} and Figure~\ref{ch1.sec4.figure 7} can be used to examine the effect of increasing the intensity value of the  white noise process originating from the natural death, $\sigma_{i}, i= S, E, I, R$, in the absence of any significant random fluctuations in the disease dynamics from the disease transmission process, that is, when $\sigma_{\beta}=O(\epsilon)$. Figure~\ref{ch1.sec4.figure 6} (s),(t),(u),(v) show that the trajectories for  $S, E, I, R$ respectively, oscillate near the disease free equilibrium $E_{0}=(9.199999, 0, 0, 0)$ over time when the intensity value is increased from $\sigma_{i}=O(\epsilon), i= S, E, I, R$ to $\sigma_{i}=0.5, i= S, E, I, R$ than is observed in the Figure~\ref{ch1.sec4.figure 2} (c), (d), (e), (f). Furthermore, the oscillations on the trajectories seem to be  small in size over time. When the intensity value,  $\sigma_{i}, i= S, E, I, R$, is  further increased to $\sigma_{i}=9, i= S, E, I, R$, the oscillations on the trajectories in Figure~\ref{ch1.sec4.figure 7} (w),(x),(y),(z), appear to have increased in size. Furthermore, Table~\ref{ch1.sec4.table3} and Remark~\ref{ch1.sec4.rem1} show that the oscillations lead to a decrease in the average values of $S, E, I, R$ over time, and  further away from the disease free state of $S^{*}_{0}=9.199999$. Moreover, the population rapidly becomes extinct over time. These observations suggest that the increase\footnote{That is, $\sigma_{i}=\theta(\frac{1}{\epsilon}), i= S, E, I, R$} in the intensity value of the white noise due to natural death in all classes, $\sigma_{i}, i= S, E, I, R$,  in the population (1.)leads to an increase in the oscillatory behavior of the system which decreases the population size averagely over time and also (2.)leads to population  extinction over time. Note that this observation is consistent with the results of Theorem~\ref{ch1.sec2-2.thm7}.
Figure~\ref{ch1.sec4.figure 2}, Figure~\ref{ch1.sec4.figure 8} and Figure~\ref{ch1.sec4.figure 9} can be used to examine the effect of increasing the intensity values, $\sigma_{i}, i= S, E, I, R,\beta$, of all the white noise processes in the system on the trajectories of the system. Figure~\ref{ch1.sec4.figure 8} (a1),(b1),(c1),(d1) show that the trajectories for $S, E, I, R$ respectively oscillate near the disease free steady state $E_{0}=(9.199999, 0, 0, 0)$ over time when the intensity value is increased  from $\sigma_{i}=O(\epsilon), i= S, E, I, R,\beta$ to $\sigma_{i}=0.5, i= S, E, I, R,\beta$ than it is observed in Figure~\ref{ch1.sec4.figure 2}(c),(d),(e),(f). Furthermore, the oscillations of the trajectories seem to be  small in size compared to the corresponding trajectories in Figure~\ref{ch1.sec4.figure 9}.  When the intensity values of $\sigma_{i}, i= S, E, I, R, \beta$ are  further increased to $\sigma_{i}=9, i= S, E, I, R$, it can be seen from  Figure~\ref{ch1.sec4.figure 9}(e1),(f1),(g1),(h1), Table~\ref{ch1.sec4.table3} and Remark~\ref{ch1.sec4.rem1} that the oscillations increase in size and lead to a sharp decrease in the average values  of $S, E, I, R$ on their trajectories over time, and also further deviating the average susceptible population size away from the disease free state of $S^{*}_{0}=9.199999$. Moreover, the population rapidly becomes extinct over time. These observations suggests that the increase in the intensity value of the white noise processes in the system due to the random fluctuations in the disease dynamics originating from the disease transmission  and natural death processes for all disease classes in the population leads to (1.) an increase in the oscillatory behavior of the system which decreases the average total population size over time, and also leads to (2.) the rapid extinction of the population over time.

It can also be observed by comparing Figure~\ref{ch1.sec4.figure 7}(w),(x),(y),(z), and  Figure~\ref{ch1.sec4.figure 9}(e1),(f1),(g1),(h1), that for a fixed value of $\sigma_{i}=9, i= S, E, I, R$,  if $\sigma_{\beta}$  increases from  $\sigma_{\beta}=O(\epsilon)$ in Figure~\ref{ch1.sec4.figure 7}(w),(x),(y),(z) to $\sigma_{\beta}=9$  in Figure~\ref{ch1.sec4.figure 9}(e1),(f1),(g1),(h1), then the population more rapidly becomes extinct than it is observed in Figure~\ref{ch1.sec4.figure 7}(w),(x),(y),(z). Indeed, in Figure~\ref{ch1.sec4.figure 7}(w),(x),(y),(z), the trajectories for the  susceptible $S$, exposed $E$, infectious $I$ and Removal $R$  states go extinct at approximately the following times $t=2, t=1.8, t=2$ and $t=1.8$ respectively. Meanwhile,  in Figure~\ref{ch1.sec4.figure 9}(e1),(f1),(g1),(h1), the trajectories for  susceptible $S$, exposed $E$, infectious $I$ and Removal $R$ go extinct earlier at the approximate times $t=1.5, t=1, t=1$ and $t=1.4$ respectively.  Note that these observations are consistent with the results of Theorem~\ref{ch1.sec2-2.thm7}.

 The following  pairs of figures:- (Figure~\ref{ch1.sec4.figure 3} (g),(h),(i),(j) \& Figure~\ref{ch1.sec4.figure 4} (k),(l),(m),(n))  and (  Figure~\ref{ch1.sec4.figure 6} (s),(t),(u),(v) \& Figure~\ref{ch1.sec4.figure 7} (w),(x),(y),(z)), can be used with reference to Figure~\ref{ch1.sec4.figure 2}, to examine and compare the two major sources of random fluctuations in the disease dynamics namely-natural death and disease transmission processes, in order to determine the source which has stronger effect on the trajectories of the system, whenever the intensity values of the white noise processes increase in value. In the absence of random fluctuations in the natural death process, that is, $\sigma_{i}=O(\epsilon), i= S, E, I, R$, as the intensity value of $\sigma_{\beta}$ is increased from $\sigma_{\beta}=0.5$ to $\sigma_{\beta}=9$,   the pair of figures (Figure~\ref{ch1.sec4.figure 3} (g),(h),(i),(j) \& Figure~\ref{ch1.sec4.figure 4} (k),(l), (m),(n)) show an increase in the oscillatory behavior on the trajectories of the system  which is more significant in size for  the $S$ and $I$ classes  over time. Furthermore, the oscillatory behavior leads to a decrease in the average susceptible and infectious populations over time than it is observed in Figure~\ref{ch1.sec4.figure 2} (c) and  Figure~\ref{ch1.sec4.figure 2}(e) respectively, as shown in Table~\ref{ch1.sec4.table3} and Remark~\ref{ch1.sec4.rem1}.   Moreover, the general disease population does not go extinct over time.

 Meanwhile, in the absence of random fluctuations in the disease transmission process, that is, $\sigma_{\beta}=O(\epsilon)$,  the increase in the intensity value of $\sigma_{i}, i= S, E, I, R$  from $\sigma_{i}=0.5, i= S, E, I, R$ to $\sigma_{i}=9, i= S, E, I, R$,   the pair of figures (  Figure~\ref{ch1.sec4.figure 6}(s),(t),(u),(v) \& Figure~\ref{ch1.sec4.figure 7}(w),(x),(y),(z)) show very strong increase in the oscillatory behavior on the trajectories of the system which is significant in all the states-  $S$, $E$, $I$ and $R$ . Furthermore, from Table~\ref{ch1.sec4.table3} and Remark~\ref{ch1.sec4.rem1}, it can be seen that the oscillatory behavior of the system leads to a rapid decrease in the average values of all the states-$S$, $E$, $I$ and $R$ over time, with the mean susceptible population size deviating much further away from the disease free steady state $S^{*}_{0}=9.199999$,  than it is observed in Figure~\ref{ch1.sec4.figure 2}.  Moreover, the disease population goes extinct over time with the increase in the intensity value of $\sigma_{i}, i= S, E, I, R$.
 \section{Conclusion}
 The presented class of stochastic SEIRS epidemic dynamic models with nonlinear incidence rates, three distributed delays and random perturbations characterizes the general dynamics of vector-borne diseases such as malaria and dengue fever, that are influenced by random environmental fluctuations from (1.) the disease transmission process between susceptible and infectious individuals, and also from (2.)  the natural death processes in the sub-categories - susceptible, exposed, infectious and removal individuals of the population. Moreover, the random fluctuations in the disease dynamics are incorporated into the epidemic dynamic models via  white noise or  Wiener processes.  Furthermore, the three delays are random variables. Whereas, two of the delays represent the incubation periods of the infectious agent in the vector and human hosts, the third delay represents the period of effective naturally acquired immunity  against the disease, that is conferred to individuals  after recovery from infection. The class of epidemic dynamic models is represented as a system of Ito-Doob type stochastic differential equations with a general nonlinear incidence function $G$. The nonlinear incidence function $G$ can be used to characterize the disease transmission rates for disease scenarios that exhibit a striking initial increase or decrease in disease transmission rates that becomes steady or bounded when the infectious population size is large.

 The existence of unique global positive solutions is validated for the stochastic dynamic system by applying the standard method of Lipschitz continuity, stopping times and energy functions. Moreover, a positive self-invariant set for the stochastic dynamic system is presented. Detailed results for the asymptotic behavior of the stochastic dynamic system are presented  namely:- (1.) the existence and asymptotic stochastic stability of a feasible disease-free equilibrium of the stochastic system, and (2.) the asymptotic oscillatory character of the solutions of the stochastic system near a potential disease-free equilibrium. The threshold values  for the stochastic stability of the disease free steady state, and for disease eradication, such as the basic reproduction number for the disease dynamics are computed.

 The analytic asymptotic results  for the epidemic dynamic system suggest that the source (disease transmission or natural death rates ) and size of the intensity values of the white noise processes in the  system exhibit direct consequences on the overall asymptotic behavior of the epidemic dynamic model with respect to the feasible or potential disease free population steady state for the epidemic dynamic model, and consequently on disease eradication. This observation leads to further thorough examination of the asymptotic properties of the stochastic epidemic dynamic system under various intensity levels of the white noise processes in the system. In addition, two numerical simulation examples are presented to justify the results from the analysis.

\end{document}